# There is Nothing Anomalous about "Anomalous" Underscreening in Concentrated Electrolytes

Sophie Baker[†,1], Gareth R. Elliott[†,1], Erica J. Wanless[1], Grant B. Webber[2], Vincent S. J. Craig[3], Alister J. Page*[1]

[1]Discipline of Chemistry, The University of Newcastle, Callaghan, New South Wales 2308, Australia

[2]Discipline of Chemical Engineering, The University of Newcastle, Callaghan, New South Wales 2308, Australia

[3]Department of Material Physics, Research School of Physics, Australian National University, Canberra, ACT 0200 Australia

[†]These authors contributed equally

*Corresponding author. Email: alister.page@newcastle.edu.au

**Abstract**

Over the last decade, experimental measurements of electrostatic screening lengths in concentrated electrolytes have exceeded theoretical predictions by orders of magnitude. This disagreement has led to a paradigm in which such screening lengths are referred to as "anomalous underscreening", while moderate screening lengths – predominantly those predicted by theory and molecular simulation – are referred to as "normal underscreening". Herein we use discrete Fourier analysis of the radial charge density obtained from molecular dynamics simulations to confirm the presence of many electrostatic screening modes present at any one time. We present a new approach for extracting screening lengths directly from the bulk charge density that reveals the origins of both normal and anomalous underscreening. These results reconcile a decades-old disagreement between experimental measurements and theoretical predictions of screening lengths in concentrated electrolytes.

**TOC Graphic**

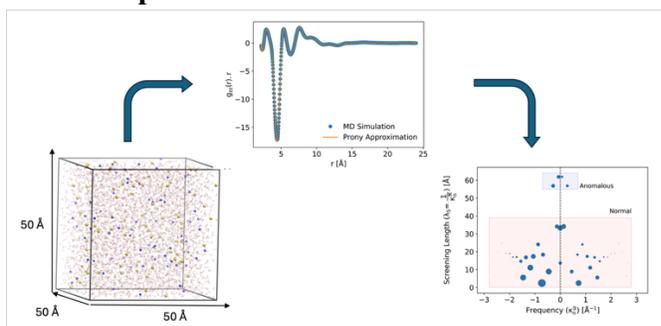

The addition of ions to a liquid causes a remarkable deviation in its behaviour and properties. Arguably the most fundamental of these is the electrostatic screening length – the length scale over which the electrostatic field around individual ions are 'screened' by the local electrolyte structure. For more than a century, this screening length has been understood via the seminal Debye-Hückel theory, which shows that in an infinitely dilute symmetric electrolyte (i.e. cations and anions are point charges with equal but opposite charge), the electrostatic potential $\psi(r)$ (interchangeably referred to as the potential of mean force) around an ion exhibits so-called 'Yukawa decay',

$$\psi(r) \sim \frac{q}{4\pi\varepsilon_0\varepsilon_r}\frac{e^{-\kappa r}}{r} \tag{1}$$

where

$$\kappa = \sqrt{\frac{q^2 \sum_j n_j Z_j^2}{\varepsilon_0 \varepsilon_r k_B T}} \tag{2}$$

Here, $\sum_j n_j Z_j^2$ is the ionic strength of the electrolyte comprising $n_j$ ions with formal charge $Z_j$ and $\varepsilon_r$ is the dielectric constant. The so-called 'Debye length' – i.e. the electrostatic screening length – is the inverse of $\kappa$, i.e. $\lambda_D \equiv \kappa^{-1}$. From Equation (2) it is immediate that, as ionic strength increases, the Debye length becomes smaller.

Debye-Hückel theory has been a staple of undergraduate chemistry curricula for decades. However, the assumptions underpinning the Debye-Hückel theory are invalid at concentrations above ~0.1-0.5 M.[1] Improvements to this theory that take into account relevant phenomena, such as ion size, and the molecular nature of the bulk solvent have therefore been sought since the 1930s. Kirkwood was the first to show that, at sufficiently high ionic strength, the electrostatic potential decay switches from the simple exponential Yukawa form shown in Equation (1), to one that is oscillatory in nature.[2] The threshold concentration at which this occurs is now known as the Kirkwood point. In some theoretical treatments, an ansatz comprising multiple decay modes for concentrations above the Kirkwood point has also been considered.[3,4] In this oscillatory regime, the screening length begins to *increase* with ionic strength, not decrease. This phenomenon is commonly referred to as 'underscreening' – the screening of the electrostatic potential in the bulk electrolyte is weaker than it should be according to the canonical Debye-Hückel theory. Using surface force balance (SFB) measurements the groups of Israelachvili[5–7] and Perkin[8–10] both demonstrated the presence of underscreening in a range of concentrated aqueous and non-aqueous electrolytes. Above a concentration of ~1 M, the screening length was observed to increase.

However, the magnitude of this increase was, in some cases, orders of magnitude larger than that predicted theoretically (Figure 1). Subsequent SFB[11–18] and fluorescence measurements[19] in electrolytes, ionic liquids and deep eutectic solvents by multiple groups, have reproduced these results. The fact that these screening lengths are anomalously large, compared to the theoretically predicted values, has led to their classification as 'anomalous underscreening'. However, atomic force microscopy (AFM),[20] small angle X-ray scattering[21] and molecular dynamics (MD) simulations of both bulk[22–24] and interfaces[24] fail to reproduce this phenomenon and yield screening lengths consistent with theory – so called 'normal underscreening'. While many phenomena have been proposed to account for the existence of anomalous underscreening, ranging from ion pairing,[25,26] particle exchange,[27,28] plasma-like collective behaviour[29,30] and the electronic effects of the solvent,[31] the origins of anomalous underscreening remains an open question.[32]

Theoretical treatments of electrostatic potential decay[4,33–37] in concentrated electrolytes have their origins in the statistical mechanical treatments of Kirkwood.[2,38,39] Using dressed ion theory, Kjellander has recently demonstrated that above the Kirkwood point the electrostatic potential is of the form,[4]

$$\psi(r) = \frac{q^{\text{eff}}}{4\pi\varepsilon_r^{\text{eff}}\varepsilon_0}\frac{e^{-\kappa r}}{r} + \frac{q'^{\text{eff}}}{4\pi\varepsilon_r'^{\text{eff}}\varepsilon_0}\frac{e^{-\kappa' r}}{r} + \text{other terms} \quad (3a)$$

while the radial charge density around individual ions takes the form,

$$\rho(r) = \frac{\kappa^2 q^{\text{eff}}\varepsilon_r}{4\pi\varepsilon_r^{\text{eff}}}\frac{e^{-\kappa r}}{r} + \frac{\kappa'^2 q'^{\text{eff}}\varepsilon_r}{4\pi\varepsilon_r'^{\text{eff}}}\frac{e^{-\kappa' r}}{r} + \text{other terms} \quad (3b)$$

In effect, the electrostatic decay and charge density within the electrolyte is multi-modal and determined by the complex effective charges and permittivity $q^{\text{eff}}$, $q'^{\text{eff}}$ and $\varepsilon_r^{\text{eff}}$, $\varepsilon_r'^{\text{eff}}$. Sufficiently high concentrations also lead to the decay parameters $\kappa$ and $\kappa'$ themselves becoming complex conjugates, i.e. $\kappa = \kappa_\Re + i\kappa_\Im$ and $\kappa' = \kappa^* = \kappa_\Re - i\kappa_\Im$, and both $\psi(r)$ and $\rho(r)$ becoming oscillatory in nature, e.g.

$$\rho(r) = \frac{|\kappa^2||q^{\text{eff}}|\varepsilon_r}{2\pi|\varepsilon_r^{\text{eff}}|}\frac{e^{-\kappa_\Re r}}{r}\cos(\kappa_\Im r + \alpha) + \cdots \quad (4a)$$

Conventionally,[8–10,19,20,22–24] the electrostatic screening length is derived from MD simulations by assuming that a single mode (i.e. term) is dominant in equation (4a), i.e.

$$\rho(r) \sim \frac{|\kappa^2||q^{\text{eff}}|\varepsilon_r}{2\pi|\varepsilon_r^{\text{eff}}|}\frac{e^{-\kappa_\Re r}}{r}\cos(\kappa_\Im r + \alpha) \quad (4b)$$

The form of $\rho(r)$ can be inferred directly from the MD total charge correlation function $g_{zz}(r)$ (Figure 1a),

$$\rho(r) \sim g_{zz}(r) = g_{++}(r) + g_{--}(r) - 2g_{+-}(r) \tag{5}$$

where $g_{++}$, $g_{--}$ and $g_{+-}$ are the cation-cation, anion-anion and cation-anion radial distribution functions respectively. The screening length $\lambda_s \equiv \kappa_\Re^{-1}$ can then be obtained simply as the inverse slope of the oscillatory component of $\ln(|g_{zz}(r)| \cdot r)$,

$$\ln(|g_{zz}(r)| \cdot r) = -\kappa_\Re r + \ln\left(\frac{|\kappa^2||q^{\text{eff}}|\varepsilon_r}{2\pi|\varepsilon_r^{\text{eff}}|} \cdot |\cos(\kappa_\Im r + \alpha)|\right) \tag{6}$$

This procedure is illustrated in Figure 1a, for an exemplar solution of ~7 M LiCl$_{(aq)}$. Screening lengths derived in this manner, from our own all-atom MD simulations[40] and others,[41–43] are compared with experimental values[8,19,20,44–47] for a series of aqueous alkali metal chloride electrolytes (~0.1 - 9 M) in Figure 1b. This figure illustrates the deviation between normal and anomalous underscreening; the predicted screening lengths are orders of magnitude smaller than those observed experimentally. For instance, $\lambda_S$ predicted for 7 M LiCl is 3.4 Å, whereas fluorescence measurements indicate that it is 75.3 Å.[19]

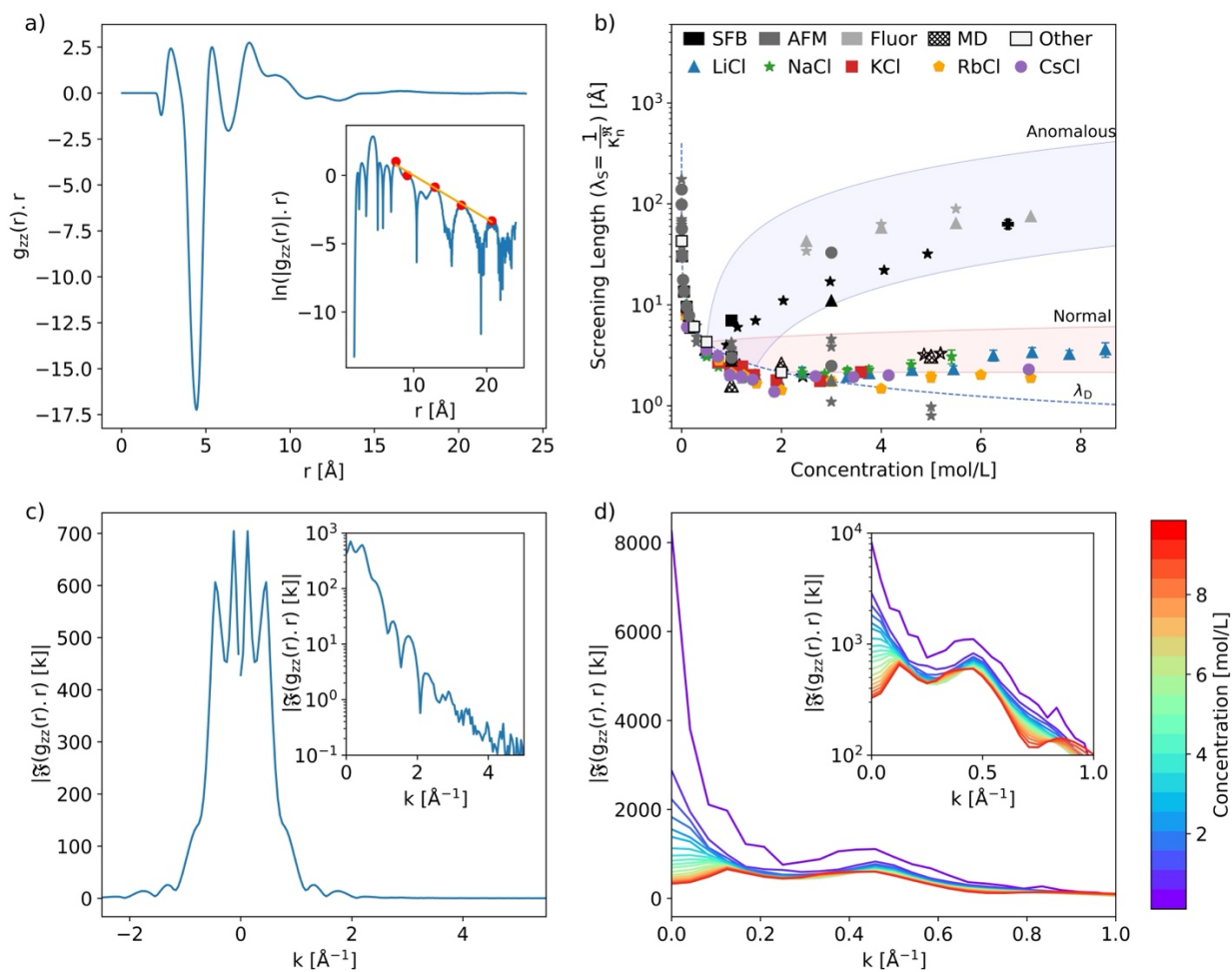

**Figure 1.** Should a single-mode ansatz be used to derive electrostatic screening lengths from molecular dynamics simulations? (a) Conventionally the electrostatic screening length can be obtained from the radial charge density $g_{zz}(r)$ by fitting the oscillatory component of $\ln(|g_{zz}(r)|\cdot r)$ as a function of radial distance $r$ (inset), illustrated here for 7M LiCl$_{(aq)}$. The red points and straight line in the inset indicate the region used to obtain $\lambda_S$ via linear regression. (b) Comparison of screening lengths obtained using the method in (a) with experimental and literature MD values for alkali metal chlorides.[8,19,20,41–47] Full details of all conventionally-derived $\lambda_S$ values presented here are provided in Supporting Information (Table S2-6, Figures S2-65). (c) The discrete Fourier transform of the radial charge density, $|\mathfrak{F}[g_{zz}(r)\cdot r]|$, demonstrates the presence of multiple decay modes in $g_{zz}(r)$; individual peaks correspond to distinct electrostatic decay modes (positive frequencies shown on a log scale inset). (d) Comparing the discrete Fourier transform of the radial charge density, $|\mathfrak{F}[g_{zz}(r)\cdot r]|$ across concentrations of LiCl$_{(aq)}$ (positive frequencies shown on a log scale inset) shows how the Yukawa modes dominate at low concentration, while oscillatory modes dominate at higher concentration.

However, is an *a priori*, single decay mode ansatz appropriate for predicting screening lengths in concentrated electrolytes? The discrete Fourier transform of the radial charge density itself, i.e.

$\mathfrak{F}[g_{zz}(r) \cdot r]|$, for 7 M LiCl$_{(aq)}$ (Figure 1c) clearly shows that the radial charge density comprises multiple oscillations, each with a distinct amplitude and spatial frequency (at least 6 peaks can unambiguously be resolved via the Figure 1c inset). It follows that the electrostatic potential in this electrolyte also comprises multiple independent decay modes, each with a characteristic screening length. Figure 1d considers the influence of electrolyte concentration on this result. At low concentrations the most prominent peak in the discrete Fourier spectrum occurs at 0 Å$^{-1}$, i.e. zero frequency, indicating that the dominant mode in the electrostatic potential is non-oscillatory Yukawa decay, consistent with theoretical prediction[2,4,33,39,48] and of course Debye-Hückel theory. As concentrations increase to those near the Kirkwood point, however, the amplitudes of the Yukawa decay mode reduces and new oscillatory decay modes, i.e. those with non-zero spatial frequencies, become apparent. At higher concentrations the prominence of these oscillatory modes over Yukawa modes continues to increase, with increasingly larger-frequency modes observed in the discrete Fourier spectrum. These results are not isolated to 7 M LiCl$_{(aq)}$; for each electrolyte considered here, the discrete Fourier spectrum of $g_{zz}(r) \cdot r$ exhibits multiple peaks, however the height, shape, and number of modes depends on both the concentration and the ions present (Figure S66-S70).

These results demonstrate therefore that multiple screening modes exist in highly concentrated electrolytes. This is unsurprising – it is predicted by exact statistical thermodynamic theories of electrostatic decay in concentrated electrolytes.[49] Models of electrolyte screening derived from all-atom MD simulations[23,24,50] have also considered the existence of multiple modes previously, however anomalous screening lengths were not predicted. It was suggested recently that the mode responsible for anomalous underscreening has a small amplitude and occurs over a large dynamic range.[12] This begs the question - which, if any, of the multiple independent modes revealed in Figure 1c,d give rise to anomalous screening lengths? Unfortunately, screening lengths cannot be unambiguously extracted from the discrete Fourier spectrum itself (see Supporting Information for a full discussion). We therefore present a different approach here and extract the screening length for each mode directly from the radial charge density itself.

Our approach assumes the ansatz of Xiao and Song,[3,51,52] which in turn is based on the dressed ion theory of Kjellander,[4] whereby the radial charge density takes the form,

$$\rho(r) \sim \sum_n \frac{\kappa_n^2 q_{\text{eff},n} \varepsilon_r}{2\pi \varepsilon_{\text{eff},n}} \frac{e^{-\kappa_n r}}{r} \qquad (7)$$

for mode-dependent effective charges ($q_{\text{eff},n}$) and permittivities ($\varepsilon_{\text{eff},n}$). The number of modes (i.e. terms) within this ansatz is infinite, however we note that the ansatz accounts for both the Yukawa decay ($\kappa_n^{\Im} = 0$) and oscillatory decay modes ($\kappa_n^{\Im} \neq 0$) that are anticipated based on Figure 1c,d. At concentrations above the Kirkwood point, each $\kappa_n$ is complex (i.e. $\kappa_n = \kappa_n^{\Re} + i\kappa_n^{\Im}$, $\kappa_n' = \kappa_n^{\Re} - i\kappa_n^{\Im}$). Under these conditions Equation (7) becomes,

$$\rho(r) \cdot r \sim \sum_n \frac{|\kappa_n|^2 |q_{\text{eff},n}^*| \varepsilon_r}{2\pi |\varepsilon_{\text{eff},n}^*|} e^{-\kappa_n^{\Re} r} \cos(\kappa_n^{\Im} r + \phi_n) \qquad (8)$$

where $\phi_n$ is an arbitrary phase shift and both the effective charge and dielectric constants of each mode also become complex. Herein we take the same assumption as in past MD studies, that is that equation 8 can be approximated by the $g_{zz}(r) \cdot r$ function from all-atom molecular dynamics, and fit to a truncated version of this exact statistical mechanical ansatz using Prony's method (full details of the Prony's method algorithm employed here are provided in Supporting Information). An example of its use to derive screening lengths is shown in Figure 2 for 7 M LiCl$_{(aq)}$. Figure 2a compares the approximate $g_{zz}(r) \cdot r$ function reconstructed using Prony's method with the original $g_{zz}(r) \cdot r$ signal obtained from MD, while the screening lengths ($\lambda_s = \kappa_{\Re}^{-1}$) for each individual mode are summarised in Figure 2b as a function of the corresponding spatial frequency ($\kappa_{\Im}$).

The reconstruction of $g_{zz}(r) \cdot r$ shown in Figure 2 is optimal according to the Bayesian information criterion (see Supporting information for a full discussion) and employs 48 terms in Equation (8). However, many have negligible amplitudes and only a handful contribute significantly to the radial charge density itself. Additionally, as anticipated from the discrete Fourier spectrum for this electrolyte (Figure 1c), the spectrum of $\kappa_n$ values obtained here is symmetric - for every mode with imaginary coefficient $+\kappa_n^{\Im}$, there is an equivalent mode with a negative imaginary coefficient $-\kappa_n^{\Im}$. Of these modes, most screening lengths are small ($\lambda_s <$ 20 Å) and are consistent with what is considered 'normal' underscreening. However, a few modes with larger oscillatory wavelengths (i.e. low spatial frequencies) have screening lengths consistent with anomalous underscreening. For instance, one mode with an oscillatory wavelength 14.5 Å exhibits a screening length of 61.7 Å, consistent with that reported by Gaddam and Ducker[19] (75.3 Å). Similarly, the non-oscillatory (i.e. zero frequency) Yukawa decay mode exhibits a screening length of 33.3 Å, which could also be considered anomalous.

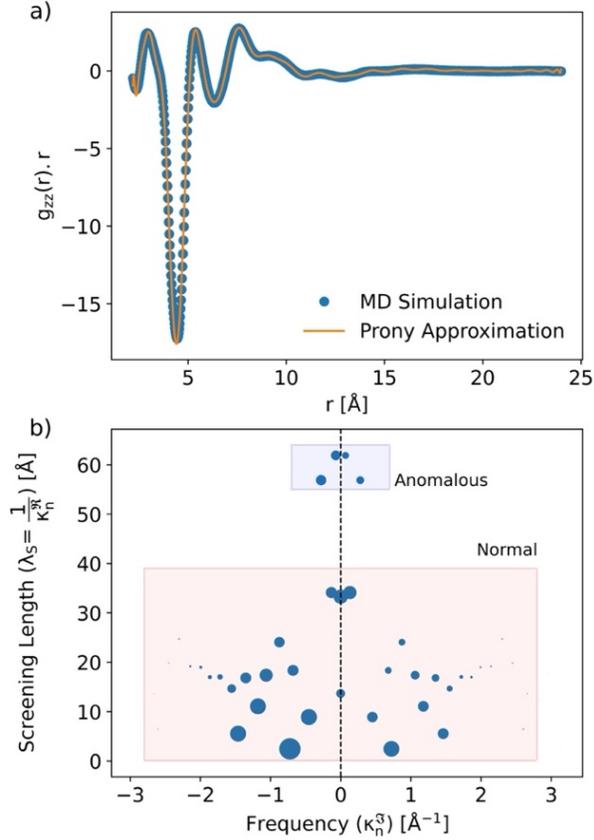

**Figure 2.** (a) Prony reconstruction of the radial charge density in 7 M LiCl$_{(aq)}$ using a 48-term expansion of Equation (7). (b) Spectrum of decay modes in 7 M LiCl$_{(aq)}$ as a function of decay mode oscillation wavelength $1/\kappa_n^{\Im}$. The size of each point reflects the relative amplitude of the corresponding mode in Equation (7) and the vertical position reflects the screening length $1/\kappa_n^{\Re}$. Note that the spectrum of screening lengths is symmetric about $\kappa_n^{\Im} = 0$, satisfying the requirement that $\kappa$ is complex at these conditions.

The results presented in Figure 2 lead us to hypothesise that anomalous underscreening in concentrated electrolytes arises from oscillations with low spatial frequencies (i.e. $1/\kappa_n^{\Im} >$ 1-2 nm) present in the radial charge density, whereas normal underscreening arises from oscillations with high spatial frequencies (e.g. $1/\kappa_n^{\Im} <$ 1 nm). It is conceivable that the former are the result of long-range collective ion correlations present in the bulk electrolyte, whereas the latter arise from correlations in ions' immediate solvation environment. However, considering the transient nature of ion-ion correlations in concentrated electrolytes[40] we do not attribute either case more specifically to a particular structural ensemble or motif that may occur in the bulk. We also refrain from interpreting the amplitudes of individual modes, which are determined by effective charges ($q^*_{\text{eff,n}}$) and dielectric constants ($\varepsilon^*_{\text{eff,n}}$), since we are unable to decouple these phenomena using our current approach. Instead we treat the $\frac{|\kappa_n|^2 |q^*_{\text{eff,n}}| \varepsilon_r}{2\pi |\varepsilon^*_{\text{eff,n}}|}$ prefactor in equation 8 as the amplitude obtained for each term using Prony's method. Interestingly, the screening lengths obtained using our approach are not strictly consistent with the wavelength of the corresponding oscillation in the radial charge density – for instance, in 7 M LiCl$_{(aq)}$, the screening length of $\lambda_s = 61.7$ Å arises from an oscillation with wavelength ($1/\kappa_n^{\Im}$) roughly one quarter that size (14.5 Å). We note that this is significantly less than $L/2$ for our simulation box employed here (L ~ 50 Å). Of course, the size of these oscillations are limited by the size of the MD simulation box employed here. However, the associated decay lengths need not be, since $\kappa_n^{\Im}$ and $\kappa_n^{\Re}$ are independent parameters in equation (8). Figure S71 demonstrates that the results reported in Figure 2 are not an artefact of the size of the simulation box employed here (50×50×50 Å³), nor arise from periodic boundary conditions enforced on the simulation box itself; this figure shows that for ~2 M LiCl$_{(aq)}$ 50×50×50 Å³ and 80×80×80 Å³ simulation boxes give essentially identical $g_{zz}(r) \cdot r$ Fourier spectra and distributions of decay mode frequencies, amplitudes and screening lengths. Thus, using a larger simulation box does not give rise to oscillations over larger length scales that are beyond that of the smaller 50×50×50 Å³ simulations.

We test the hypothesis proposed above for a series of aqueous alkali metal chloride electrolytes in Figure 3. Decay lengths, oscillatory wavelength and amplitudes of the decay modes elucidated using Prony's method are also listed in Supporting Information, along with full details of the corresponding fitting parameters. Figure 3 also compares our predicted screening lengths with experimental values where possible. We have considered concentrations from 2.0 M up to each respective experimental solubility limit, with the exception of LiCl$_{(aq)}$ (which is considered up to 10 M). Concentrations below 2.0 M are excluded to ensure that Equation (7), which strictly applies only to concentrated electrolytes, remains valid.

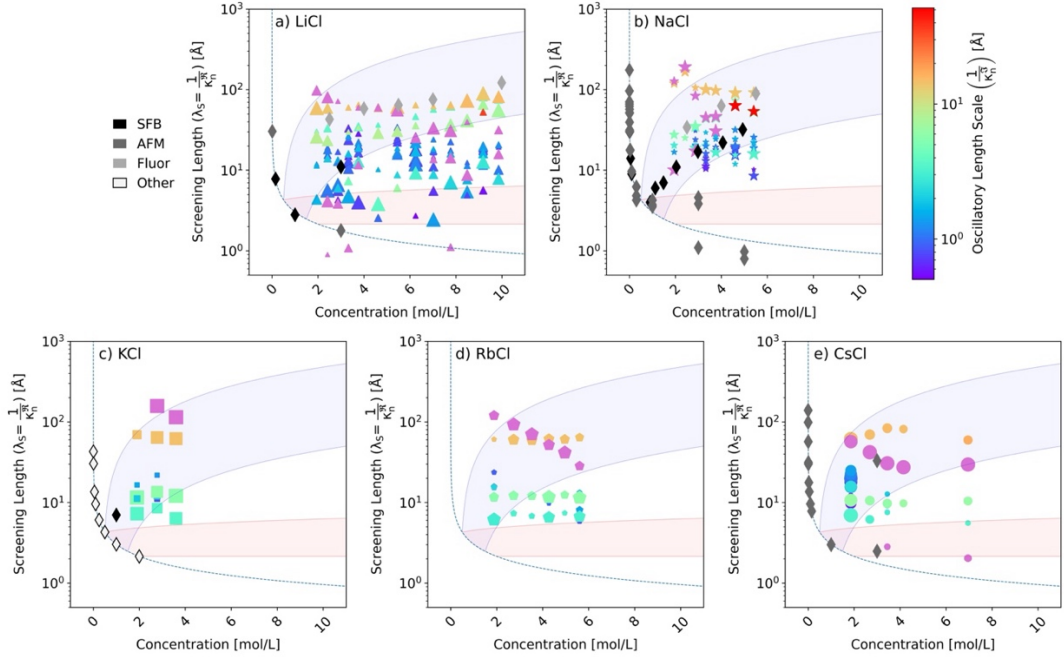

**Figure 3.** Comparison of experimentally observed screening lengths (greyscale) in concentrated electrolytes and those obtained from the Prony reconstruction of the radial charge density observed in all-atom MD simulations (colour). Anomalous screening lengths arise from large (> 5-15 Å) oscillatory wavelength in the corresponding decay mode. Symbols in (a) - (e) indicate the electrolyte: ▲ $LiCl_{(aq)}$, ★$NaCl_{(aq)}$, ■ $KCl_{(aq)}$, ⬟ $RbCl_{(aq)}$ and ● $CsCl_{(aq)}$. Predicted screening lengths are coloured according to the oscillatory wavelength of the corresponding decay mode. Oscillatory length scales > 100 Å are considered to approximate monotonic Yukawa modes (since $\kappa_n^{\Im} < 0.01 \sim 0$ Å$^{-1}$) and are presented in pink. Marker size indicates the normalised amplitude of the corresponding mode. Only modes with amplitudes on the same order of magnitude of the largest amplitude are shown. Experimental data is adapted from references [8,19,20,41–47] and presented as in Figure 1.

For each of the electrolytes considered here, Figure 3 supports our hypothesis; irrespective of the electrolyte; screening lengths typical of anomalous underscreening arise from oscillations with wavelengths of $1/\kappa_n^{\Im} \sim$ 5-15 Å or from modes exhibiting purely monotonic decay (i.e. $\kappa_n^{\Im} = 0$), while oscillatory wavelengths $1/\kappa_n^{\Im} < 5$ Å are responsible for shorter-range decay lengths, consistent with normal underscreening. This result is consistently observed across all electrolytes considered here, in the concentration range considered. Interestingly, the screening lengths associated with Yukawa (non-oscillatory) decay consistently decrease with concentration for all electrolytes considered, whereas those associated with oscillatory decay are less consistent and don't follow a distinct trend with concentration. Perhaps the most striking trend across the series of electrolytes considered here is the density of modes for $LiCl_{(aq)}$ and $NaCl_{(aq)}$, compared to that observed for the larger cations $K^+$, $Rb^+$ and $Cs^+$. Unsurprisingly, this is consistent with the respective discrete Fourier spectra presented in Figures S66-S70, which shows many more peaks can be discerned for $LiCl_{(aq)}$ and $NaCl_{(aq)}$ than $KCl_{(aq)}$, $RbCl_{(aq)}$, and $CsCl_{(aq)}$, which each generally only exhibit ~2 peaks. We attribute this result to the charge density asymmetry in each respective cation/anion pair, as quantified by the charge density descriptor þ.[54] For instance, both $Li^+$ and $Na^+$ are considerably more charge dense than $Cl^-$ (þ($Li^+$):þ($Cl^-$) ~ -5.1, þ($Na^+$):þ($Cl^-$) ~ -2.4), whereas $K^+$, $Rb^+$ and $Cs^+$ all have charge density comparable to $Cl^-$ (þ($K^+$):þ($Cl^-$) ~ -1.5; þ($Rb^+$):þ($Cl^-$) ~ -1.2; þ($Cs^+$):þ($Cl^-$) ~ -1.0). This asymmetry in charge density will influence ion packing in highly concentrated electrolytes, leading to a more pronounced long-range solvation structure, and thereby more distinct oscillations in the radial charge density (i.e. modes); this is evident from comparison of Figures S72-76. We note that an equivalent trend is observed for ion correlation lifetimes in concentrated $LiCl_{(aq)}$ and $NaCl_{(aq)}$, which are significantly longer-lived, compared to those observed in $KCl_{(aq)}$, $RbCl_{(aq)}$, and $CsCl_{(aq)}$ at equivalent concentrations.[40] The relative sizes of the cation and anion have also been invoked to explain asymptotic screening lengths in $LiCl_{(aq)}$ and $NaCl_{(aq)}$ derived from recent optical second harmonic scattering experiments.[55]

Figure 3 shows all reported experimental anomalous screening lengths are consistent with at least one of the screening lengths obtained via the reconstructed $g_{zz}(r) \cdot r$ using Prony's method. There is no apparent trend between a decay mode's amplitude and its screening length, however; modes with small and large amplitudes alike give rise to normal and anomalous screening lengths. Generally, the anomalous screening lengths for $LiCl_{(aq)}$ and $NaCl_{(aq)}$ reported using fluorescence spectroscopy[19] – which are consistently larger than those reported using SFB – coincide with those observed here for modes with the largest oscillatory wavelengths ($1/\kappa_n^\Im$ ~ 10-15) Å. On the other hand, SFB screening lengths[8,44,45] are consistent with those observed here for modes with oscillatory wavelengths of $1/\kappa_n^\Im$ ~ 3-5 Å. For $NaCl_{(aq)}$ there is considerable agreement between SFB screening lengths and those observed here for some of the Yukawa modes present ($\kappa_\Im = 0$). Similarly, for $CsCl_{(aq)}$, the two screening lengths reported at 3 M (both normal and anomalous) agree with those associated with different monotonic Yukawa decay modes. However, it is unclear if these consistencies are genuine, or coincidental. We note there are generally more 'normal' modes than 'anomalous' modes for the electrolytes considered here. While the electrostatic potential in a concentrated electrolyte is comprised of multiple decay modes, it is conceivable that only the most prominent modes can be measured experimentally. Or, alternatively, experimental techniques may be measuring the average, or superposition, of multiple modes simultaneously. This could explain why recent scattering experiments[21] have only shown the need to fit a single mode (resulting from a single complex pole in their analytical derivation). SFB force/distance curves consistently produce anomalous screening lengths, while AFM force/distance curves do not,[20,44] with one exception at elevated temperatures.[56] The presence of multiple modes in the electrolyte however explains this disagreement between seemingly similar experiments; the mode(s) each technique are detecting, and hence the screening lengths, may simply be different. It has been proposed that SFB is potentially capable of resolving modes with much longer screening lengths[12,32] because the active interface in an SFB experiment has a much larger surface area, compared to that present in AFM.[12,32] This proposal, however does not account for fluorescence measurements[19] and inferences made from re-entrant colloidal and soft matter behaviour[46,57–59] that suggest anomalous underscreening occurs without the presence of an interface. Results presented here confirm this - both normal and anomalous underscreening are simultaneous bulk phenomena. The fact that applied potentials influence screening lengths in AFM experiments[17] indicates that an interface simply modulates this bulk phenomenon in some way that depends on the surface potential. Understanding the effect of surface chemistry and the geometric nature of the confinement on screening length is still an open area of enquiry,[32] and we suggest here it should be a priority for the field.

To conclude, we have presented a new approach to the analysis of radial charge densities in bulk electrolytes that reveals the origins of both normal and anomalous underscreening in concentrated electrolytes. Discrete Fourier analysis shows the radial charge density, and hence the electrostatic potential, in concentrated electrolytes comprises multiple independent screening modes. By applying the least-squares version of Prony's method to a multi-modal ansatz for the radial charge density, $\kappa_\Re$ and $\kappa_\Im$ values for each decay mode can be extracted directly from the radial charge density itself. Anomalous screening lengths obtained using this method agree with experiment, and are consistently associated with decay modes exhibiting oscillatory wavelengths of ~5-15 Å. This length scale suggests anomalous underscreening may be a product of ion-ion correlations beyond ions' first or second solvation shells. Monotonic (non-oscillatory) Yukawa decay is also observed to give rise to anomalously large screening lengths. On the other hand, normal underscreening corresponds to decay modes with shorter oscillatory wavelengths (e.g. 3-5 Å), i.e. on the scale of short-range ion-ion correlations. We reiterate here that these results have been obtained in the absence of any *a priori* assumptions beyond the exact statistical mechanics ansatz employed; they arise naturally from the description of the electrolyte structure obtained in our all-atom MD simulations. These results effectively reconcile the disagreement between experimental and theoretical screening lengths in concentrated electrolytes that, over the last decade, has led to the paradigm of 'normal' and 'anomalous' underscreening. Instead, we suggest here that, ultimately, there is in fact nothing anomalous about anomalous underscreening.

**Computational Methods**

Full simulation details are in the Supporting Information. Polarisable molecular dynamics (MD) simulations were performed on 50×50×50 Å$^3$ cubic boxes prepared using Packmol[60]. Aqueous systems employed AMOEBA[61–64] polarisable force field, as implemented in the OpenMM code.[65] All MD simulations were first equilibrated under the isobaric-isothermal (NPT) ensemble, prior to another under the canonical (NVT) ensemble, before sampling. Analysis of radial distribution functions was performed using the MDAnalysis[66,67] package. Calculation of the Fourier transform was completed using the fast Fourier transform module in SciPy,[68,69] and Prony's method approximations were conducted using an in-house code.

**Acknowledgements**

The authors acknowledge Australian Research Council funding (ARC DP190100788, DP230102030). Sophie Baker acknowledges an Australian Government Research Training Program Scholarship. The authors thank Mr Lachlann Howard and A/Prof Jeff Hogan for useful discussions. This research was undertaken with the assistance of resources provided at the NCI National Facility systems at the Australian National University, through the National Computational Merit Allocation Scheme supported by the Australian Government.

**Supporting Information Available:**

Molecular dynamics simulation details and settings, further discussion of the Fourier transform, details and discussion of the Prony's method algorithm, conventionally derivations of $\lambda_S$, radial distribution functions, $g_{zz}(r) \cdot r$, simulation box size comparison and discrete Fourier transforms and Prony's method analysis for each system considered in the main text.

# There is Nothing Anomalous about "Anomalous" Underscreening in Concentrated Electrolytes

Sophie Baker[†,1], Gareth R. Elliott[†,1], Erica J. Wanless[1], Grant B. Webber[2], Vincent S. J. Craig[3], Alister J. Page*[1]

[1]Discipline of Chemistry, The University of Newcastle, Callaghan, New South Wales 2308, Australia

[2]Discipline of Chemical Engineering, The University of Newcastle, Callaghan, New South Wales 2308, Australia

[3]Department of Material Physics, Research School of Physics, Australian National University, Canberra, ACT 0200 Australia

## *Supporting Information*

## Table of Contents

# Molecular Dynamics Simulation Details

Polarisable molecular dynamics calculations of aqueous systems have been detailed previously.[1] Briefly, periodic simulation boxes were prepared with a volume of approximately 50×50×50 Å with 4167 water molecules, and the required number of ion pairs to create a given concentration (listed below). Four independent simulations at each concentration were prepared with randomly initialised coordinates by the packmol[2] tool. After an energy minimisation, simulations were then run using the Atomic Multipole Optimized Energetics for Biomolecular Applications (AMOEBA) force field[3] as implemented in OpenMM.[4] Each of the independent simulations used the isobaric-isothermal (NpT) ensemble to relax and equilibrate the density. After equilibration, the average volume was then determined from the last 500 ps and the box fixed at this size. The coordinates were then equilibrated under the canonical (NVT) ensemble prior to sampling, which was done for 2 ns for each replica (totally 8 ns of sampling). All stages of the simulations used a timestep of 1 fs to propagate the equations of motion. All simulations employed the water parameters of Ren et al.,[5] while the ion parameters were obtained from the thesis of Wang.[6] Exact simulation settings/values used are presented in Table S1.

**Table S1.** Settings/values used for simulations in OpenMM.

| Setting | Value |
| --- | --- |
| Non-bonded Method | Particle Mesh Ewald (PME) |
| Non-bonded Cutoff | 14.0 Å |
| Polarisation | mutual |
| Mutual Induced Target Epsilon | 0.00001 |
| Remove Centre of Mass Motion | True (default) |
| Platform | CUDA |
| CUDA Precision | mixed |
| Integrator/Thermostat Type | Nóse-Hoover Integrator (LFMiddle) |
| Thermostat Temperature | 300 K |
| Thermostat Collision Frequency | 0.1 ps$^{-1}$ |
| Integrator Step Size | 1 fs |
| Nóse-Hoover Chain Length | 3 (default) |
| Number of Chain Propagation Steps | 3 (default) |
| Number of Yoshida Suzuki Terms | 7 (default) |
| Barostat (NpT only) | Monte Carlo Barostat |
| Barostat temperature | 300 K |
| Barostat pressure | 1 atm |
| Barostat frequency | 0.1 ps$^{-1}$ |

# Fourier Transform

Here we present an alternative analysis of the radial charge density $\rho(r)$ that reveals the presence of multiple screening lengths in concentrated electrolytes. Instead of restricting the expansion of $\rho(r)$ to a single mode, however, we apply an analytical Fourier transform $\mathfrak{F}$ to the complete expansion resulting in the equation:

$$\mathcal{F}\left\{\sum_{j=1}^{N}\frac{|\kappa|^2|q_i^*|\varepsilon_r}{2\pi|\varepsilon_r^*(\kappa)|}e^{-\kappa_{\Re_j}r}\cos\left(\kappa_{\Im_j}r+\alpha_j\right)\right\}= \tag{S1}$$

$$\sum_{j=1}^{N}\frac{A\cos(\alpha_j)\left(2\kappa_{\Re_j}^3+2\kappa_{\Re_j}\left(k^2+\kappa_{\Im_j}^2\right)\right)+A\sin(\alpha_j)(2\kappa_{\Re_j}^2\kappa_{\Im_j}+(k+\kappa_{\Im_j})(k-\kappa_{\Im_j})^2-(k-\kappa_{\Im_j})(k+\kappa_{\Im_j})^2)}{2\kappa_{\Re_j}^4+4\kappa_{\Re_j}^2\left(k^2+\kappa_{\Im_j}^2\right)+2(k+\kappa_{\Im_j})^2(k-\kappa_{\Im_j})^2}$$

$$+i\frac{A\cos(\alpha_j)(2\kappa_{\Re_j}^2 k+(k-\kappa_{\Im_j})(k+\kappa_{\Im_j})^2+\left(k+\kappa_{\Im_j}\right)\left(k-\kappa_{\Im_j}\right)^2)+A\sin(\alpha_j)(4\kappa_{\Re_j}k\kappa_{\Im_j})}{+2\kappa_{\Re_j}^4+4\kappa_{\Re_j}^2\left(k^2+\kappa_{\Im_j}^2\right)+2(k+\kappa_{\Im_j})^2(k-\kappa_{\Im_j})^2}$$

The modulus of Equation 2.13 is plotted in Figure S1, for a one term and three term expansion as examples, resulting in a Fourier spectrum with peaks at each value of $-\kappa_{\Im_j}$ and $\kappa_{\Im_j}$ for every value of $j$; i.e. for the three term expansion there are three peaks for positive $k$ values and three for negative $k$ values. Additionally, the height and width of each peak is modulated by $\kappa_{\Re_j}$.

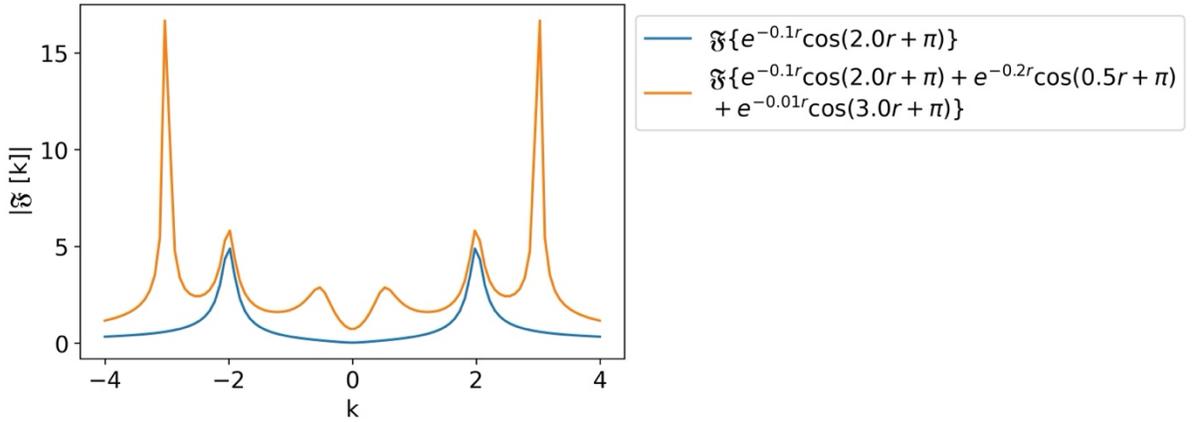

**Figure S1.** Analytical Fourier transform examples from Equation S1.

Therefore, taking the same assumption as above that $\rho(r)$ can be inferred from $g_{zz}(r)$, it is possible to determine if the resulting signal has multiple modes by examining its discrete Fourier transform, presented in Figures S66-70 for all systems considered in the main text. It is difficult to directly fit the discrete Fourier transform as $\kappa_{\Re_j}$ and $\frac{|\kappa|^2|q_i^*|\varepsilon_r}{2\pi|\varepsilon_r^*(\kappa)|}$ (A) are coupled, with both affecting the height of the individuals peaks, making it difficult to determine their individual values.

# Prony's Method Algorithm

Prony's method is an algorithm[7] developed in 1795 to fit a function:[8]

$$f(x) = \sum_{n=1}^{N} a_n e^{\gamma_n x} \tag{S2}$$

to an observed signal, $\hat{f}(x)$ consisting of $M$ evenly spaced samples. If $\hat{f}(x)$ involves oscillatory decay, then $e^{\gamma_n x}$ in Equation S2 will be made up of complex exponential pairs and $a_n$ and $\gamma_n$ can be defined as:

$$a_n = \frac{1}{2} A_i e^{\phi_i j} + \frac{1}{2} A_i e^{-\phi_i j} \tag{S3}$$

$$\gamma_n = \sigma_i \pm j\omega_i \tag{S4}$$

Combining Equations S2-S4 results in the equation:

$$f(x) = \sum_{i=1}^{N} A_i e^{\sigma_i x} \cos(\omega_i x + \phi_i) \tag{S5}$$

which takes the same form as that of $\rho(r)$. The summation of complex exponentials is the homogeneous solution to a linear difference equation; therefore, the signal can be expressed as the difference equation:[9,10]

$$\hat{f}(x) = \sum_{n=1}^{N} \hat{f}[\Delta_t(m-n)] P_n \tag{S6}$$

where $N$ is the number of complex exponentials, and $P_n$ are the coefficients of the polynomial:

$$z^N - P_1 z^{N-1} - \cdots - P_n = \prod_{n=1}^{N}(z - e^{\gamma_n}) \tag{S7}$$

where $e^{\gamma_n}$ are the roots of the polynomial and can be related to decay and frequency components of equation S5. Therefore, $\kappa_\Re$ and $\kappa_\Im$ can be calculated via the roots of Equation S6. The Prony's method algorithm was performed using an in-house code using SciPy[11] and NumPy[12] modules, consisting of the following steps:

1. Construct and solve the matrix equation for the coefficients $P_n$ of Equation S7:

$$\begin{pmatrix} F_N \\ \vdots \\ F_{M-1} \end{pmatrix} = \begin{pmatrix} F_{N-1} & \cdots & F_0 \\ \vdots & \ddots & \vdots \\ F_{M-2} & \cdots & F_{M-N-1} \end{pmatrix} \begin{pmatrix} P_1 \\ \vdots \\ P_N \end{pmatrix} \qquad (S8)$$

where $F_n$ is the $m^{th}$ of $M$ samples of the signal. When $M = 2N$ the system can be solved exactly using a matrix inverse, however when $M > 2N$ the linear system of equations is overdetermined, and a least-squares algorithm is needed to solve Equation S8.[13] This is referred to as the Prony's least squares method,[14] an extension of the original method which was developed before least squares and consisted of a $N \times N$ matrix, neglecting to include any information in $\hat{f}(x)$ past the $2N^{th}$ sample.

2. Find the roots of the polynomial in Equation S7; the $n^{th}$ root is equal to $e^{\gamma_n}$.

3. Extract the decay rates and frequencies from $\gamma_m$ as the real and imaginary components respectively.

4. Construct and solve the matrix equation to determine $a_n$:[15]

$$\begin{pmatrix} F_0 \\ \vdots \\ F_M \end{pmatrix} = \begin{pmatrix} (e^{\gamma_1})^0 & \cdots & (e^{\gamma_N})^0 \\ \vdots & \ddots & \vdots \\ (e^{\gamma_1})^M & \cdots & (e^{\gamma_N})^M \end{pmatrix} \begin{pmatrix} a_1 \\ \vdots \\ a_N \end{pmatrix} \qquad (S9)$$

5. Extract the amplitude $A_i$ and phase $\phi_i$ for the $i^{th}$ mode from $a_n$ using the formulas:

$$A_i = |a_n| \qquad (S10)$$

$$\phi_i = \arctan(a_n) \qquad (S11)$$

6. Reconstruct the signal using $A_i$, $\phi_i$, $\sigma_i$ and $\omega_i$.

To determine the number of modes needed to fit $g_{zz}(r) \cdot r$ the Bayesian information criterion is used:

$$BIC = -2 \ln(\hat{L}) + k \ln(n) \qquad (S12)$$

where $k$ is the number of modes in the approximation, $n$ is the number of data points and $ln(\hat{L})$ is the log-likelihood function, defined for a normal distribution as follows:

$$ln(\hat{L}) = \frac{-nln(2\pi\sigma^2)}{2} - \frac{\sum_{i=1}^{n}(x_i - \mu_i)^2}{2\sigma^2} \quad \text{(S13)}$$

where $\sigma^2$ is the variance and $\mu_i$ is the original signal

To find the minimum BIC, the BIC is calculated for Prony approximations for $n$=2-200 and then fitted with a cubic spline interpolation curve and the minimum found using a Nelder-Mead simplex algorithm (see Figures S71-S75).[16]

# Conventional Derivation of $\lambda_s$

**LiCl$_{(aq)}$**

**Table S2.** Concentrations, number of ion pairs (# IP), the number of solvent molecules (# Solvent) and conventionally-derived $\lambda_S$ values for aqueous LiCl simulations (see Figures S2-S20).

| Conc (m) | Conc (M) | # IP | # Solvent | Decay Length (Å) |
|---|---|---|---|---|
| 0.1 | 0.11 | 8 | 4167 | 8.13 ± 0.17 |
| 0.5 | 0.50 | 38 | 4167 | 3.38 ± 0.39 |
| 0.75 | 0.74 | 56 | 4167 | 2.68 ± 0.10 |
| 1.0 | 0.99 | 75 | 4167 | 2.09 ± 0.04 |
| 1.25 | 1.23 | 94 | 4167 | 2.00 ± 0.08 |
| 1.5 | 1.47 | 113 | 4167 | 2.04 ± 0.11 |
| 2.0 | 1.94 | 150 | 4167 | 1.82 ± 0.10 |
| 2.5 | 2.42 | 188 | 4167 | 2.18 ± 0.24 |
| 3.0 | 2.87 | 225 | 4167 | 1.87 ± 0.07 |
| 3.5 | 3.32 | 263 | 4167 | 1.95 ± 0.16 |
| 4.0 | 3.76 | 300 | 4167 | 2.10 ± 0.14 |
| 5.0 | 4.62 | 375 | 4167 | 2.27 ± 0.09 |
| 6.0 | 5.45 | 450 | 4167 | 2.30 ± 0.18 |
| 7.0 | 6.25 | 525 | 4167 | 3.16 ± 0.39 |
| 8.0 | 7.02 | 600 | 4167 | 3.38 ± 0.39 |
| 9.0 | 7.77 | 675 | 4167 | 3.25 ± 0.30 |
| 10.0 | 8.49 | 750 | 4167 | 3.62 ± 0.59 |

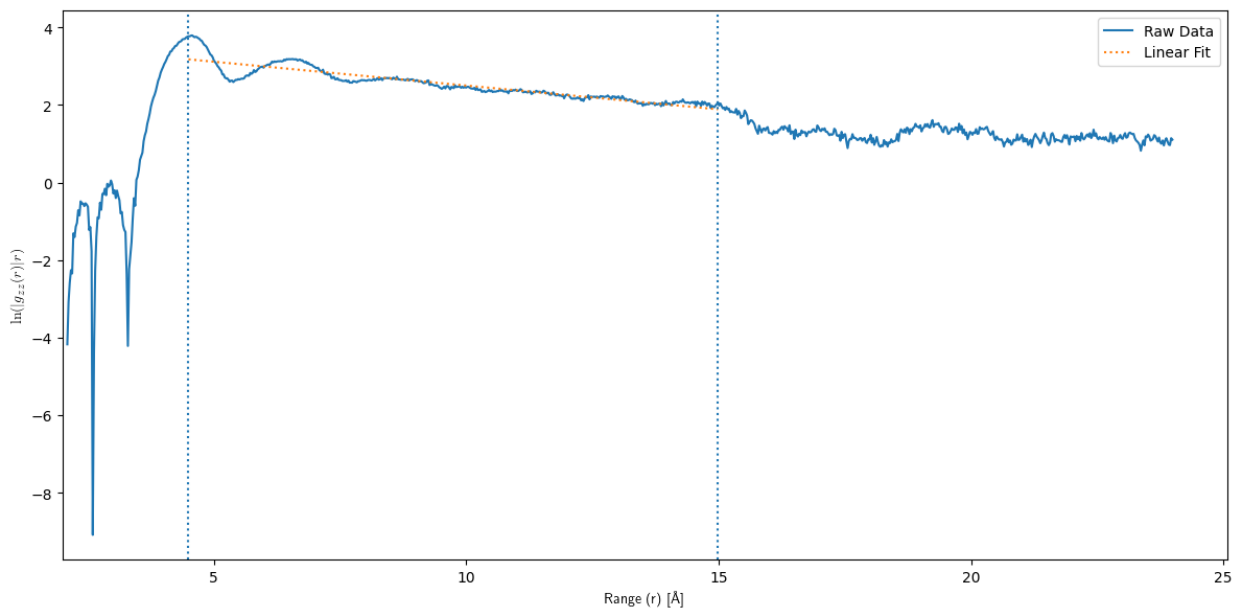

**Figure S2.** Fitting of aqueous LiCl at a concentration of 0.11 M. In this instance, because of the presence of a plateau in the data, a single straight line was used to fit to the points, rather than fitting the peaks. The vertical lines indicate the extrema that fitting was carried out over.

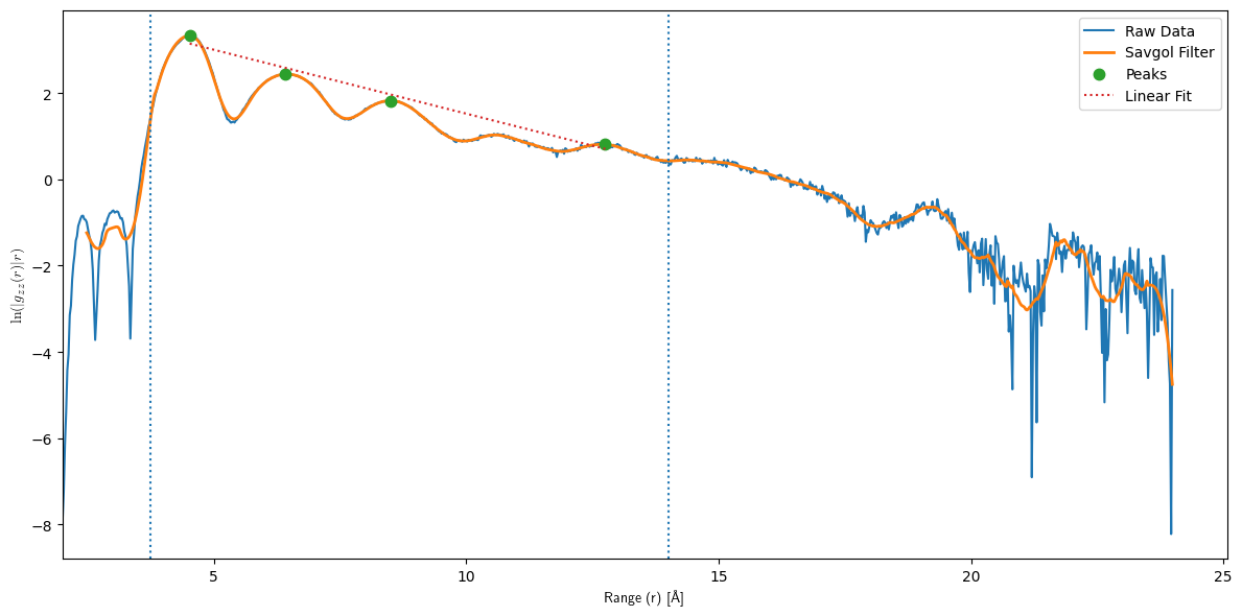

**Figure S3.** Fitting of aqueous LiCl at a concentration of 0.50 M. The envelope of the fit is determined by fitting through points at the top of each of the marked peaks. A Savgol filter was applied to smooth the data to assist peak finding, particularly at higher values of $r$. The vertical lines indicate the extrema that fitting was carried out over.

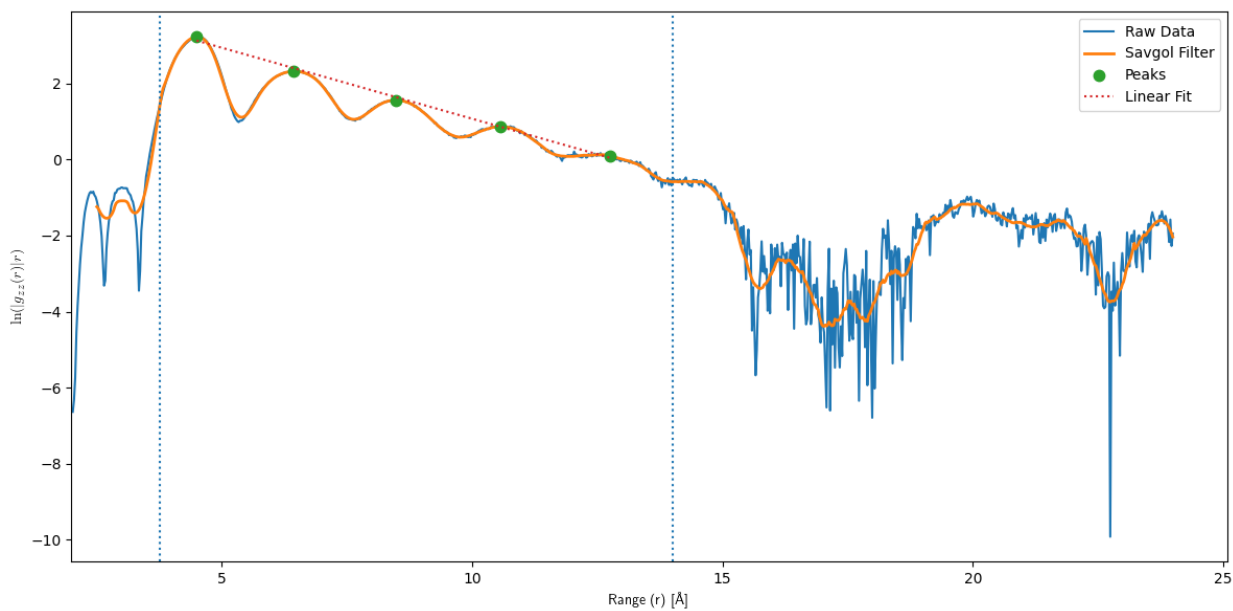

**Figure S4.** Fitting of aqueous LiCl at a concentration of 0.74 M. The envelope of the fit is determined by fitting through points at the top of each of the marked peaks. A Savgol filter was applied to smooth the data to assist peak finding, particularly at higher values of $r$. The vertical lines indicate the extrema that fitting was carried out over.

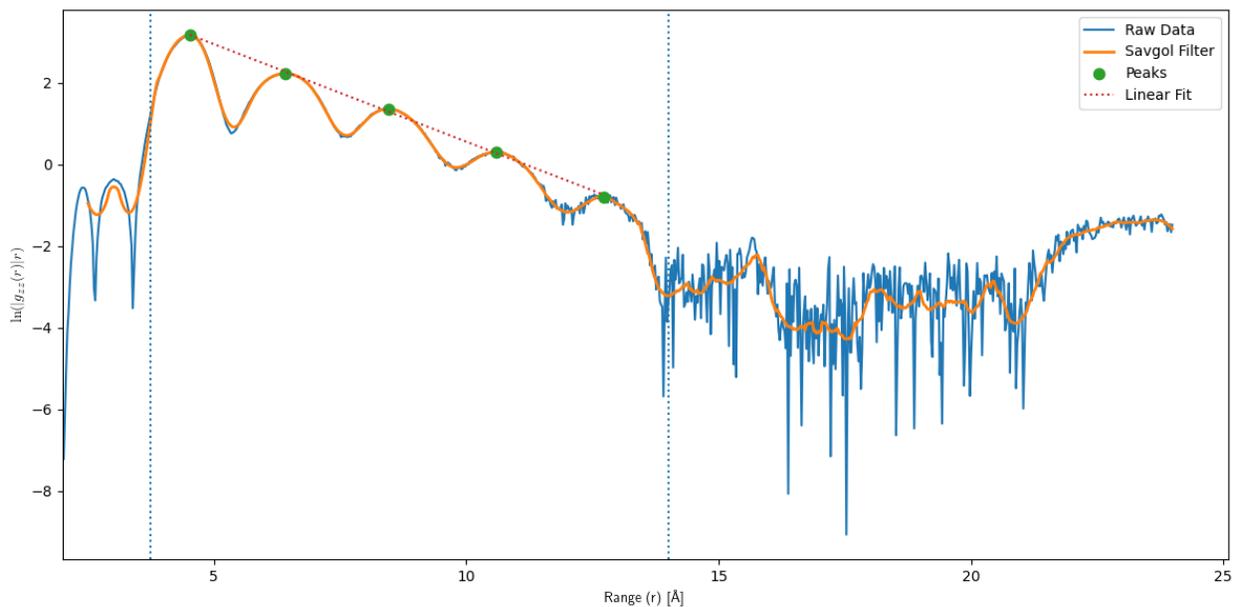

**Figure S5.** Fitting of aqueous LiCl at a concentration of 0.99 M. The envelope of the fit is determined by fitting through points at the top of each of the marked peaks. A Savgol filter was applied to smooth the data to assist peak finding, particularly at higher values of $r$. The vertical lines indicate the extrema that fitting was carried out over.

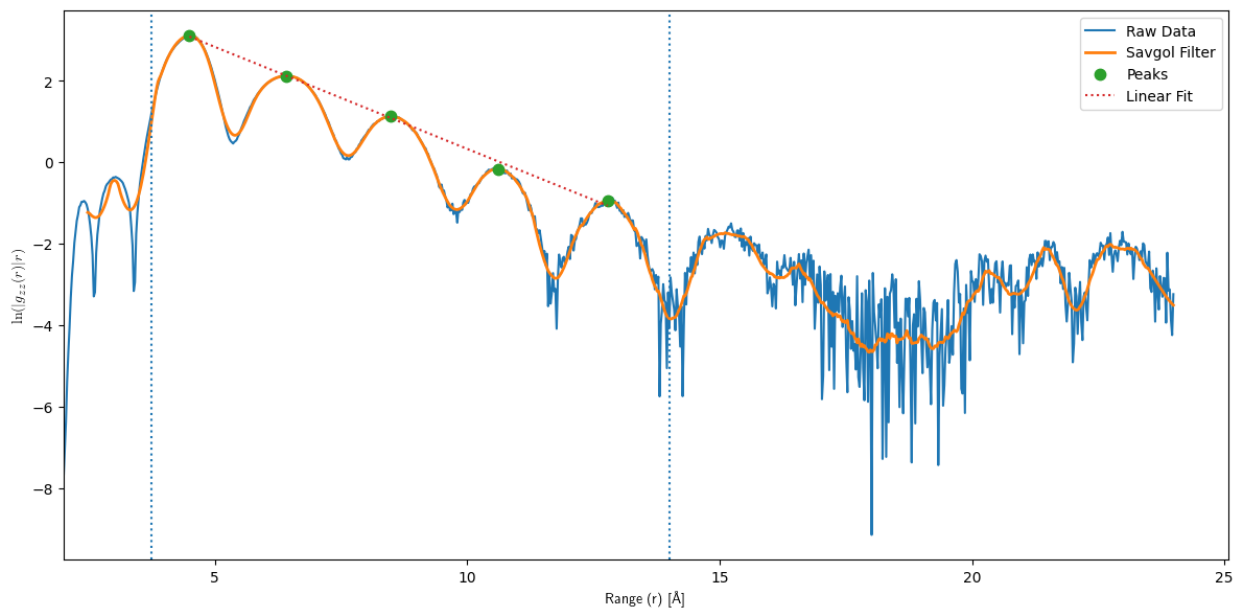

**Figure S6.** Fitting of aqueous LiCl at a concentration of 1.23 M. The envelope of the fit is determined by fitting through points at the top of each of the marked peaks. A Savgol filter was applied to smooth the data to assist peak finding, particularly at higher values of $r$. The vertical lines indicate the extrema that fitting was carried out over.

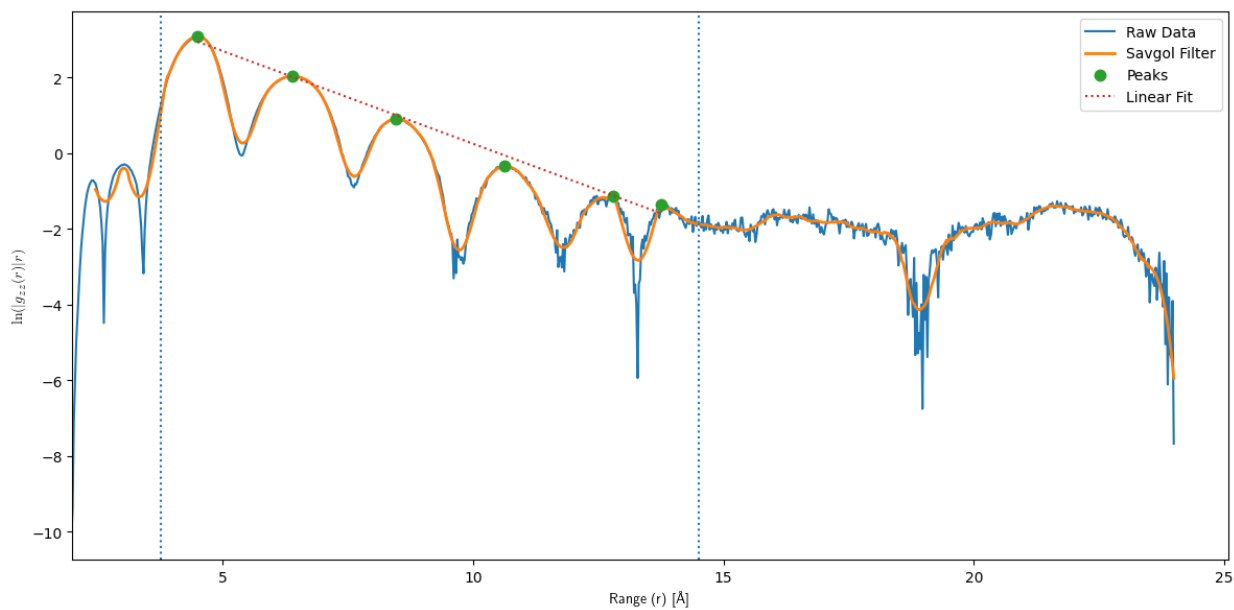

**Figure S7.** Fitting of aqueous LiCl at a concentration of 1.47 M. The envelope of the fit is determined by fitting through points at the top of each of the marked peaks. A Savgol filter was applied to smooth the data to assist peak finding, particularly at higher values of $r$. The vertical lines indicate the extrema that fitting was carried out over.

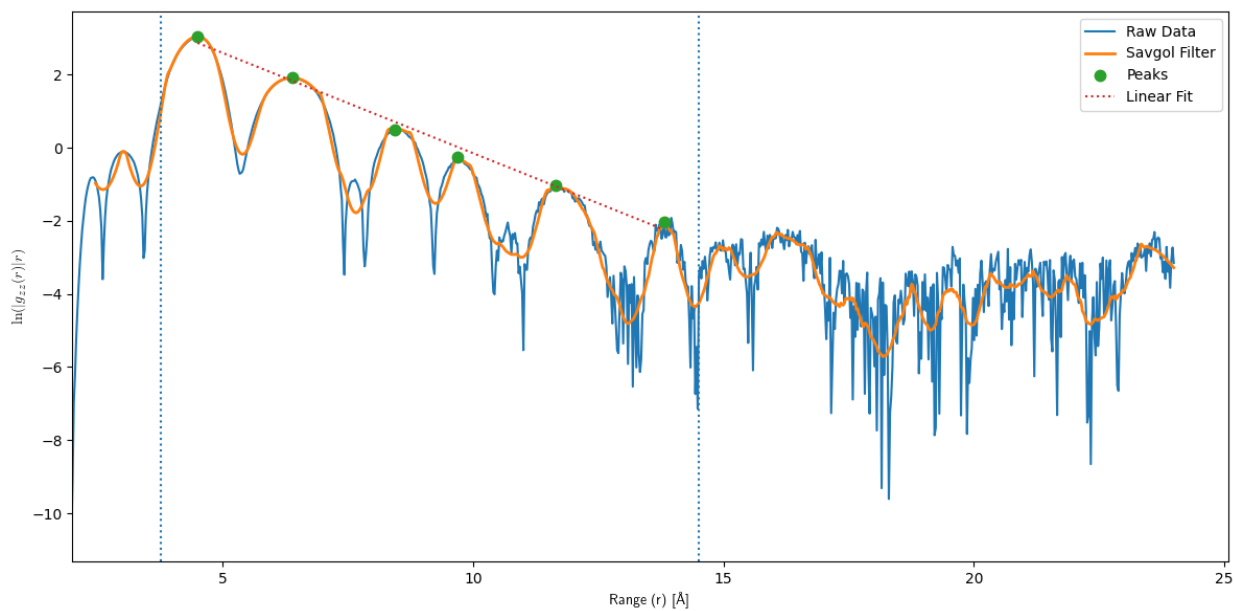

**Figure S8.** Fitting of aqueous LiCl at a concentration of 1.94 M. The envelope of the fit is determined by fitting through points at the top of each of the marked peaks. A Savgol filter was applied to smooth the data to assist peak finding, particularly at higher values of $r$. The vertical lines indicate the extrema that fitting was carried out over.

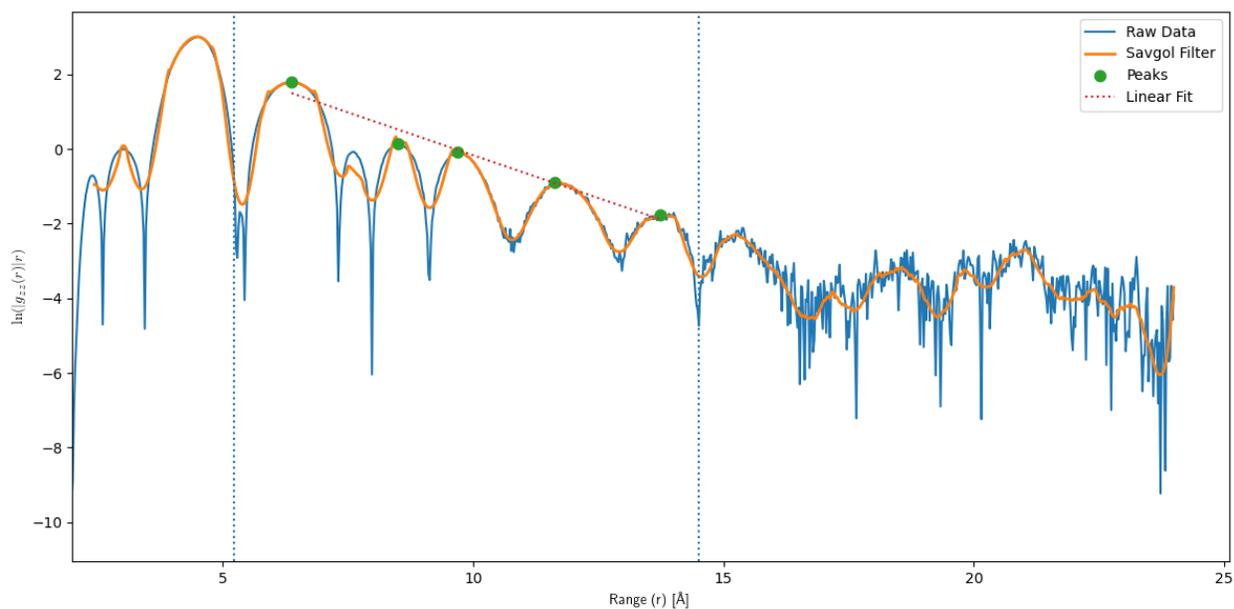

**Figure S9.** Fitting of aqueous LiCl at a concentration of 2.42 M The envelope of the fit is determined by fitting through points at the top of each of the marked peaks. A Savgol filter was applied to smooth the data to assist peak finding, particularly at higher values of $r$. The vertical lines indicate the extrema that fitting was carried out over.

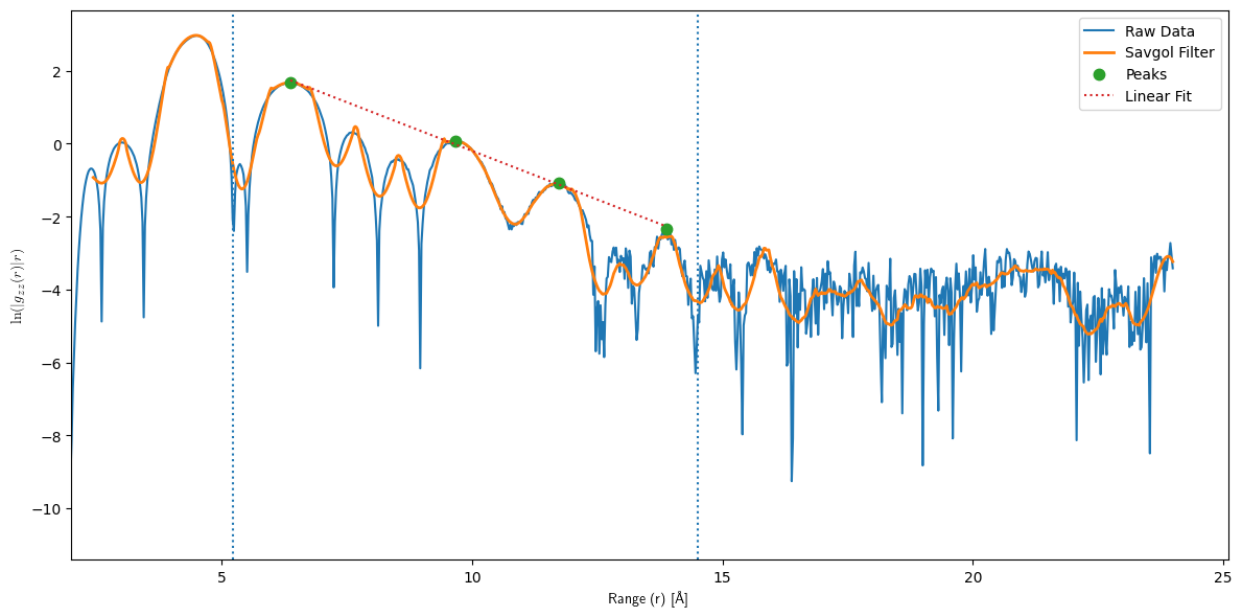

**Figure S10.** Fitting of aqueous LiCl at a concentration of 2.87 M. The envelope of the fit is determined by fitting through points at the top of each of the marked peaks. A Savgol filter was applied to smooth the data to assist peak finding, particularly at higher values of $r$. The vertical lines indicate the extrema that fitting was carried out over.

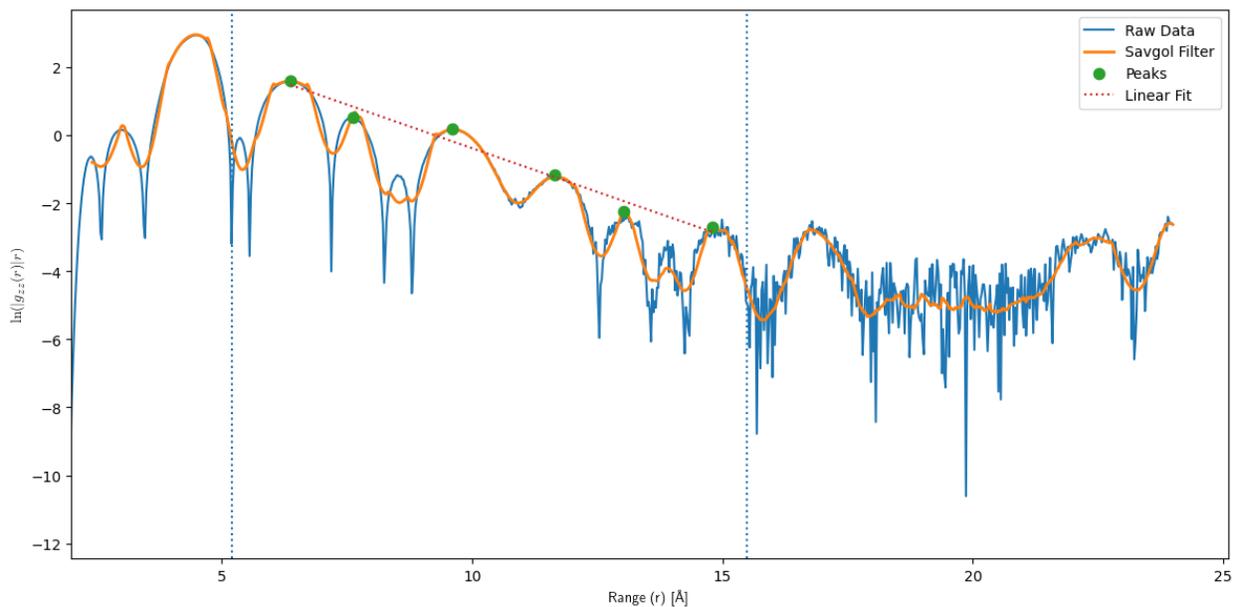

**Figure S11.** Fitting of aqueous LiCl at a concentration of 3.32 M. The envelope of the fit is determined by fitting through points at the top of each of the marked peaks. A Savgol filter was applied to smooth the data to assist peak finding, particularly at higher values of $r$. The vertical lines indicate the extrema that fitting was carried out over.
.

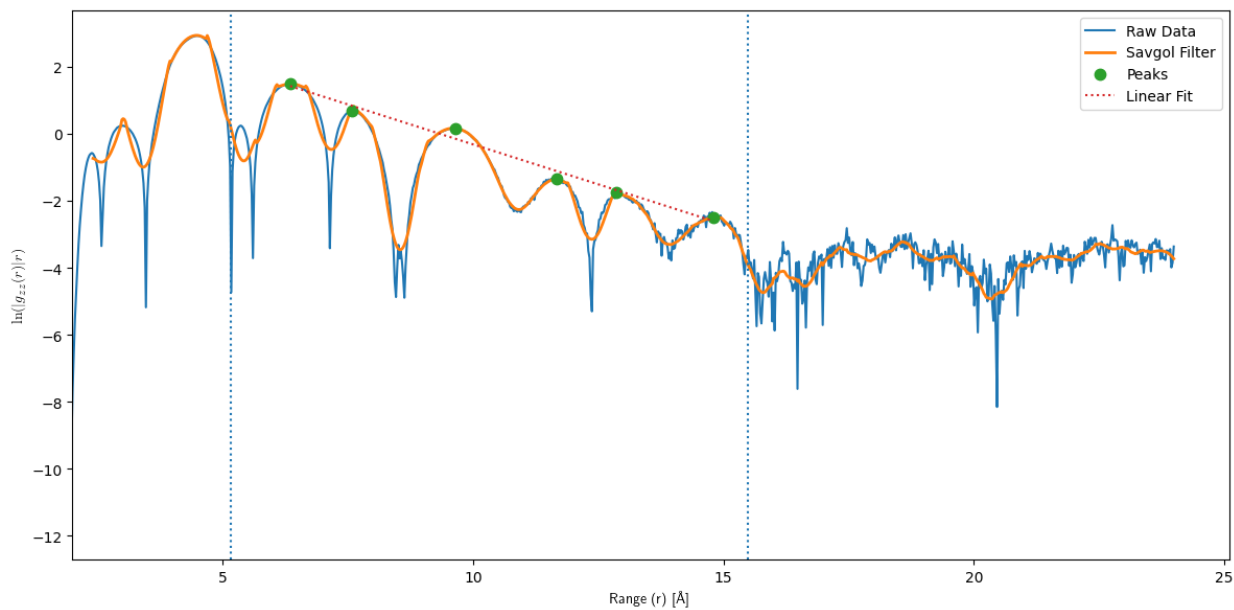

**Figure S12.** Fitting of aqueous LiCl at a concentration of 3.76 M. The envelope of the fit is determined by fitting through points at the top of each of the marked peaks. A Savgol filter was applied to smooth the data to assist peak finding, particularly at higher values of $r$. The vertical lines indicate the extrema that fitting was carried out over.

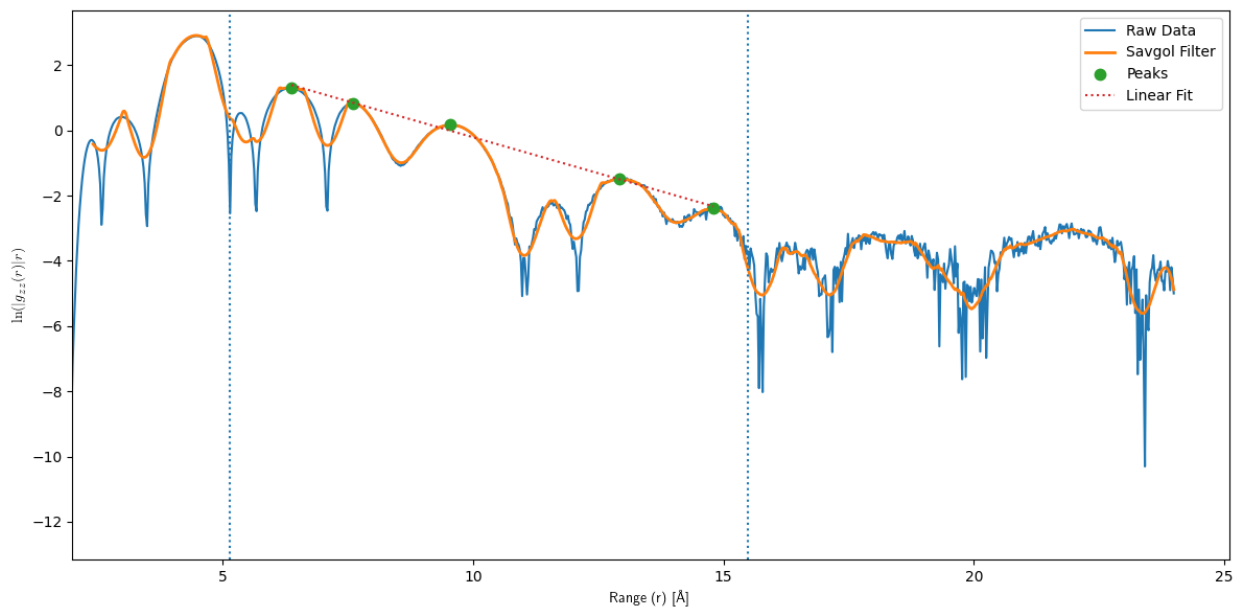

**Figure S13.** Fitting of aqueous LiCl at a concentration of 4.62 M. The envelope of the fit is determined by fitting through points at the top of each of the marked peaks. A Savgol filter was applied to smooth the data to assist peak finding, particularly at higher values of $r$. The vertical lines indicate the extrema that fitting was carried out over.

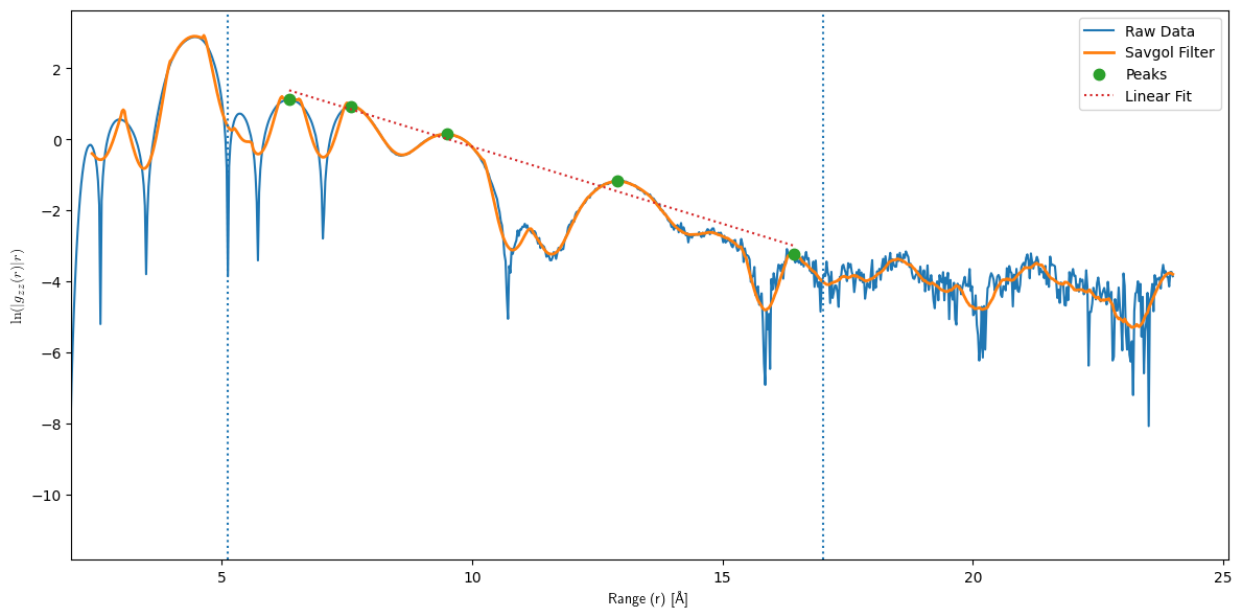

**Figure S14.** Fitting of aqueous LiCl at a concentration of 5.45 M The envelope of the fit is determined by fitting through points at the top of each of the marked peaks. A Savgol filter was applied to smooth the data to assist peak finding, particularly at higher values of $r$. The vertical lines indicate the extrema that fitting was carried out over.

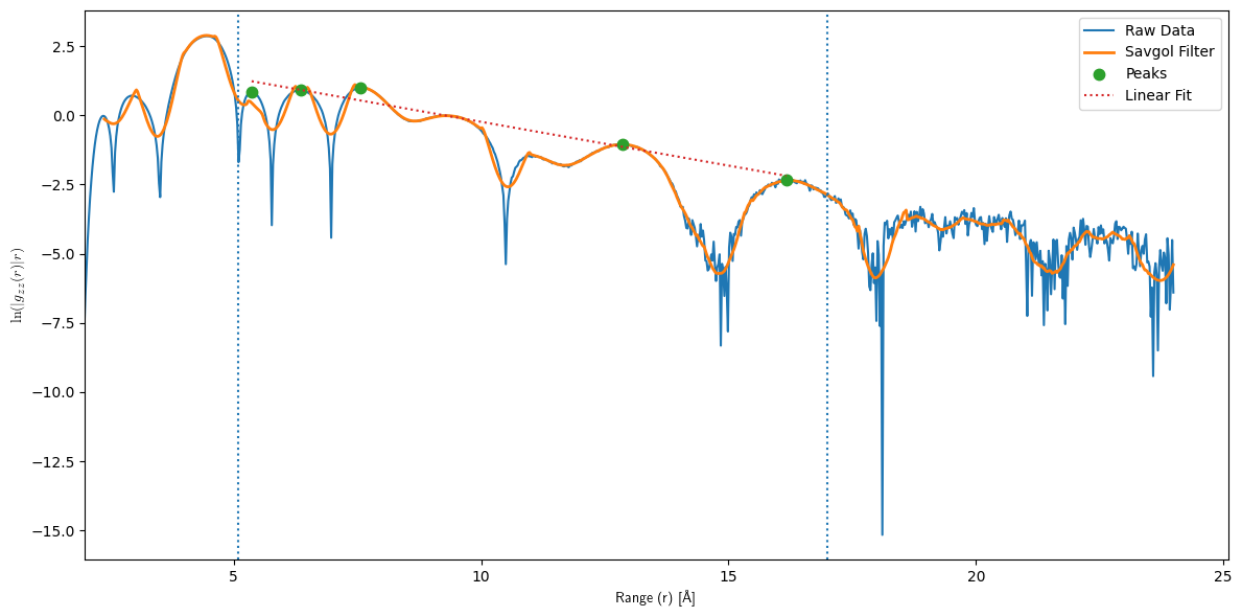

**Figure S15.** Fitting of aqueous LiCl at a concentration of 6.25 M. The envelope of the fit is determined by fitting through points at the top of each of the marked peaks. A Savgol filter was applied to smooth the data to assist peak finding, particularly at higher values of $r$. The vertical lines indicate the extrema that fitting was carried out over.

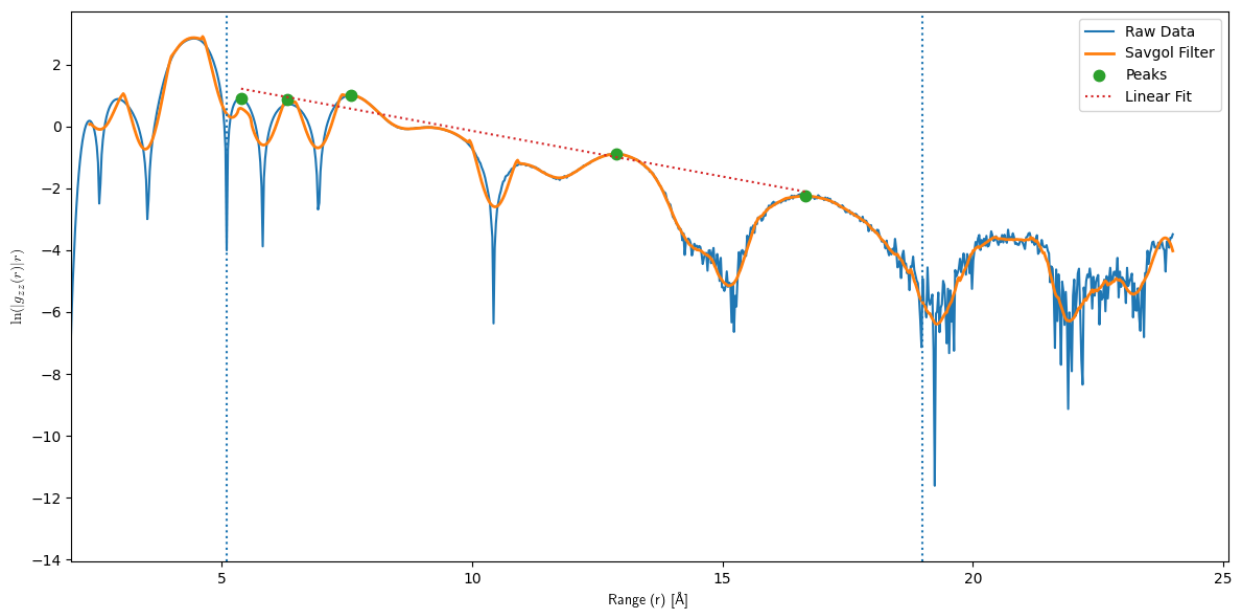

**Figure S16.** Fitting of aqueous LiCl at a concentration of 7.02 M. The envelope of the fit is determined by fitting through points at the top of each of the marked peaks. A Savgol filter was applied to smooth the data to assist peak finding, particularly at higher values of $r$. The vertical lines indicate the extrema that fitting was carried out over.

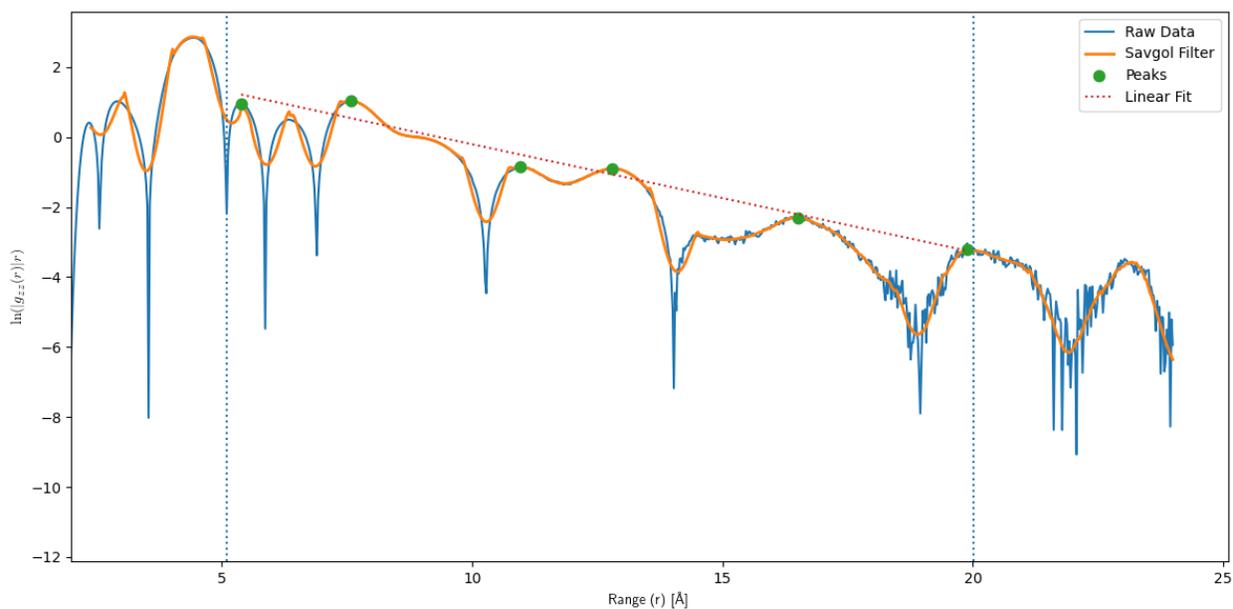

**Figure S17.** Fitting of aqueous LiCl at a concentration of 7.77 M. The envelope of the fit is determined by fitting through points at the top of each of the marked peaks. A Savgol filter was applied to smooth the data to assist peak finding, particularly at higher values of $r$. The vertical lines indicate the extrema that fitting was carried out over.

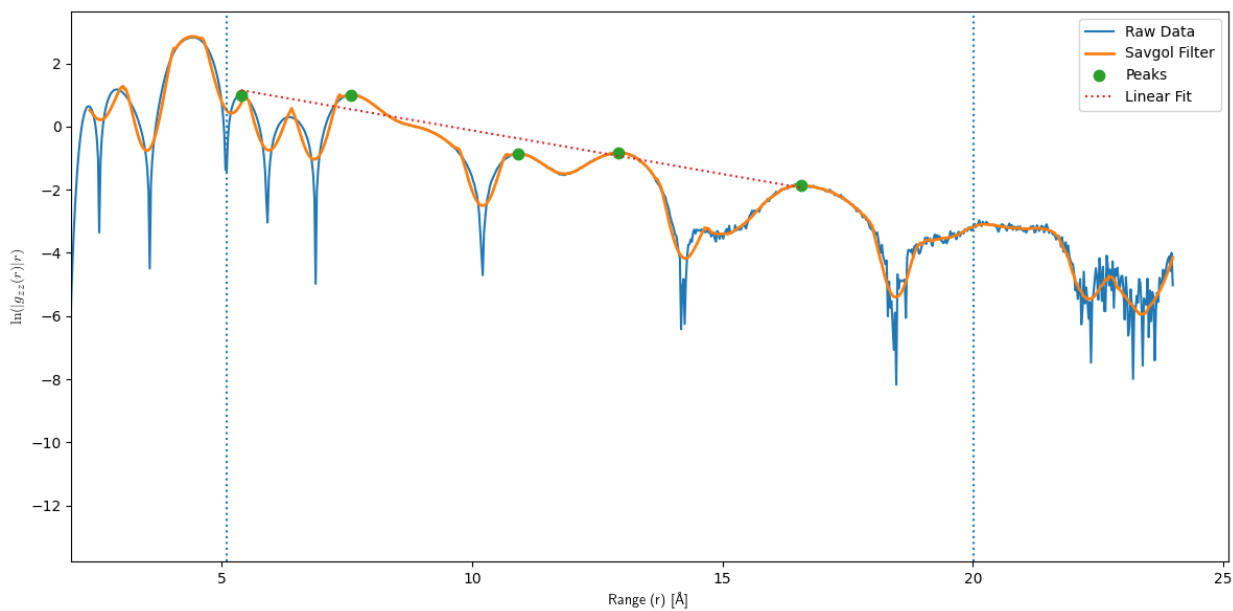

**Figure S18.** Fitting of aqueous LiCl at a concentration of 8.49 M. The envelope of the fit is determined by fitting through points at the top of each of the marked peaks. A Savgol filter was applied to smooth the data to assist peak finding, particularly at higher values of $r$. The vertical lines indicate the extrema that fitting was carried out over.

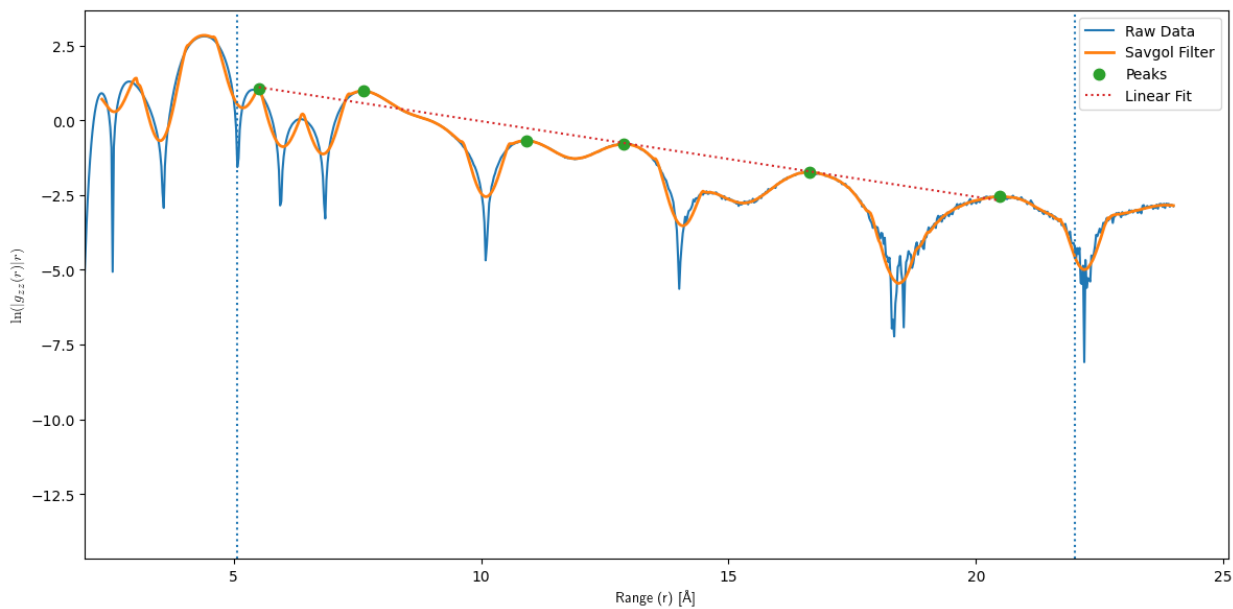

**Figure S19.** Fitting of aqueous LiCl at a concentration of 9.19 M. The envelope of the fit is determined by fitting through points at the top of each of the marked peaks. A Savgol filter was applied to smooth the data to assist peak finding, particularly at higher values of $r$. The vertical lines indicate the extrema that fitting was carried out over.

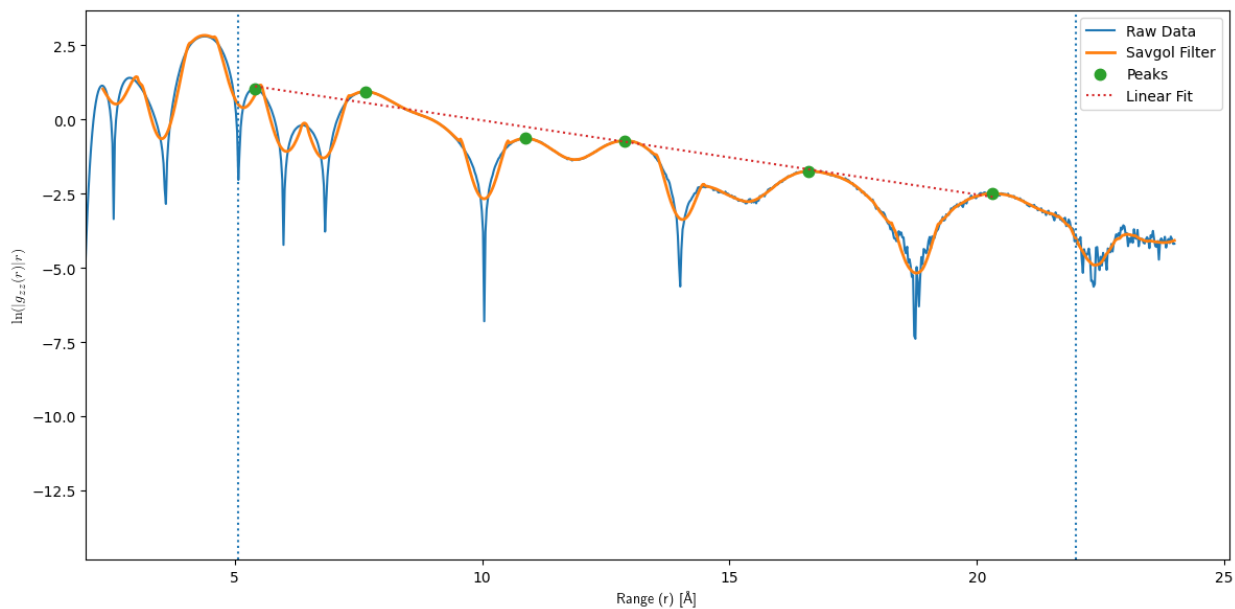

**Figure S20.** Fitting of aqueous LiCl at a concentration of 9.86 M. The envelope of the fit is determined by fitting through points at the top of each of the marked peaks. A Savgol filter was applied to smooth the data to assist peak finding, particularly at higher values of $r$. The vertical lines indicate the extrema that fitting was carried out over.

# NaCl$_{(aq)}$

**Table S3.** Concentrations, number of ion pairs (# IP), the number of solvent molecules (# Solvent) and conventionally-derived $\lambda_S$ values for aqueous NaCl simulations (see Figures S21-S33).

| Conc (m) | Conc (M) | #IP | # Solvent | Decay Length (Å) |
|---|---|---|---|---|
| 0.1 | 0.11 | 8 | 4167 | 10.17 ± 0.24 |
| 0.5 | 0.50 | 38 | 4167 | 3.07 ± 0.15 |
| 0.75 | 0.74 | 56 | 4167 | 2.43 ± 0.05 |
| 1.0 | 0.99 | 75 | 4167 | 2.07 ± 0.05 |
| 1.25 | 1.23 | 94 | 4167 | 1.87 ± 0.08 |
| 1.5 | 1.48 | 113 | 4167 | 1.95 ± 0.24 |
| 2.0 | 1.95 | 150 | 4167 | 1.95 ± 0.15 |
| 2.5 | 2.42 | 188 | 4167 | 2.12 ± 0.25 |
| 3.0 | 2.87 | 225 | 4167 | 2.08 ± 0.10 |
| 3.5 | 3.32 | 263 | 4167 | 2.24 ± 0.14 |
| 4.0 | 3.75 | 300 | 4167 | 2.28 ± 0.19 |
| 5.0 | 4.60 | 375 | 4167 | 2.56 ± 0.28 |
| 6.0 | 5.41 | 450 | 4167 | 3.08 ± 0.49 |

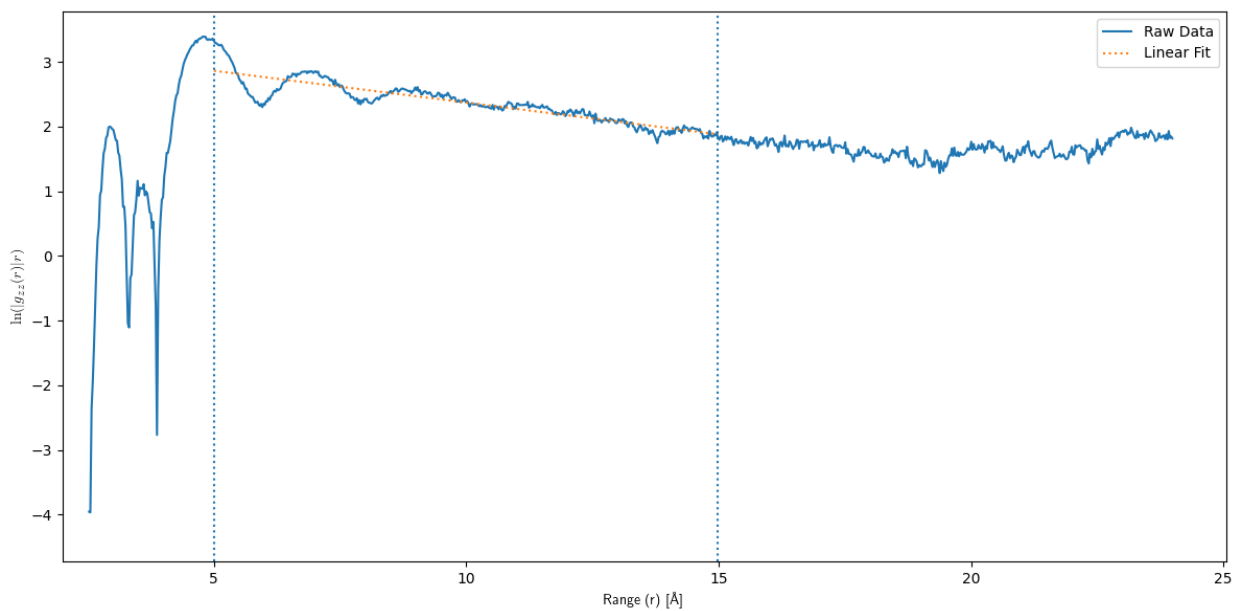

**Figure S21.** Fitting of aqueous NaCl at a concentration of 0.11 M. In this instance, because of the presence of a plateau in the data, a single straight line was used to fit to the points, rather than fitting the peaks. The vertical lines indicate the extrema that fitting was carried out over.

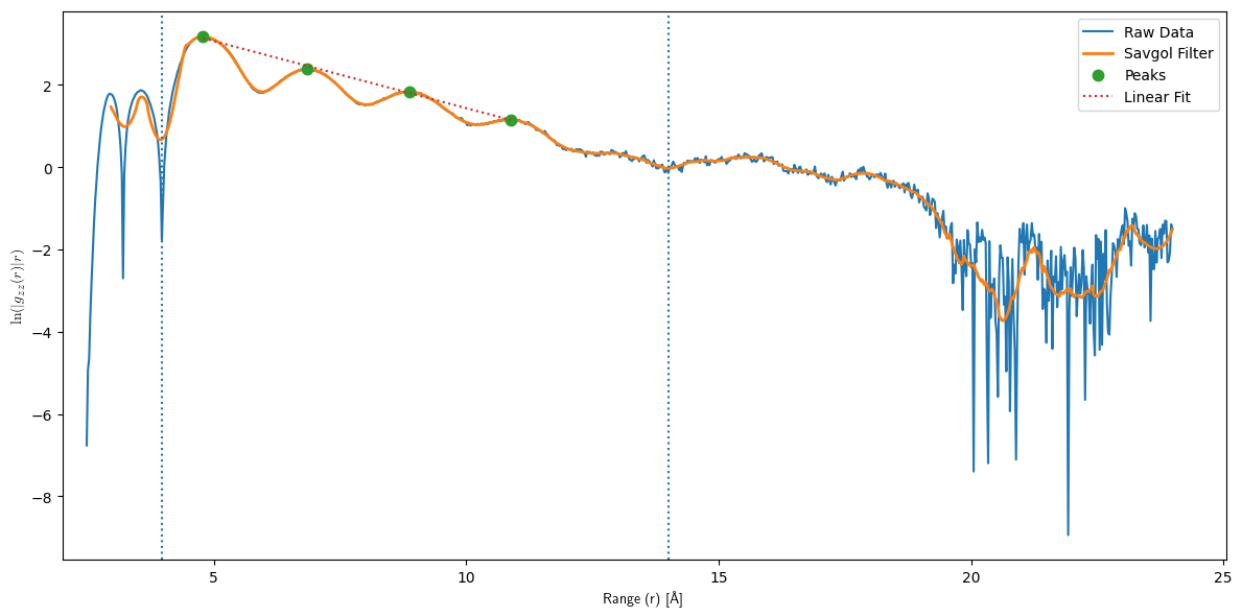

**Figure S22.** Fitting of aqueous NaCl at a concentration of 0.50 M. The envelope of the fit is determined by fitting through points at the top of each of the marked peaks. A Savgol filter was applied to smooth the data to assist peak finding, particularly at higher values of $r$. The vertical lines indicate the extrema that fitting was carried out over.

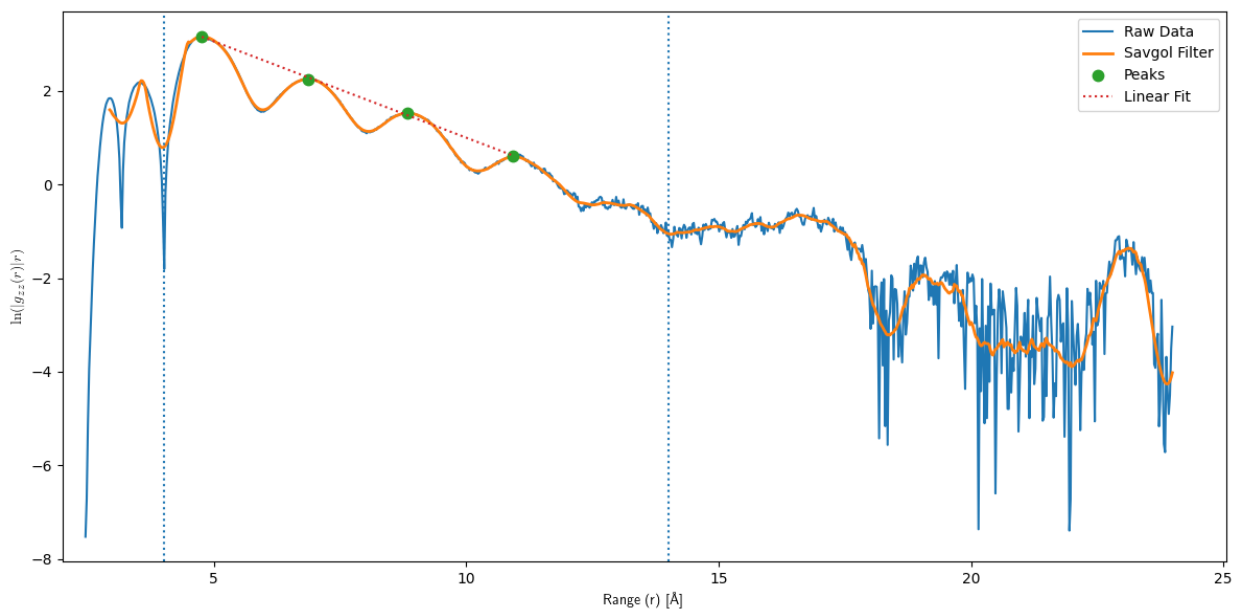

**Figure S23.** Fitting of aqueous NaCl at a concentration of 0.74 M. The envelope of the fit is determined by fitting through points at the top of each of the marked peaks. A Savgol filter was applied to smooth the data to assist peak finding, particularly at higher values of $r$. The vertical lines indicate the extrema that fitting was carried out over.

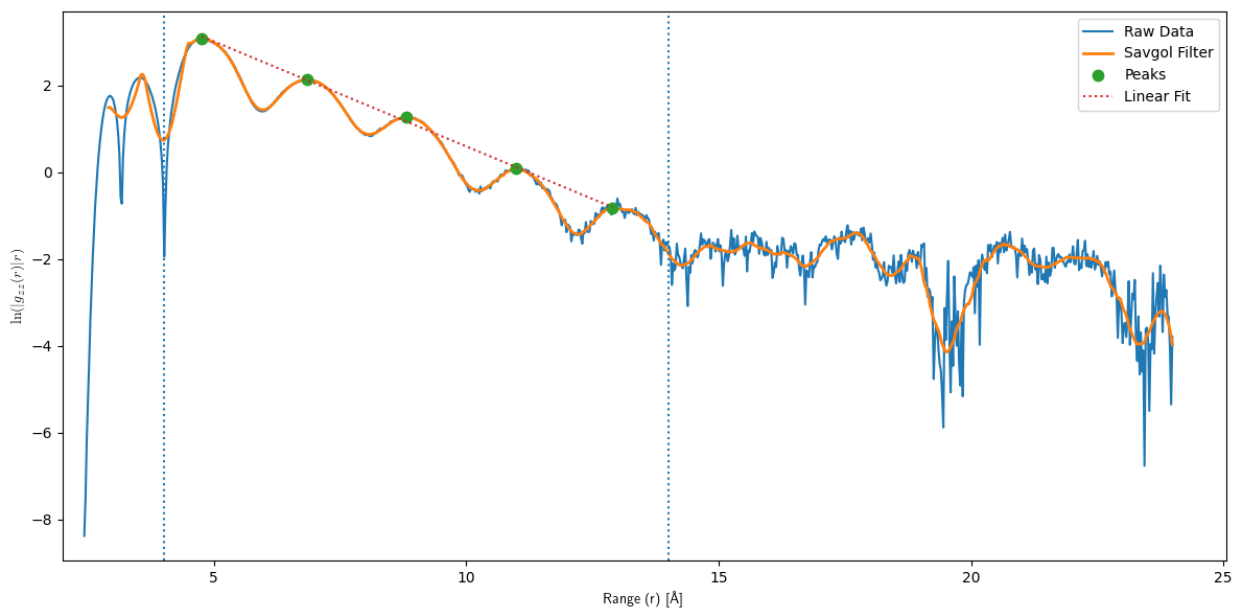

**Figure S24.** Fitting of aqueous NaCl at a concentration of 0.99 M. The envelope of the fit is determined by fitting through points at the top of each of the marked peaks. A Savgol filter was applied to smooth the data to assist peak finding, particularly at higher values of $r$. The vertical lines indicate the extrema that fitting was carried out over.

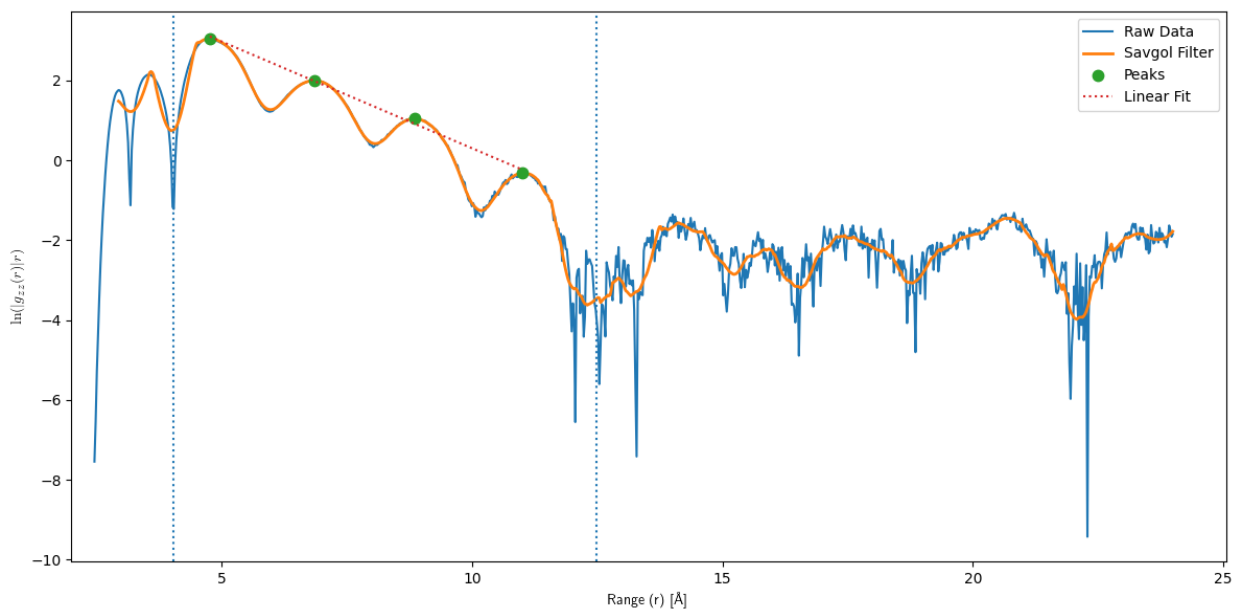

**Figure S25.** Fitting of aqueous NaCl at a concentration of 1.23 M. The envelope of the fit is determined by fitting through points at the top of each of the marked peaks. A Savgol filter was applied to smooth the data to assist peak finding, particularly at higher values of $r$. The vertical lines indicate the extrema that fitting was carried out over.

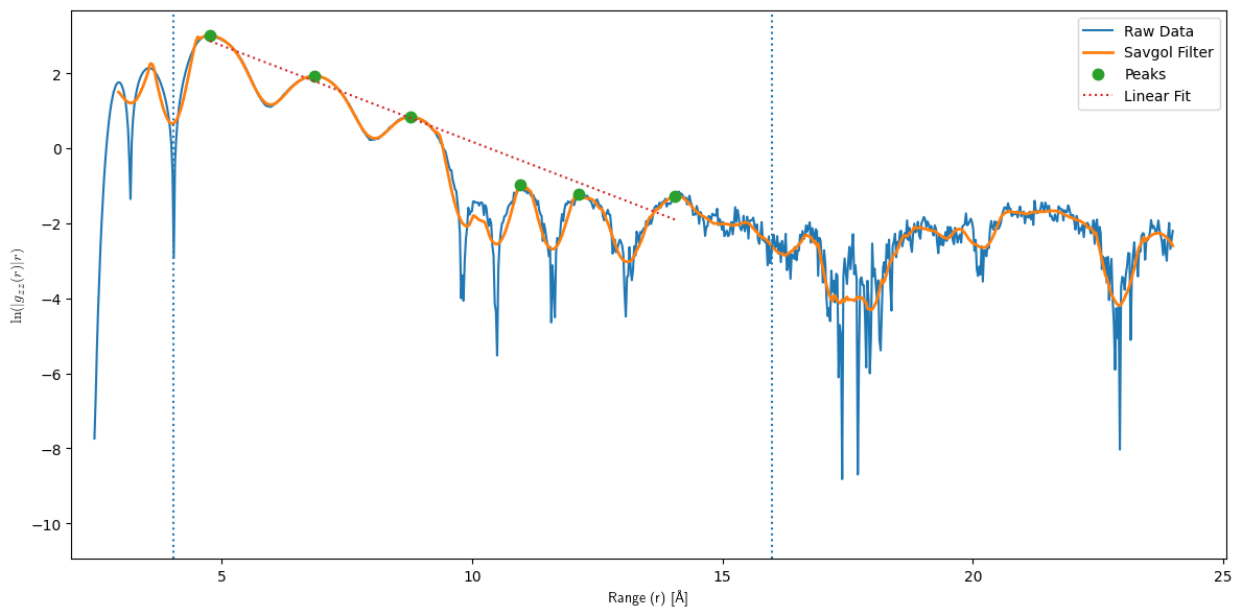

**Figure S26.** Fitting of aqueous NaCl at a concentration of 1.48 M. The envelope of the fit is determined by fitting through points at the top of each of the marked peaks. A Savgol filter was applied to smooth the data to assist peak finding, particularly at higher values of $r$. The vertical lines indicate the extrema that fitting was carried out over.

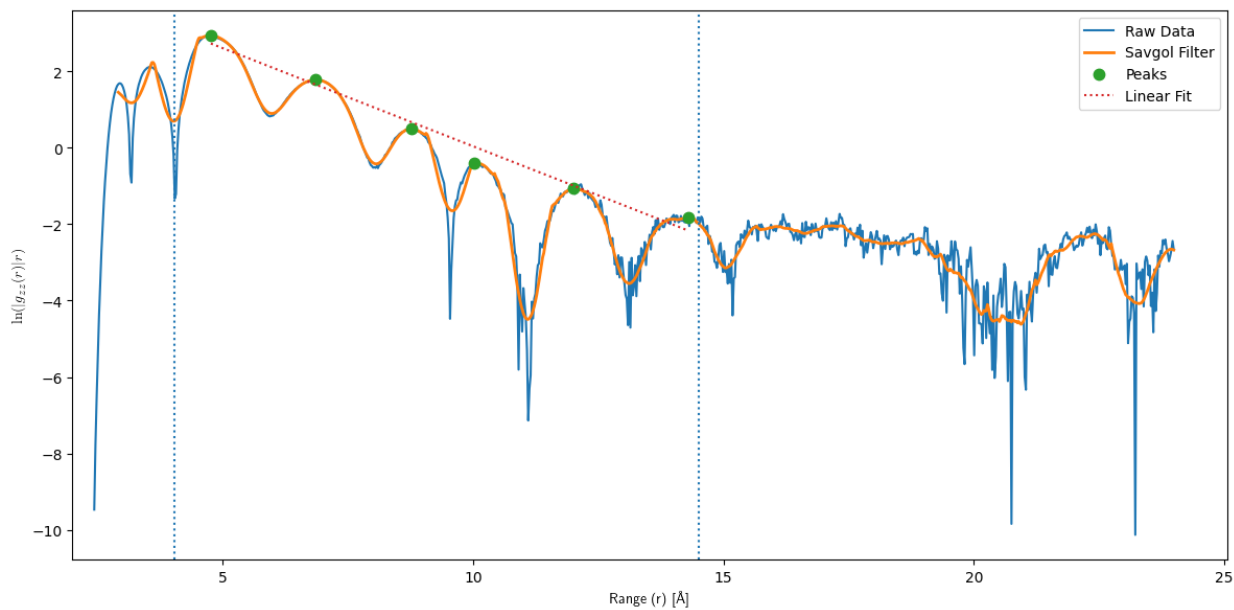

**Figure S27.** Fitting of aqueous NaCl at a concentration of 1.95 M. The envelope of the fit is determined by fitting through points at the top of each of the marked peaks. A Savgol filter was applied to smooth the data to assist peak finding, particularly at higher values of $r$. The vertical lines indicate the extrema that fitting was carried out over.

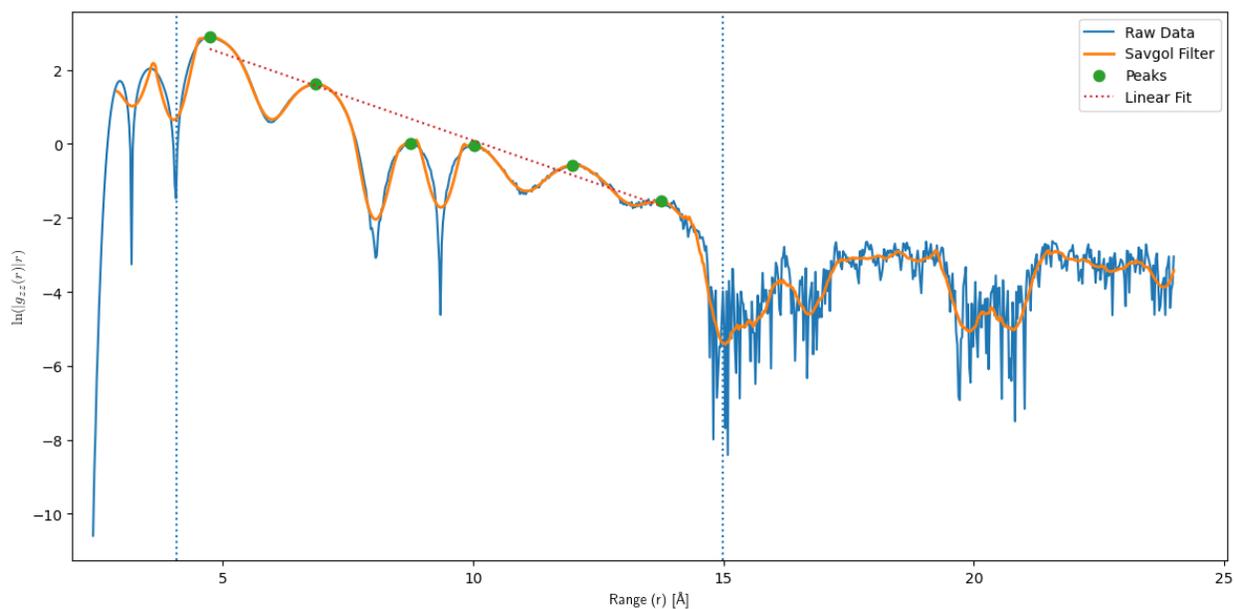

**Figure S28.** Fitting of aqueous NaCl at a concentration of 2.42 M. The envelope of the fit is determined by fitting through points at the top of each of the marked peaks. A Savgol filter was applied to smooth the data to assist peak finding, particularly at higher values of $r$. The vertical lines indicate the extrema that fitting was carried out over.
24

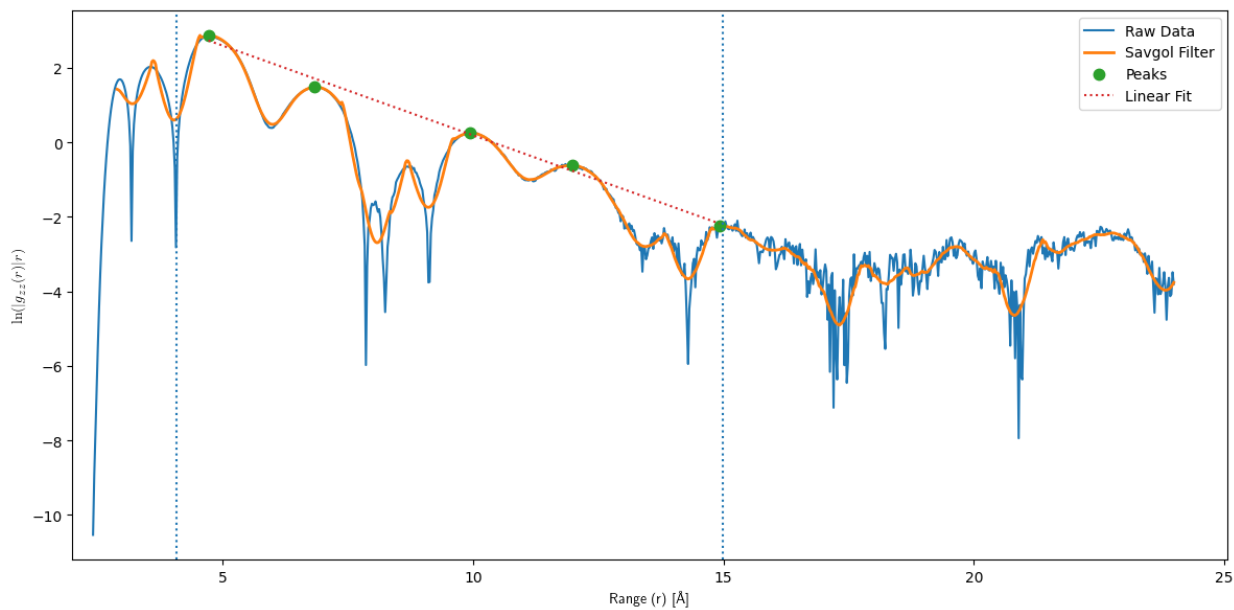

**Figure S29.** Fitting of aqueous NaCl at a concentration of 2.87 M. The envelope of the fit is determined by fitting through points at the top of each of the marked peaks. A Savgol filter was applied to smooth the data to assist peak finding, particularly at higher values of $r$. The vertical lines indicate the extrema that fitting was carried out over.

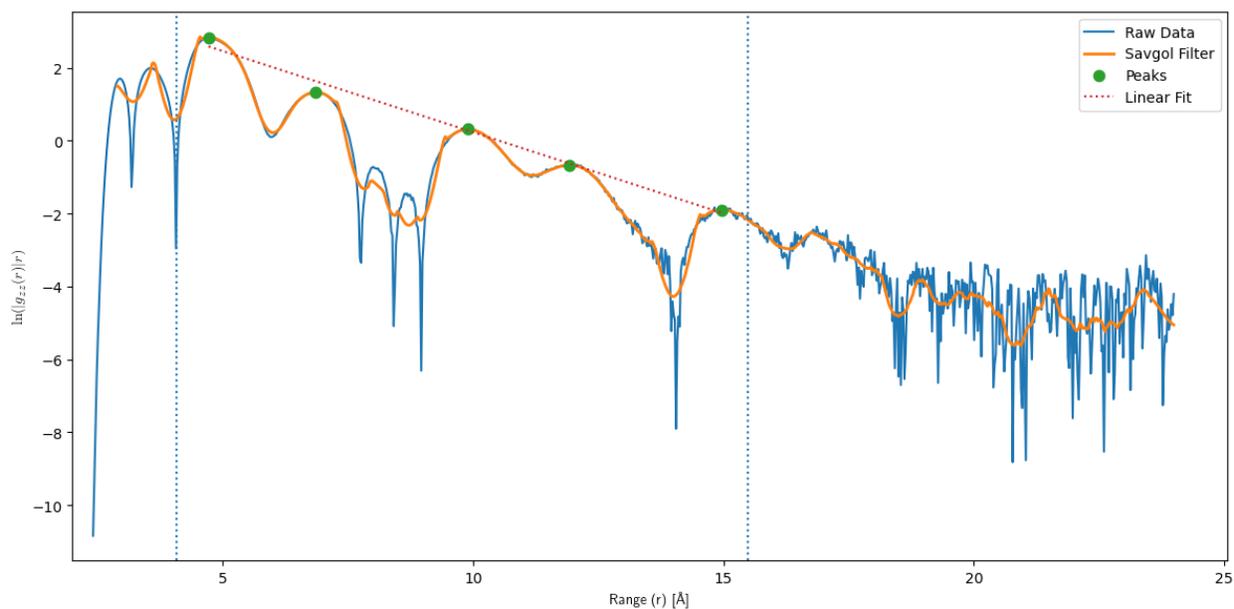

**Figure S30.** Fitting of aqueous NaCl at a concentration of 3.32 M. The envelope of the fit is determined by fitting through points at the top of each of the marked peaks. A Savgol filter was applied to smooth the data to assist peak finding, particularly at higher values of $r$. The vertical lines indicate the extrema that fitting was carried out over.

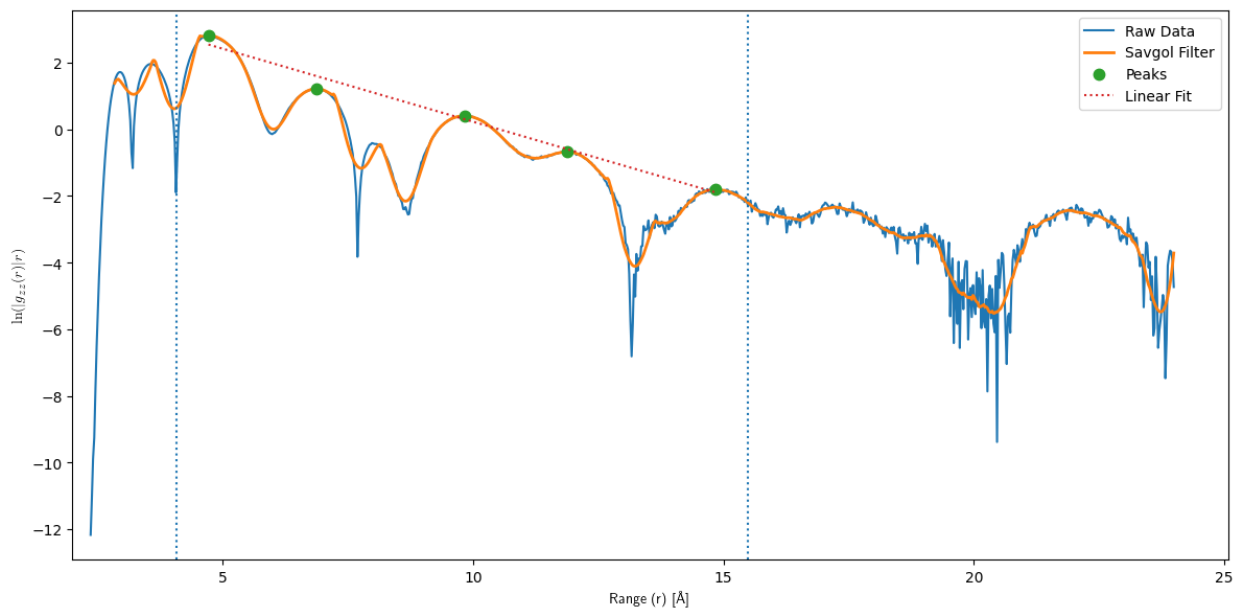

**Figure S31.** Fitting of aqueous NaCl at a concentration of 3.75 M. The envelope of the fit is determined by fitting through points at the top of each of the marked peaks. A Savgol filter was applied to smooth the data to assist peak finding, particularly at higher values of $r$. The vertical lines indicate the extrema that fitting was carried out over.

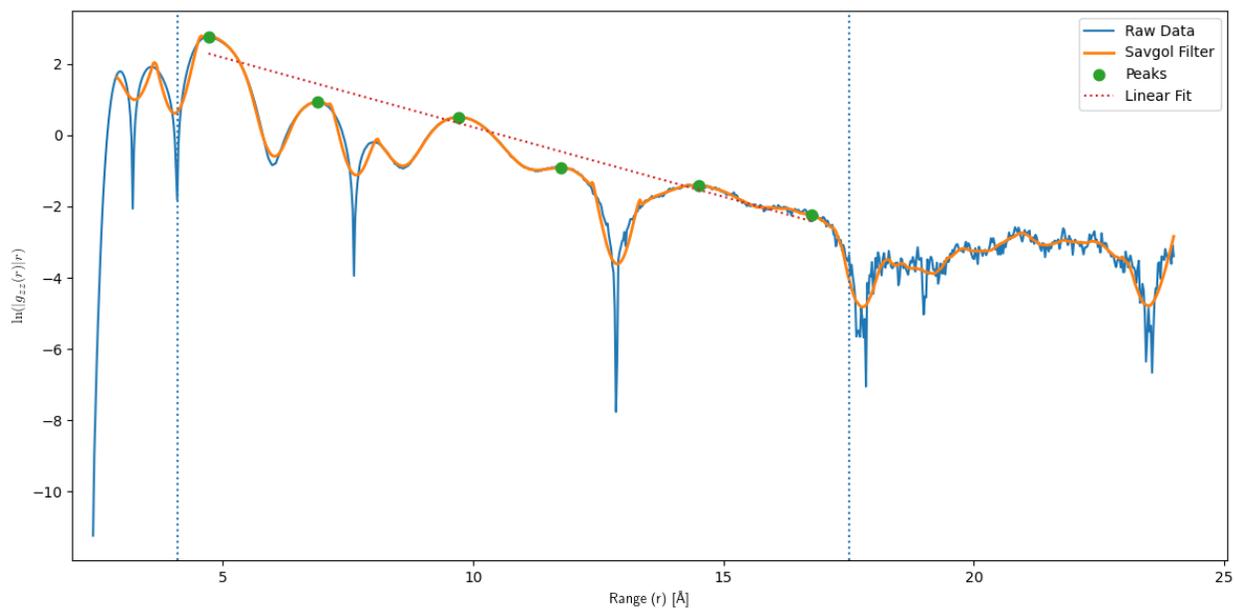

**Figure S32.** Fitting of aqueous NaCl at a concentration of 4.60 M. The envelope of the fit is determined by fitting through points at the top of each of the marked peaks. A Savgol filter was applied to smooth the data to assist peak finding, particularly at higher values of $r$. The vertical lines indicate the extrema that fitting was carried out over.

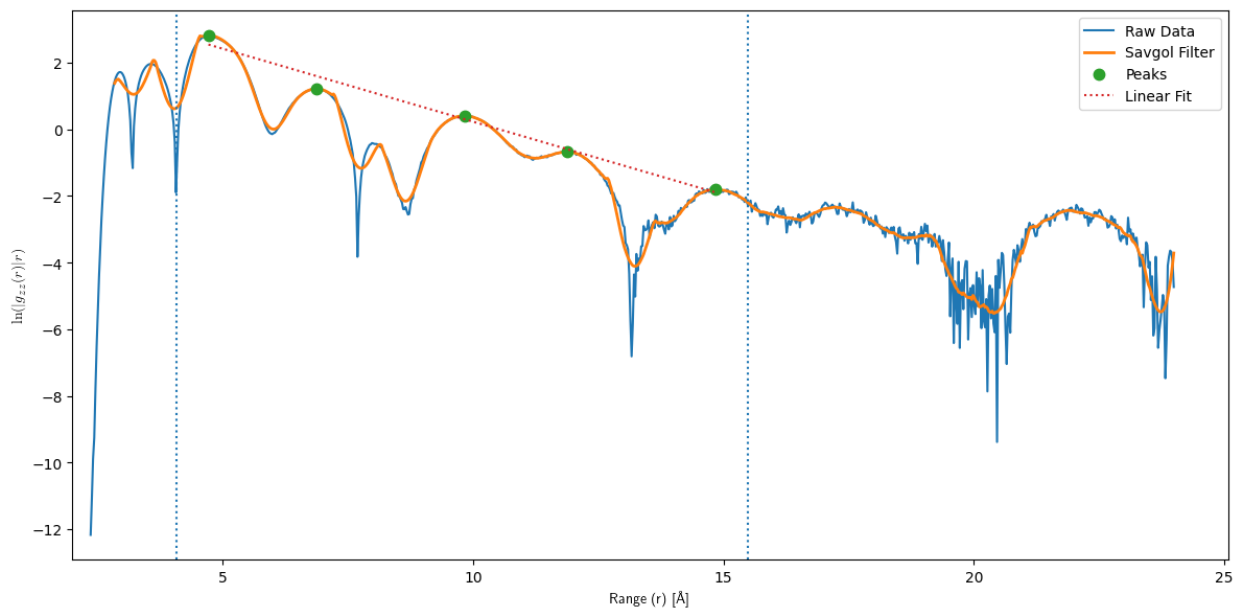

**Figure S31.** Fitting of aqueous NaCl at a concentration of 3.75 M. The envelope of the fit is determined by fitting through points at the top of each of the marked peaks. A Savgol filter was applied to smooth the data to assist peak finding, particularly at higher values of $r$. The vertical lines indicate the extrema that fitting was carried out over.

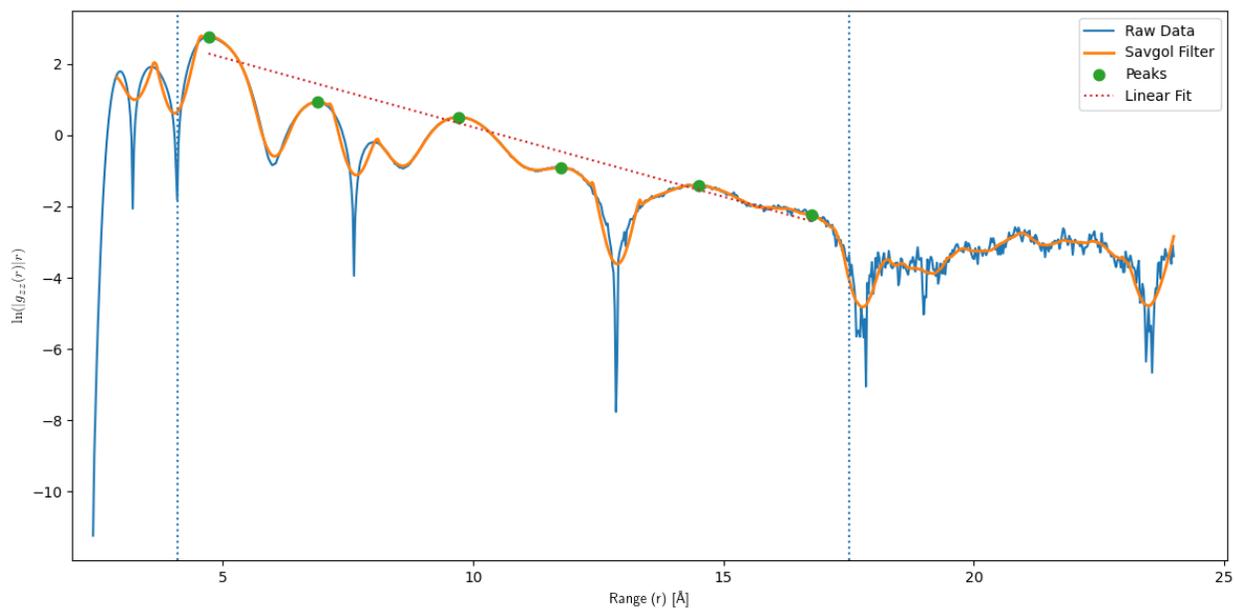

**Figure S32.** Fitting of aqueous NaCl at a concentration of 4.60 M. The envelope of the fit is determined by fitting through points at the top of each of the marked peaks. A Savgol filter was applied to smooth the data to assist peak finding, particularly at higher values of $r$. The vertical lines indicate the extrema that fitting was carried out over.

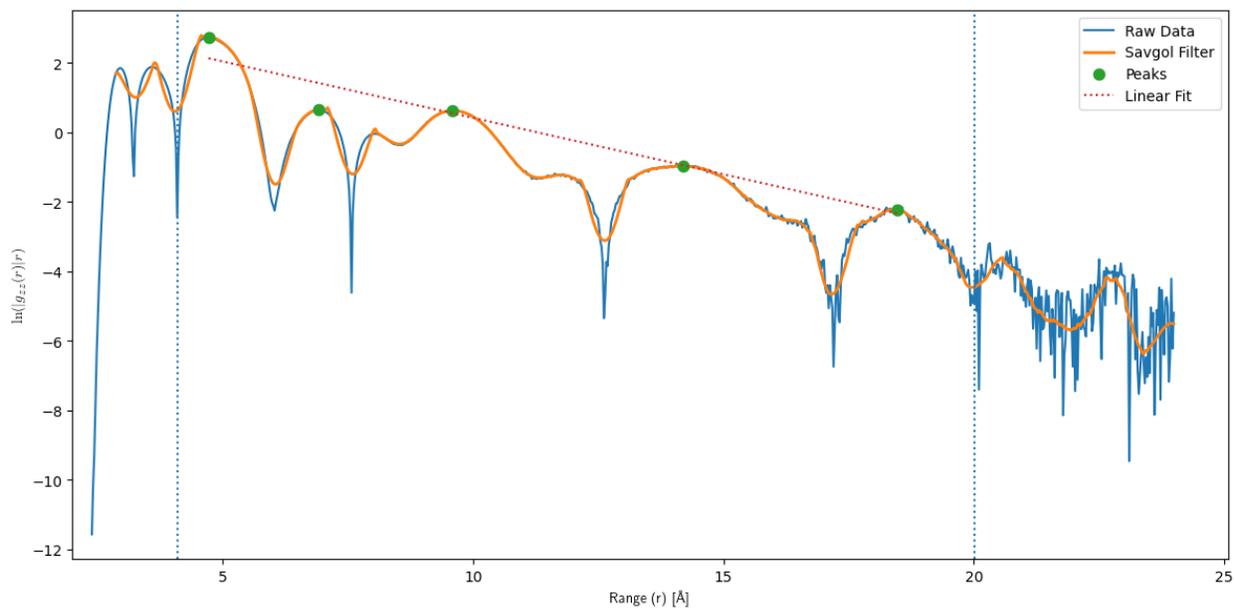

**Figure S33.** Fitting of aqueous NaCl at a concentration of 5.41 M. The envelope of the fit is determined by fitting through points at the top of each of the marked peaks. A Savgol filter was applied to smooth the data to assist peak finding, particularly at higher values of $r$. The vertical lines indicate the extrema that fitting was carried out over.

# KCl$_{(aq)}$

**Table S4.** Concentrations, number of ion pairs (# IP), the number of solvent molecules (# Solvent) and conventionally-derived $\lambda_S$ values for aqueous KCl simulations (see Figures S34-S42).

| Conc (m) | Conc (M) | # IP | # Solvent | Decay Length (Å) |
|---|---|---|---|---|
| 0.1 | 0.11 | 8 | 4167 | 8.60 ± 0.13 |
| 0.5 | 0.50 | 38 | 4167 | 3.76 ± 0.33 |
| 0.75 | 0.73 | 56 | 4167 | 2.70 ± 0.08 |
| 1.0 | 0.98 | 75 | 4167 | 2.35 ± 0.04 |
| 1.25 | 1.22 | 94 | 4167 | 2.45 ± 0.05 |
| 1.5 | 1.45 | 113 | 4167 | 2.02 ± 0.16 |
| 2.0 | 1.90 | 150 | 4167 | 1.79 ± 0.06 |
| 3.0 | 2.78 | 225 | 4167 | 1.76 ± 0.09 |
| 4.0 | 3.60 | 300 | 4167 | 2.15 ± 0.07 |

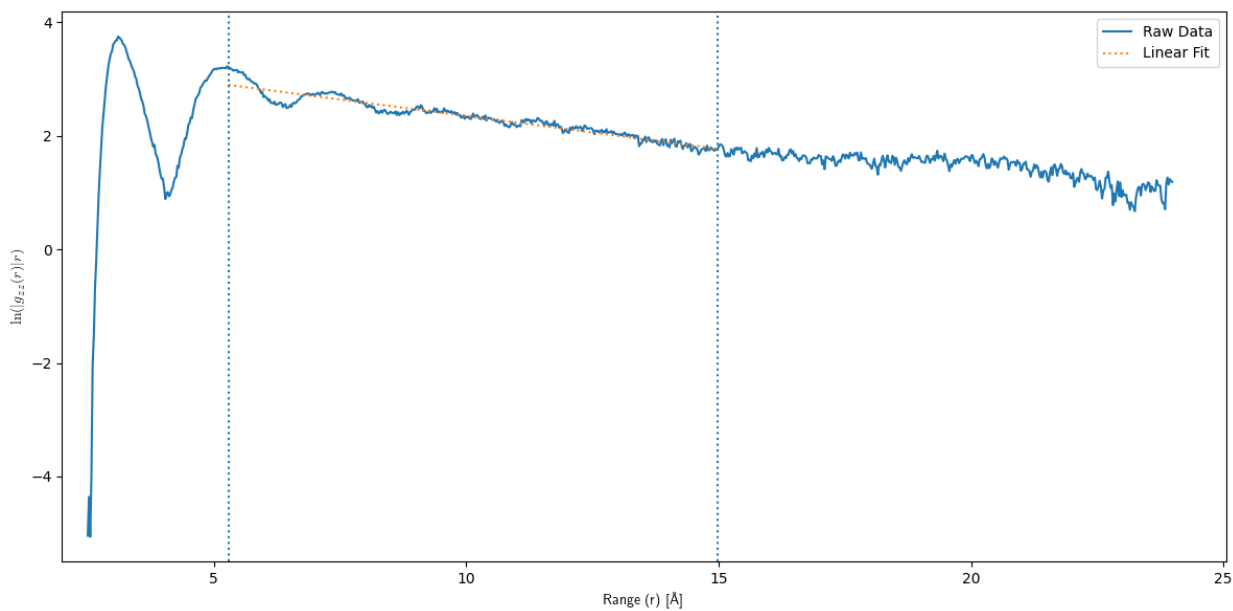

**Figure S34.** Fitting of aqueous KCl at a concentration of 0.11 M. In this instance, because of the presence of a plateau in the data, a single straight line was used to fit to the points, rather than fitting the peaks. The vertical lines indicate the extrema that fitting was carried out over.

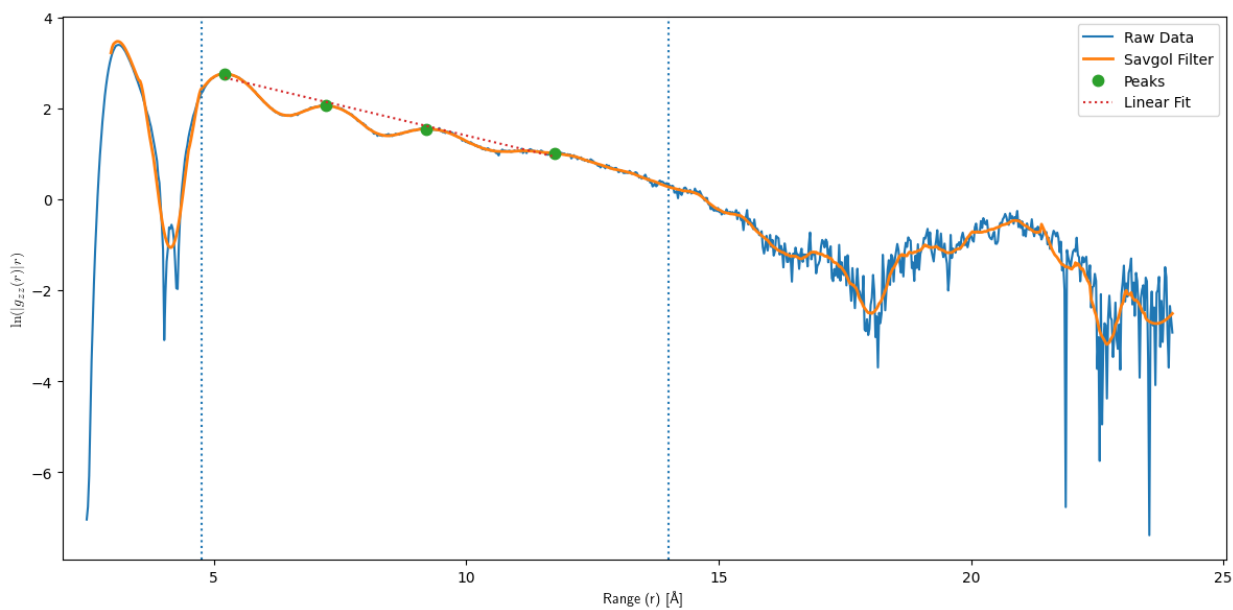

**Figure S35.** Fitting of aqueous KCl at a concentration of 0.50 M. The envelope of the fit is determined by fitting through points at the top of each of the marked peaks. A Savgol filter was applied to smooth the data to assist peak finding, particularly at higher values of $r$. The vertical lines indicate the extrema that fitting was carried out over.

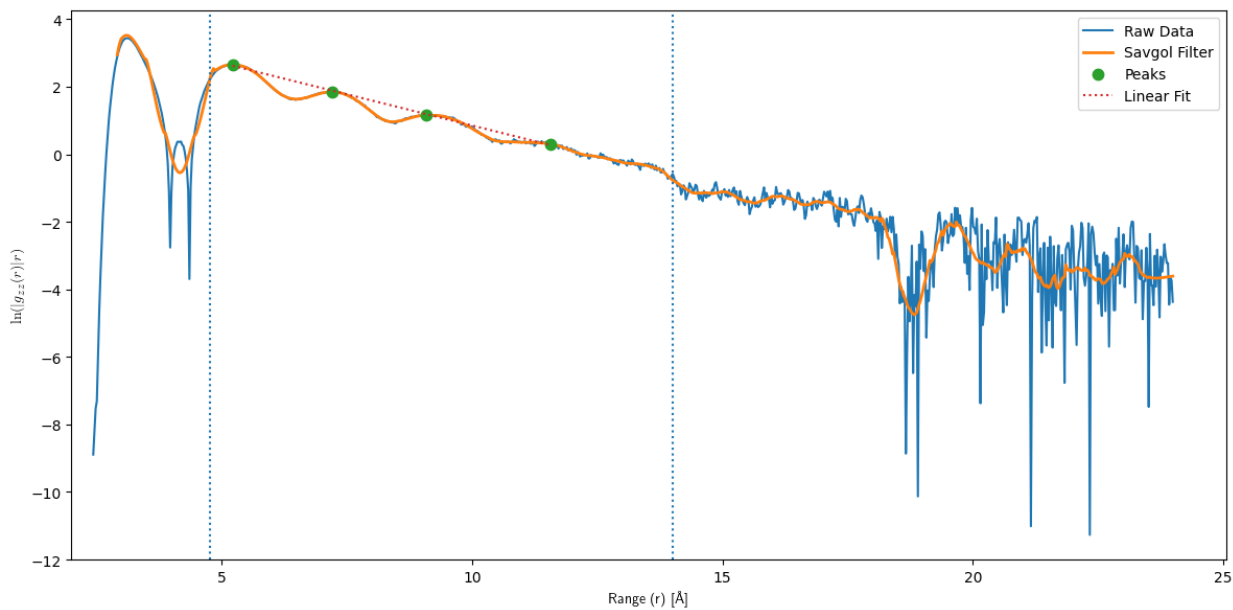

**Figure S36.** Fitting of aqueous KCl at a concentration of 0.73 M. The envelope of the fit is determined by fitting through points at the top of each of the marked peaks. A Savgol filter was applied to smooth the data to assist peak finding, particularly at higher values of $r$. The vertical lines indicate the extrema that fitting was carried out over.

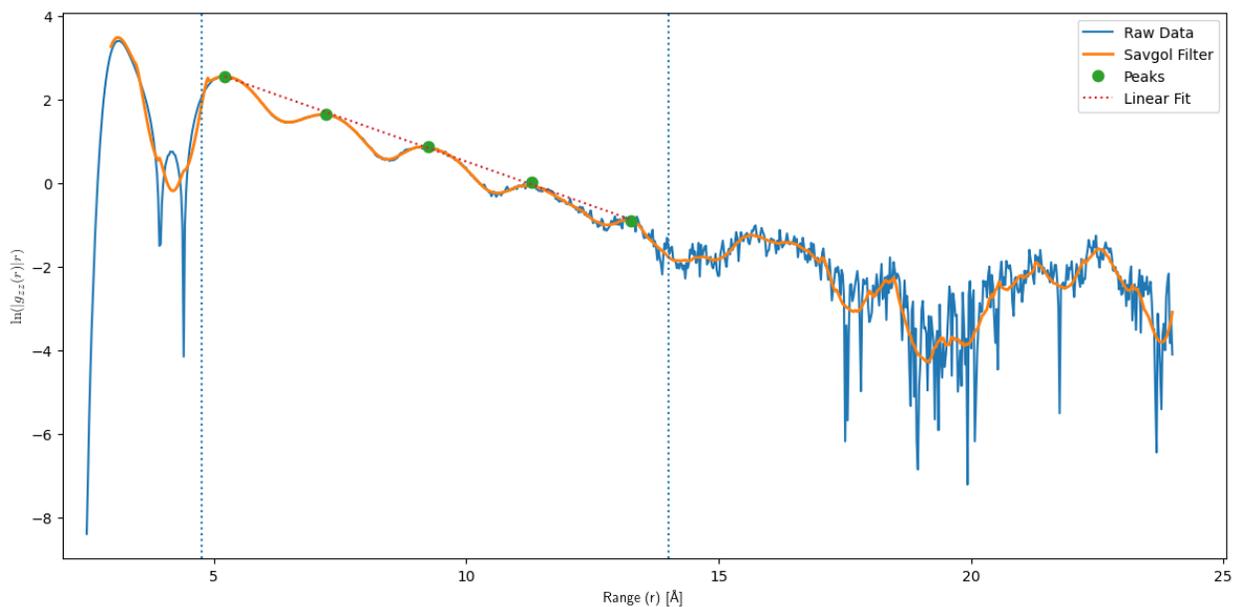

**Figure S37.** Fitting of aqueous KCl at a concentration of 0.98 M. The envelope of the fit is determined by fitting through points at the top of each of the marked peaks. A Savgol filter was applied to smooth the data to assist peak finding, particularly at higher values of $r$. The vertical lines indicate the extrema that fitting was carried out over.

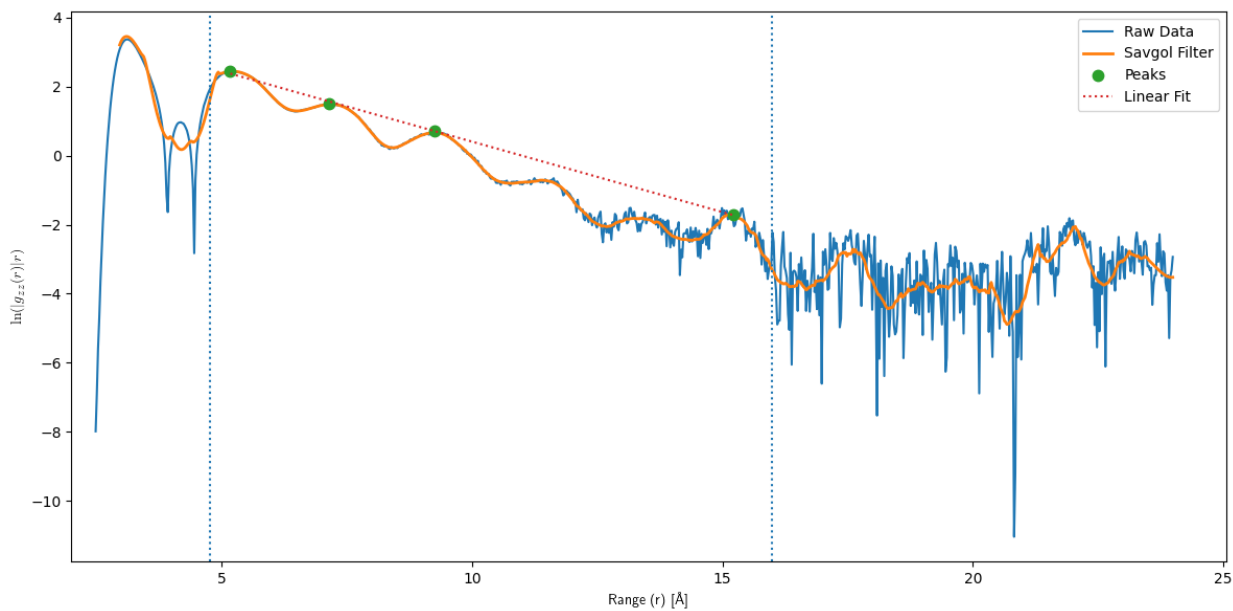

**Figure S38.** Fitting of aqueous KCl at a concentration of 1.22 M. The envelope of the fit is determined by fitting through points at the top of each of the marked peaks. A Savgol filter was applied to smooth the data to assist peak finding, particularly at higher values of $r$. The vertical lines indicate the extrema that fitting was carried out over.

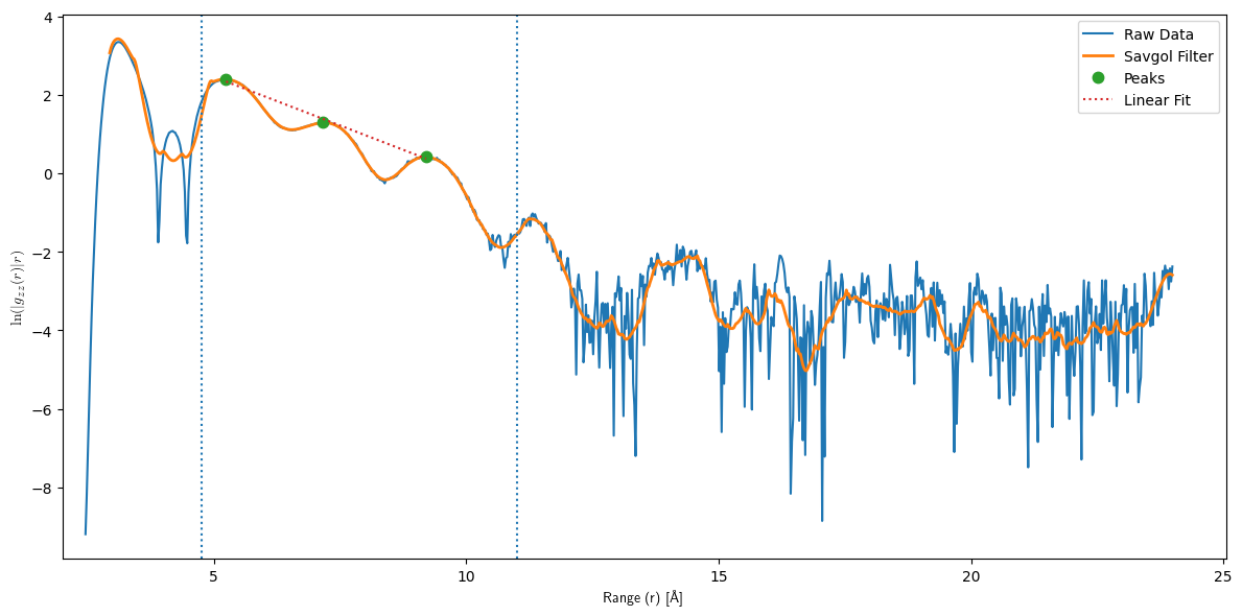

**Figure S39.** Fitting of aqueous KCl at a concentration of 1.45 M. The envelope of the fit is determined by fitting through points at the top of each of the marked peaks. A Savgol filter was applied to smooth the data to assist peak finding, particularly at higher values of $r$. The vertical lines indicate the extrema that fitting was carried out over.

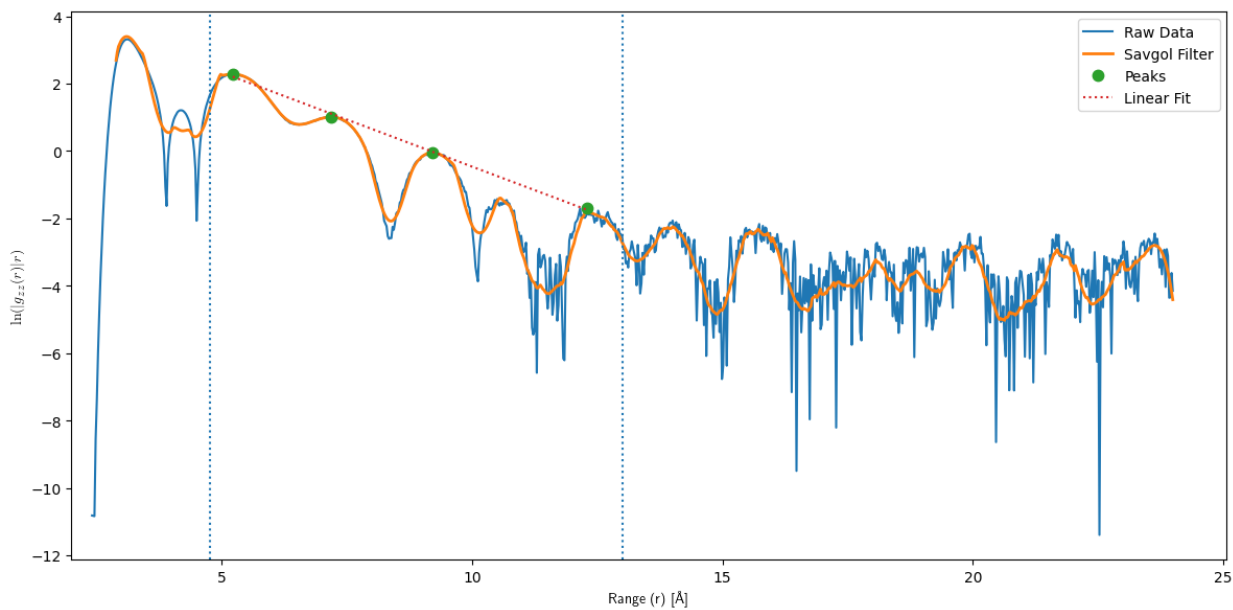

**Figure S40.** Fitting of aqueous KCl at a concentration of 1.90 M. The envelope of the fit is determined by fitting through points at the top of each of the marked peaks. A Savgol filter was applied to smooth the data to assist peak finding, particularly at higher values of $r$. The vertical lines indicate the extrema that fitting was carried out over.

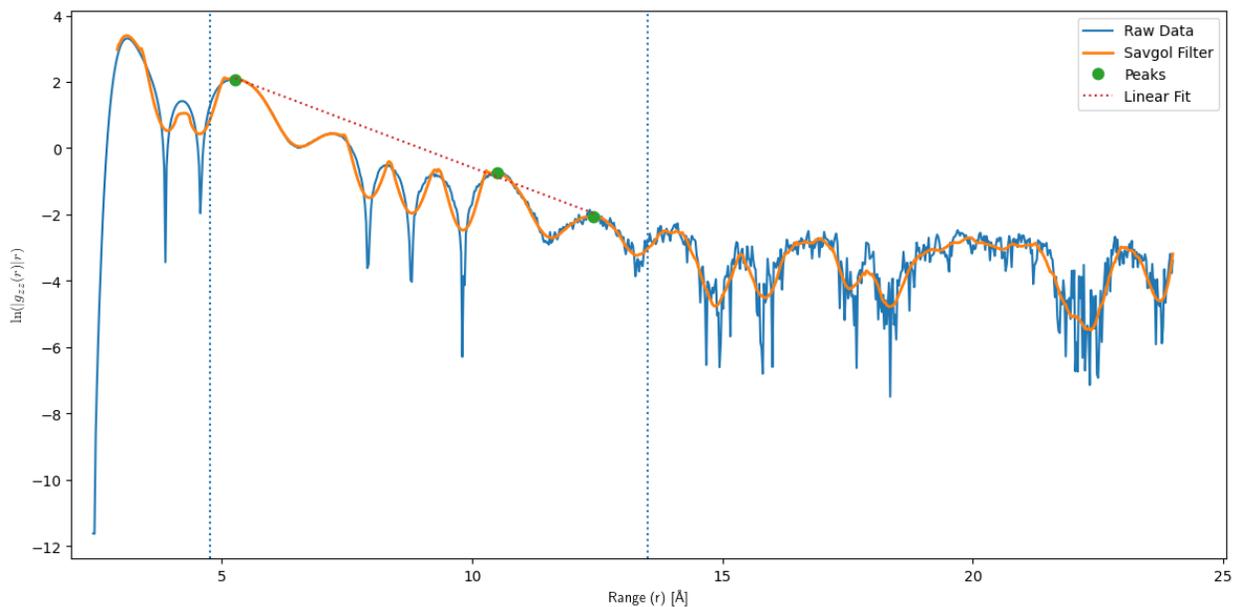

**Figure S41.** Fitting of aqueous KCl at a concentration of 2.78 M. The envelope of the fit is determined by fitting through points at the top of each of the marked peaks. A Savgol filter was applied to smooth the data to assist peak finding, particularly at higher values of $r$. The vertical lines indicate the extrema that fitting was carried out over.

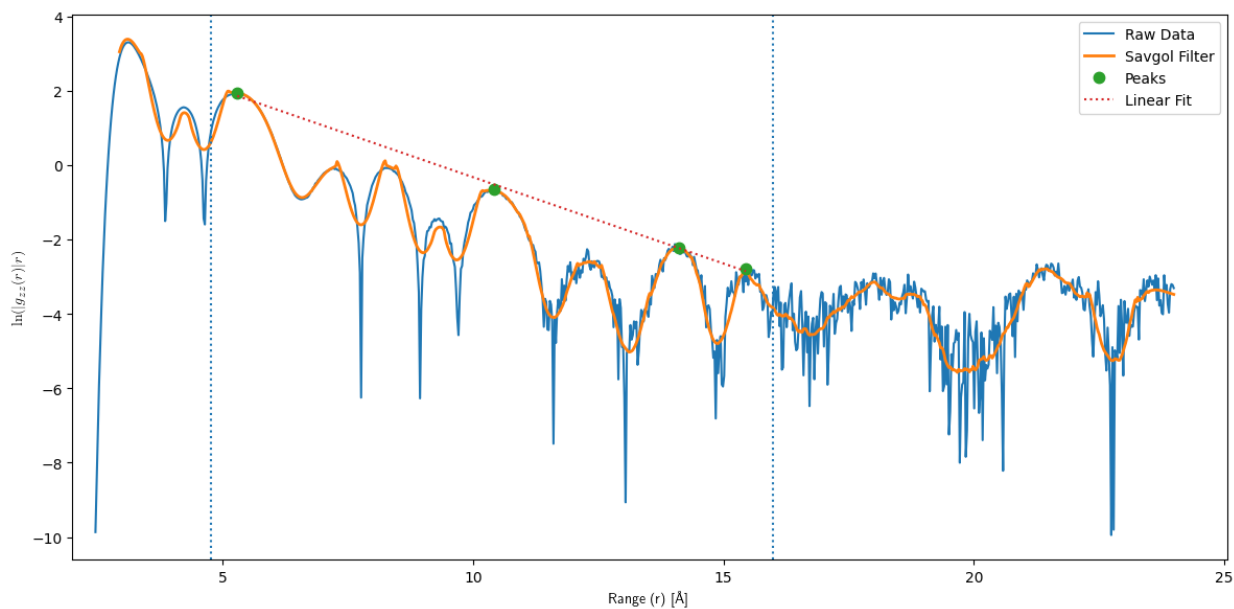

**Figure S42.** Fitting of aqueous KCl at a concentration of 3.60 M. The envelope of the fit is determined by fitting through points at the top of each of the marked peaks. A Savgol filter was applied to smooth the data to assist peak finding, particularly at higher values of $r$. The vertical lines indicate the extrema that fitting was carried out over.

# RbCl$_{(aq)}$

**Table S5.** Concentrations, number of ion pairs (# IP), the number of solvent molecules (# Solvent) and conventionally-derived $\lambda_S$ values for aqueous RbCl simulations (see Figures S43-S54).

| Conc (m) | Conc (M) | # IP | # Solvent | Decay Length (Å) |
|---|---|---|---|---|
| 0.1 | 0.11 | 8 | 4167 | 7.36 ± 0.11 |
| 0.5 | 0.50 | 38 | 4167 | 3.58 ± 0.33 |
| 0.75 | 0.73 | 56 | 4167 | 2.80 ± 0.11 |
| 1.0 | 0.97 | 75 | 4167 | 2.10 ± 0.04 |
| 1.25 | 1.21 | 94 | 4167 | 2.01 ± 0.12 |
| 1.5 | 1.44 | 113 | 4167 | 1.66 ± 0.03 |
| 2.0 | 1.88 | 150 | 4167 | 1.43 ± 0.03 |
| 3.0 | 2.74 | 225 | 4167 | 1.75 ± 0.19 |
| 4.0 | 3.53 | 300 | 4167 | 1.48 ± 0.09 |
| 5.0 | 4.27 | 375 | 4167 | 1.93 ± 0.22 |
| 6.0 | 4.96 | 450 | 4167 | 2.02 ± 0.15 |
| 7.0 | 5.61 | 525 | 4167 | 1.88 ± 0.09 |

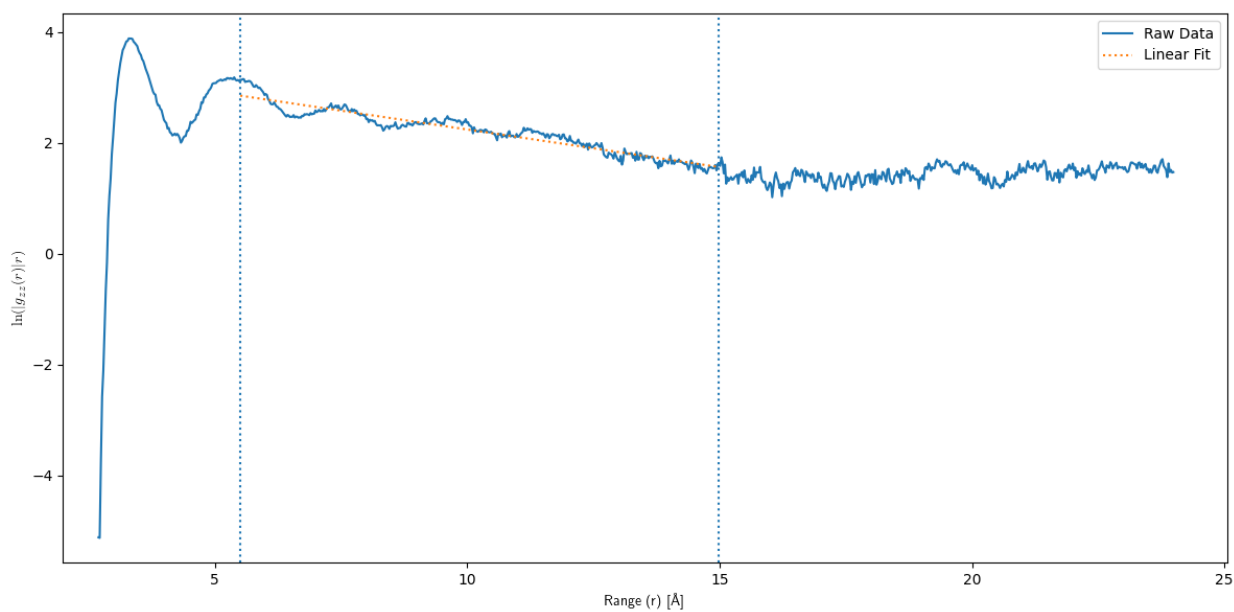

**Figure S43.** Fitting of aqueous RbCl at a concentration of 0.11 M. In this instance, because of the presence of a plateau in the data, a single straight line was used to fit to the points, rather than fitting the peaks. The vertical lines indicate the extrema that fitting was carried out over.

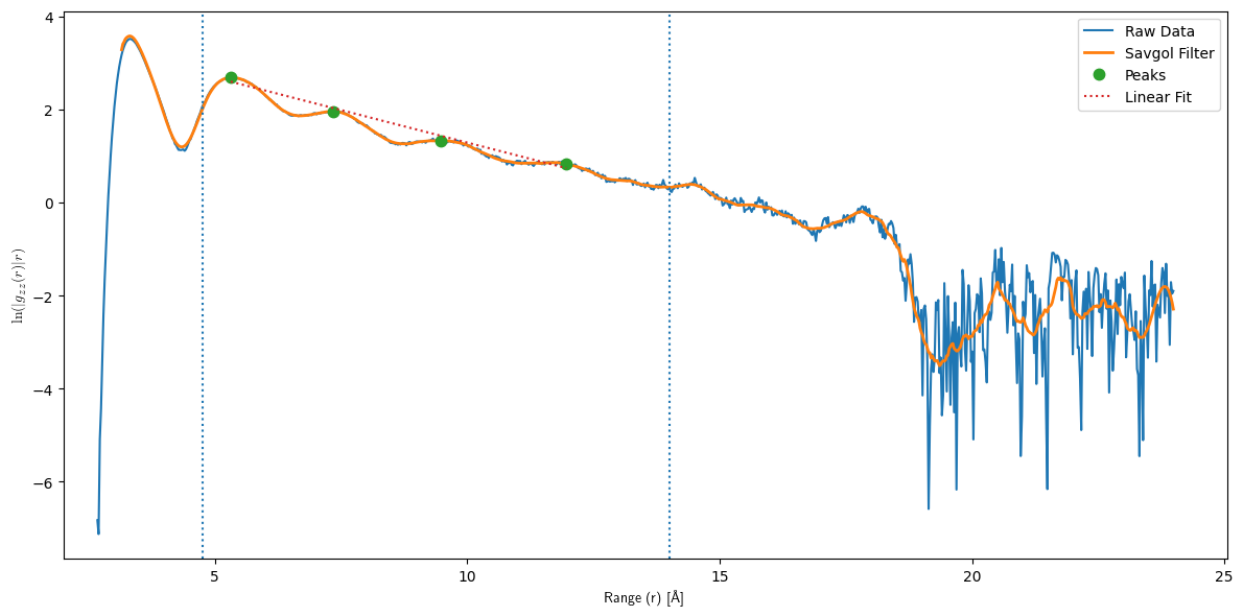

**Figure S44.** Fitting of aqueous RbCl at a concentration of 0.50 M. The envelope of the fit is determined by fitting through points at the top of each of the marked peaks. A Savgol filter was applied to smooth the data to assist peak finding, particularly at higher values of $r$. The vertical lines indicate the extrema that fitting was carried out over.

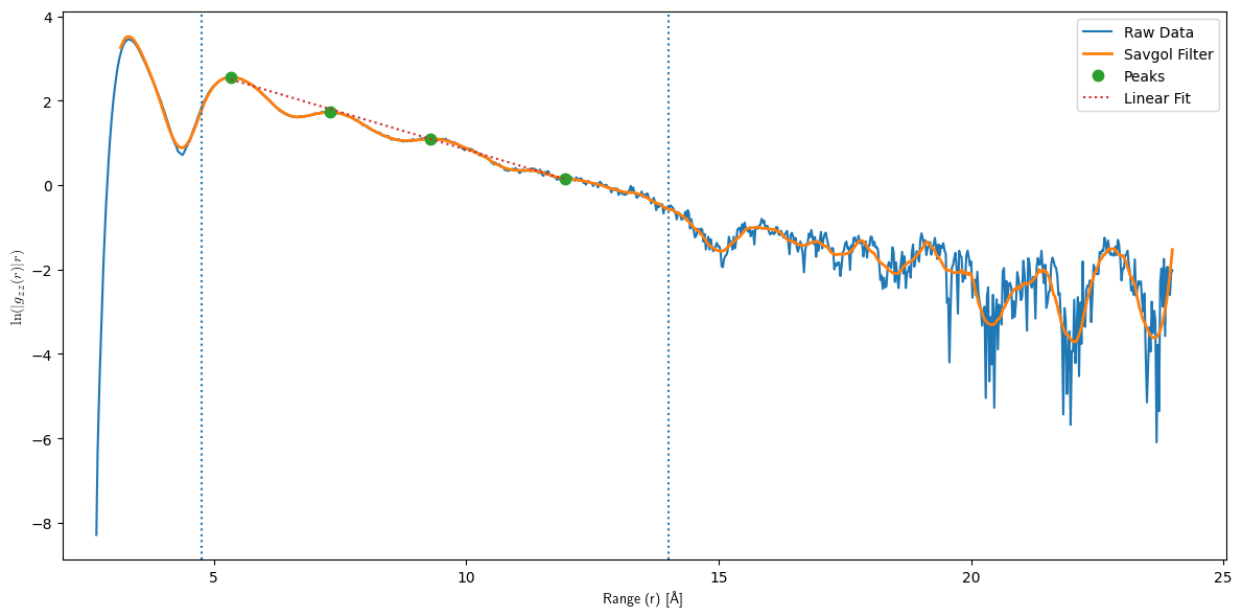

**Figure S45.** Fitting of aqueous RbCl at a concentration of 0.73 M. The envelope of the fit is determined by fitting through points at the top of each of the marked peaks. A Savgol filter was applied to smooth the data to assist peak finding, particularly at higher values of $r$. The vertical lines indicate the extrema that fitting was carried out over.

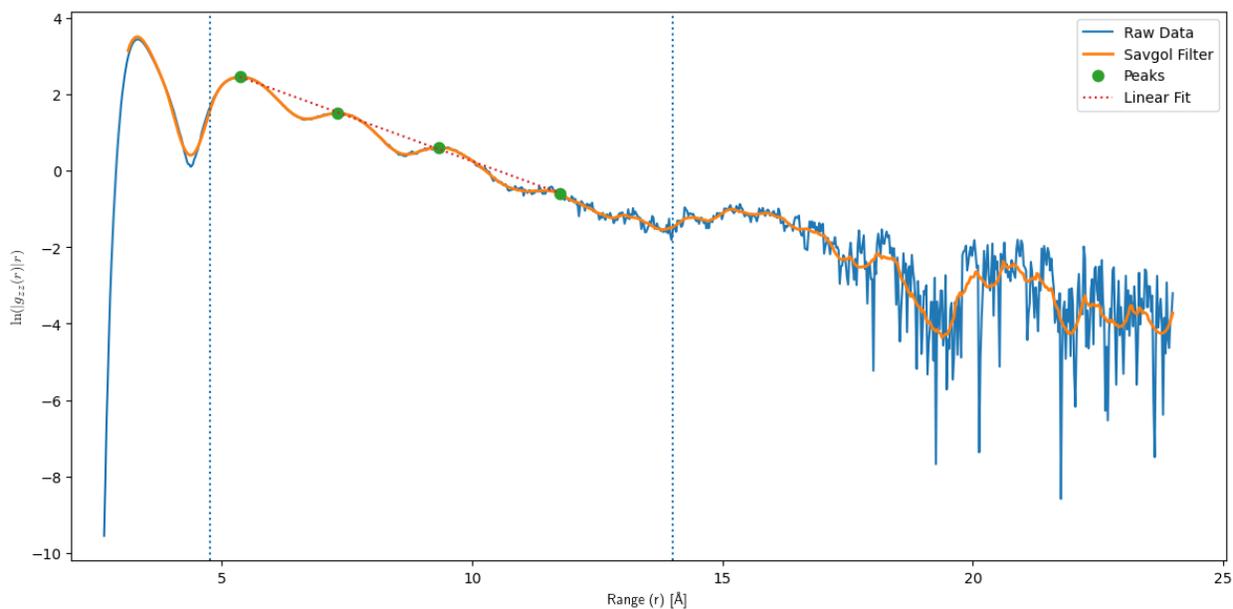

**Figure S46.** Fitting of aqueous RbCl at a concentration of 0.97 M. The envelope of the fit is determined by fitting through points at the top of each of the marked peaks. A Savgol filter was applied to smooth the data to assist peak finding, particularly at higher values of $r$. The vertical lines indicate the extrema that fitting was carried out over.

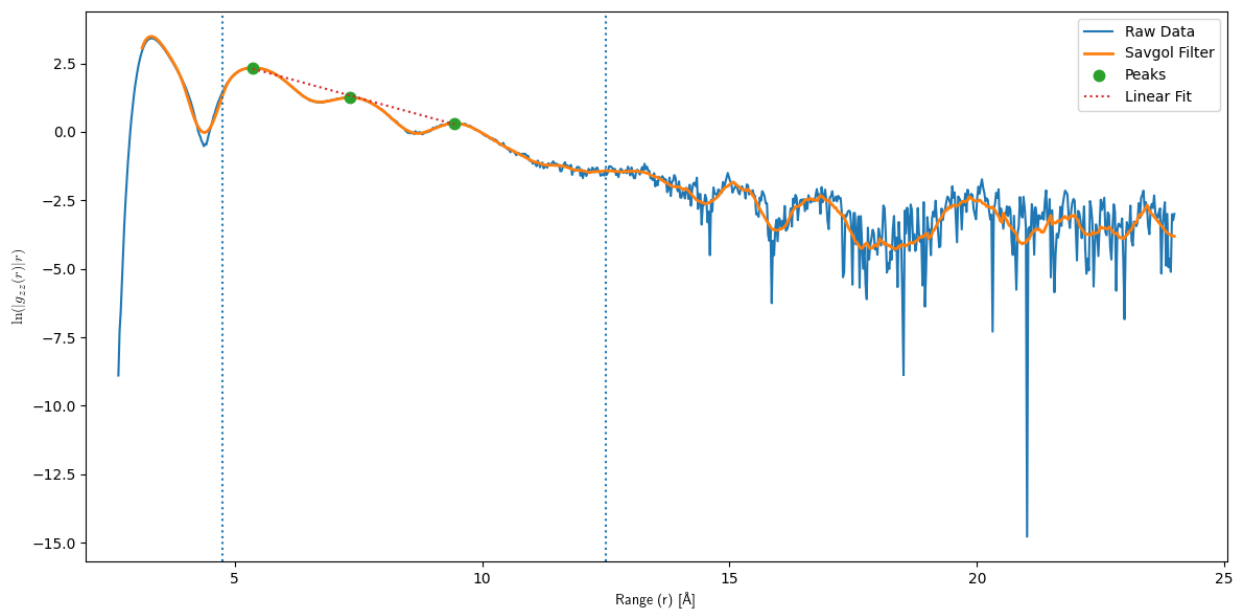

**Figure S47.** Fitting of aqueous RbCl at a concentration of 1.21 M. The envelope of the fit is determined by fitting through points at the top of each of the marked peaks. A Savgol filter was applied to smooth the data to assist peak finding, particularly at higher values of $r$. The vertical lines indicate the extrema that fitting was carried out over.

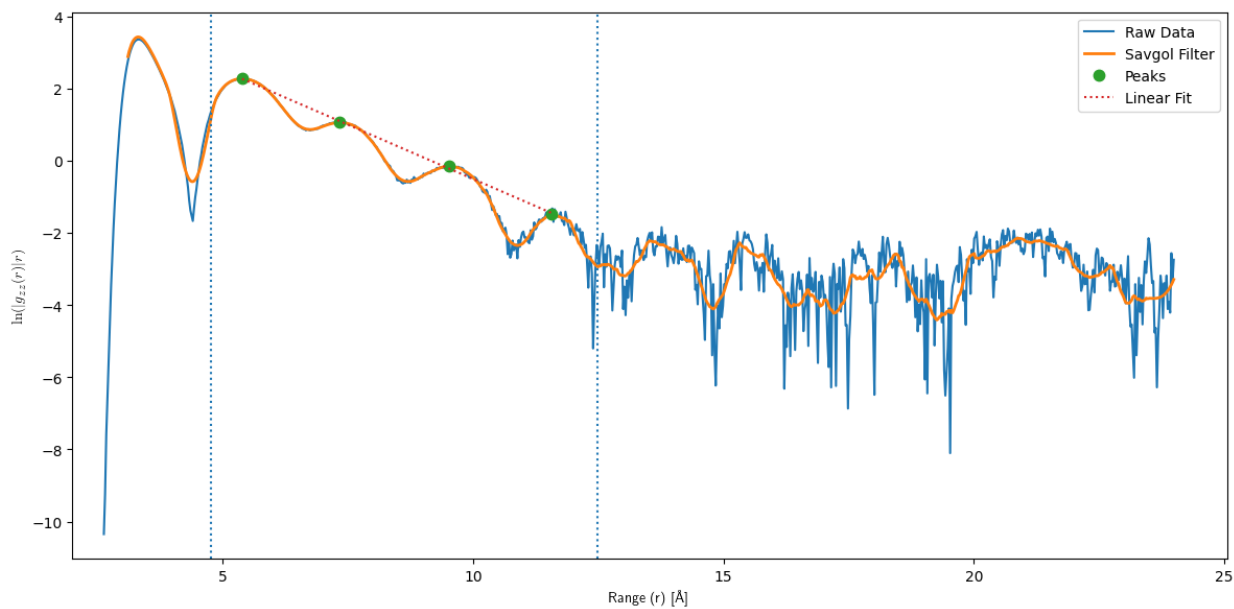

**Figure S48.** Fitting of aqueous RbCl at a concentration of 1.44 M. The envelope of the fit is determined by fitting through points at the top of each of the marked peaks. A Savgol filter was applied to smooth the data to assist peak finding, particularly at higher values of $r$. The vertical lines indicate the extrema that fitting was carried out over.

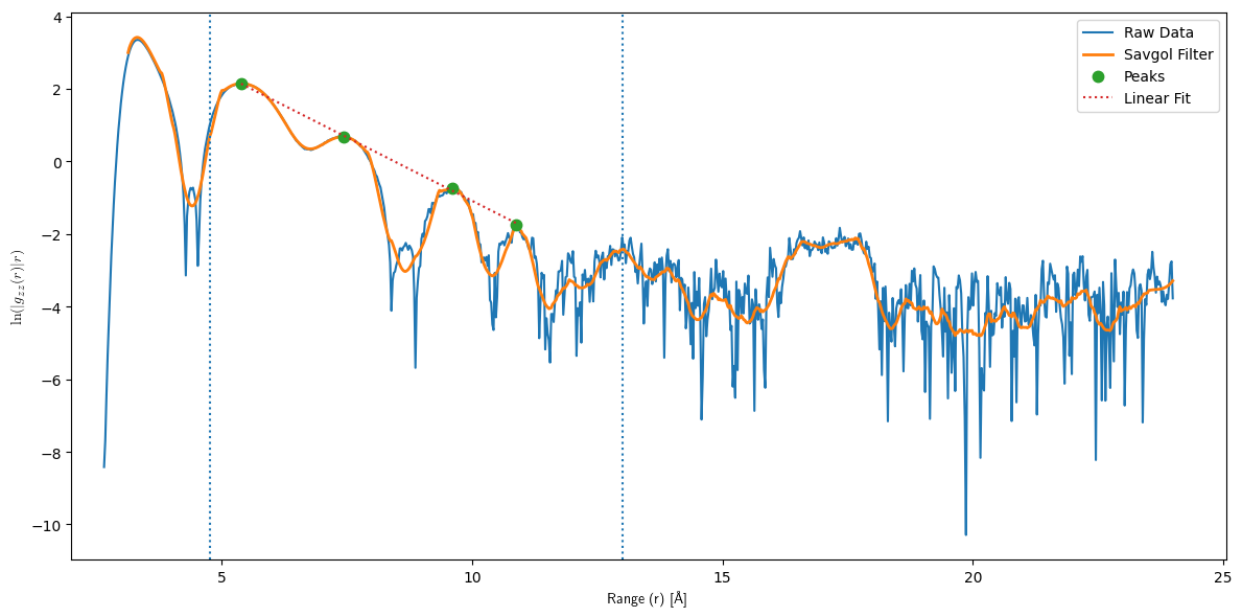

**Figure S49.** Fitting of aqueous RbCl at a concentration of 1.88 M. The envelope of the fit is determined by fitting through points at the top of each of the marked peaks. A Savgol filter was applied to smooth the data to assist peak finding, particularly at higher values of $r$. The vertical lines indicate the extrema that fitting was carried out over.

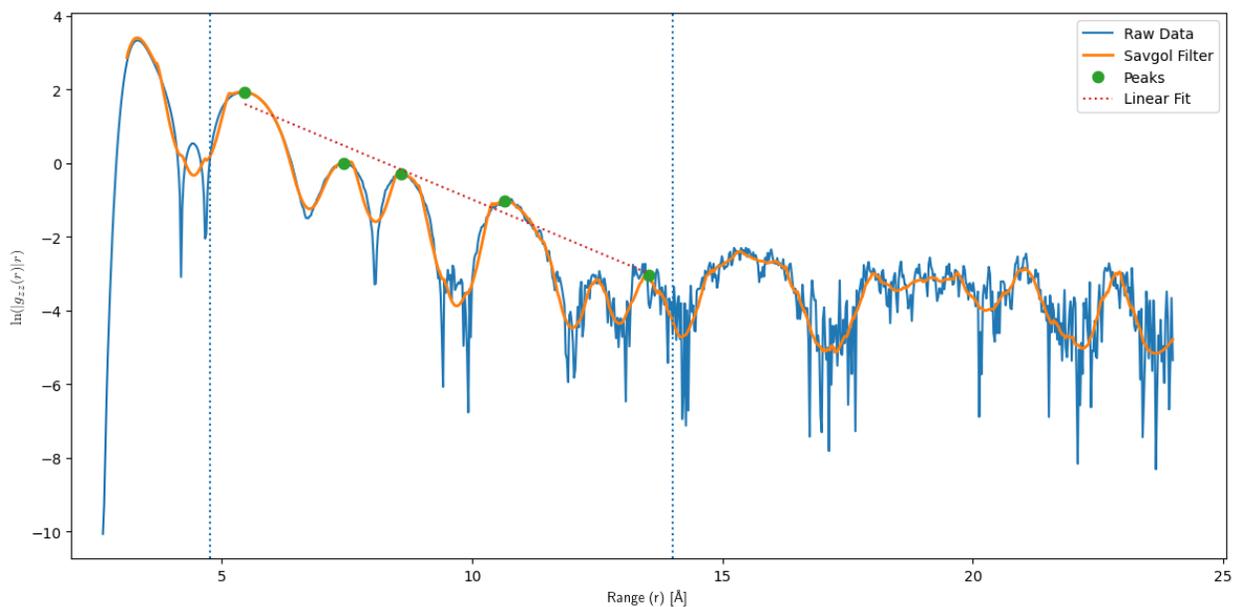

**Figure S50.** Fitting of aqueous RbCl at a concentration of 2.74 M. The envelope of the fit is determined by fitting through points at the top of each of the marked peaks. A Savgol filter was applied to smooth the data to assist peak finding, particularly at higher values of $r$. The vertical lines indicate the extrema that fitting was carried out over.

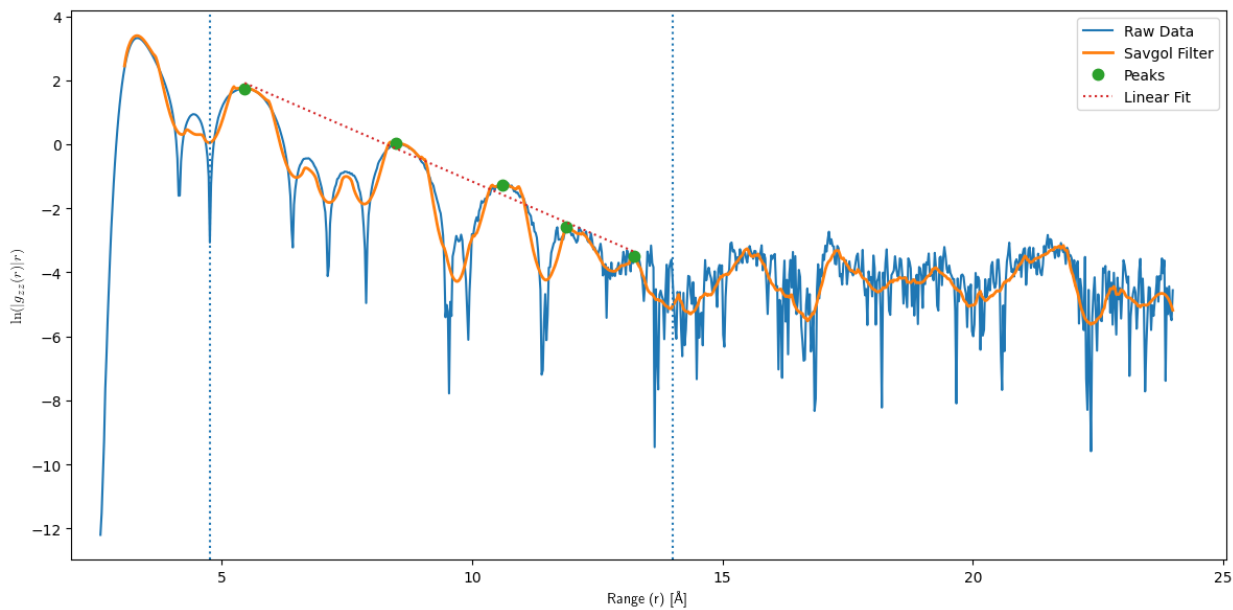

**Figure S51.** Fitting of aqueous RbCl at a concentration of 3.53 M. The envelope of the fit is determined by fitting through points at the top of each of the marked peaks. A Savgol filter was applied to smooth the data to assist peak finding, particularly at higher values of $r$. The vertical lines indicate the extrema that fitting was carried out over.

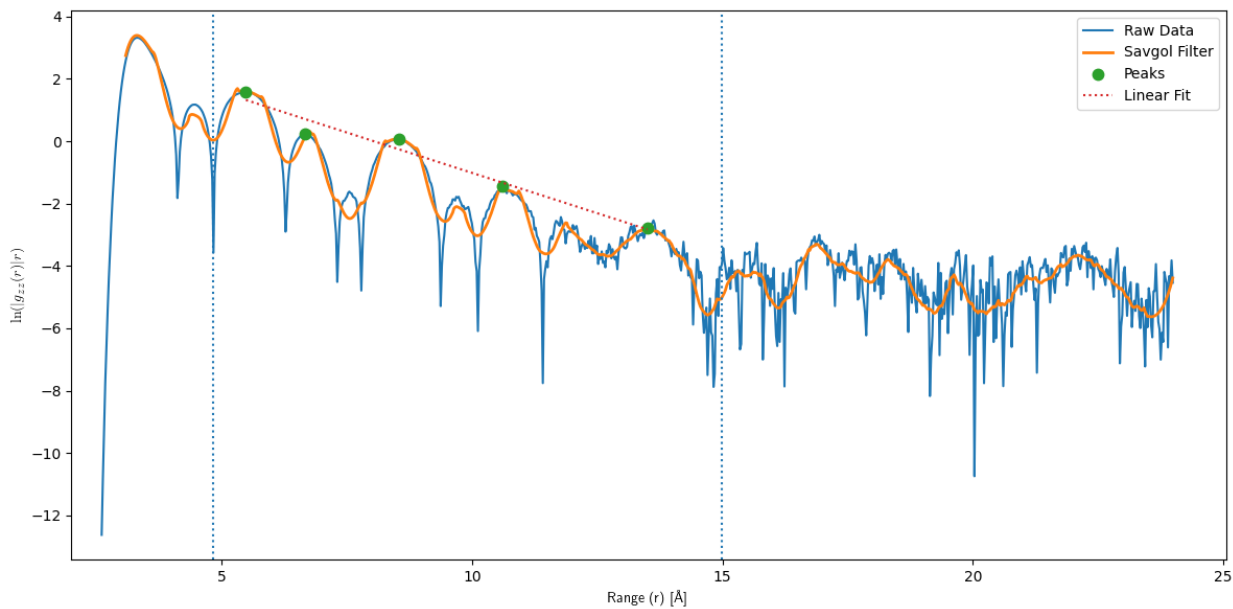

**Figure S52.** Fitting of aqueous RbCl at a concentration of 4.27 M. The envelope of the fit is determined by fitting through points at the top of each of the marked peaks. A Savgol filter was applied to smooth the data to assist peak finding, particularly at higher values of $r$. The vertical lines indicate the extrema that fitting was carried out over.

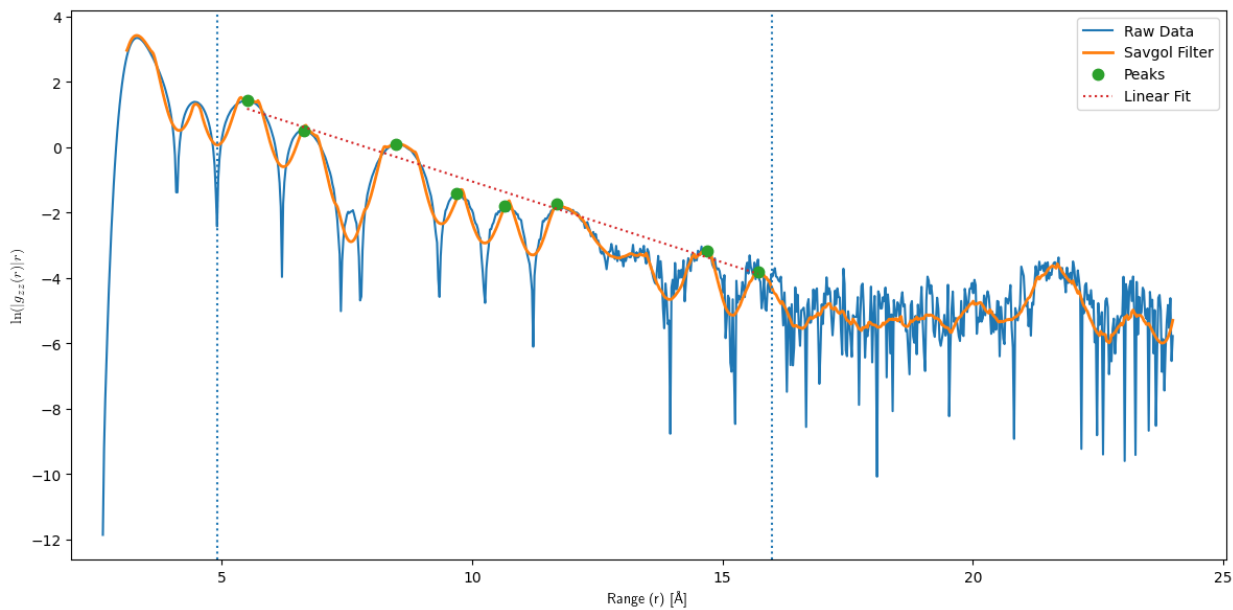

**Figure S53.** Fitting of aqueous RbCl at a concentration of 4.96 M. The envelope of the fit is determined by fitting through points at the top of each of the marked peaks. A Savgol filter was applied to smooth the data to assist peak finding, particularly at higher values of $r$. The vertical lines indicate the extrema that fitting was carried out over.

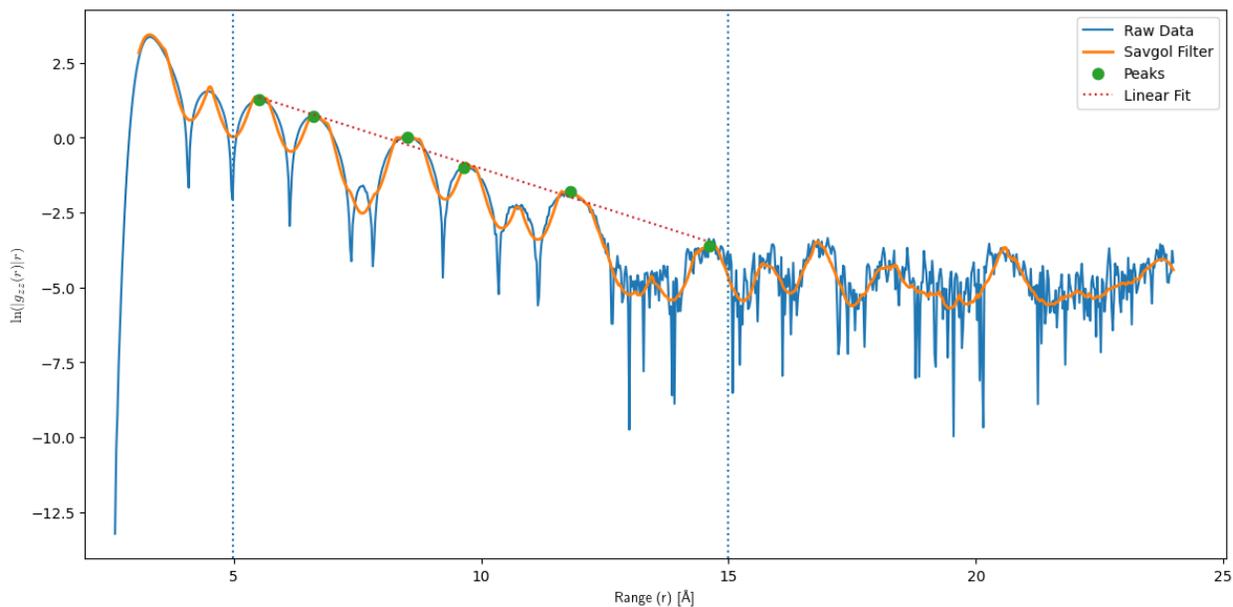

**Figure S54.** Fitting of aqueous RbCl at a concentration of 5.61 M. The envelope of the fit is determined by fitting through points at the top of each of the marked peaks. A Savgol filter was applied to smooth the data to assist peak finding, particularly at higher values of $r$. The vertical lines indicate the extrema that fitting was carried out over.

# CsCl$_{(aq)}$

**Table S6.** Concentrations, number of ion pairs (# IP), the number of solvent molecules (# Solvent) and conventionally-derived $\lambda_S$ values for aqueous CsCl simulations (see Figures S55-S65).

| Conc (m) | Conc (M) | # IP | # Solvent | Decay Length (Å) |
|---|---|---|---|---|
| 0.1 | 0.11 | 8 | 4167 | 6.02 ± 0.08 |
| 0.5 | 0.50 | 38 | 4167 | 3.55 ± 0.23 |
| 0.75 | 0.73 | 56 | 4167 | 3.07 ± 0.42 |
| 1.0 | 0.96 | 75 | 4167 | 2.04 ± 0.26 |
| 1.25 | 1.20 | 94 | 4167 | 1.94 ± 0.26 |
| 1.5 | 1.42 | 113 | 4167 | 1.81 ± 0.17 |
| 2.0 | 1.86 | 150 | 4167 | 1.38 ± 0.14 |
| 3.0 | 2.68 | 225 | 4167 | 1.96 ± 0.11 |
| 4.0 | 3.44 | 300 | 4167 | 1.93 ± 0.11 |
| 5.0 | 4.15 | 375 | 4167 | 2.01 ± 0.14 |
| 10.0 | 6.96 | 750 | 4167 | 2.29 ± 0.13 |

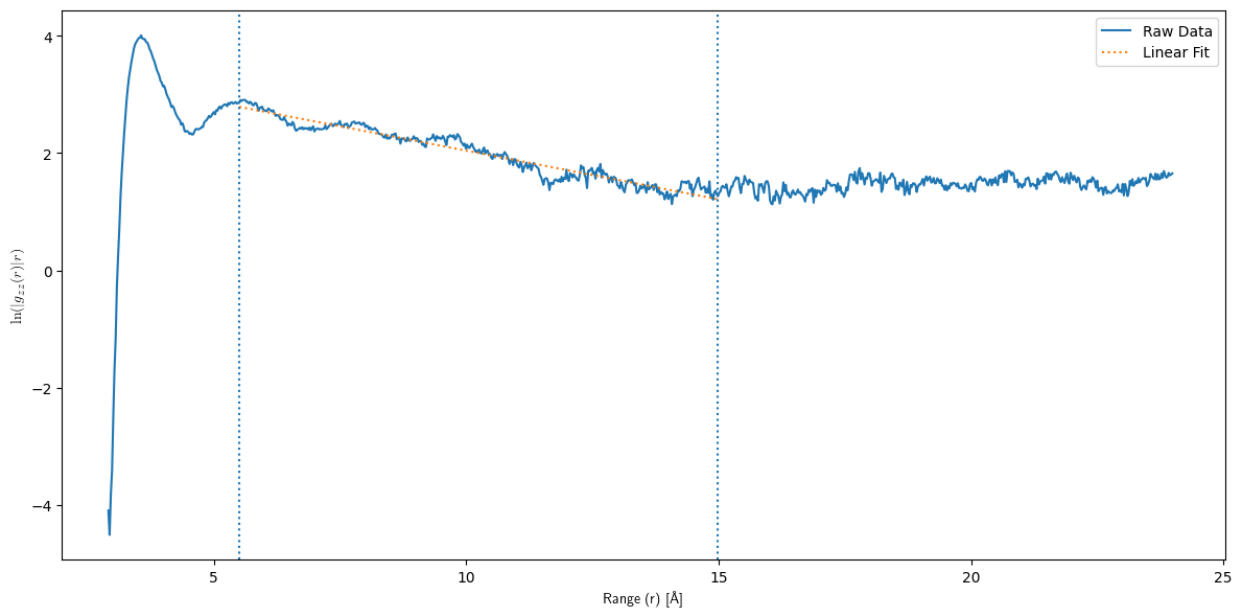

**Figure S55.** Fitting of aqueous CsCl at a concentration of 0.11 M. In this instance, because of the presence of a plateau in the data, a single straight line was used to fit to the points, rather than fitting the peaks. The vertical lines indicate the extrema that fitting was carried out over.

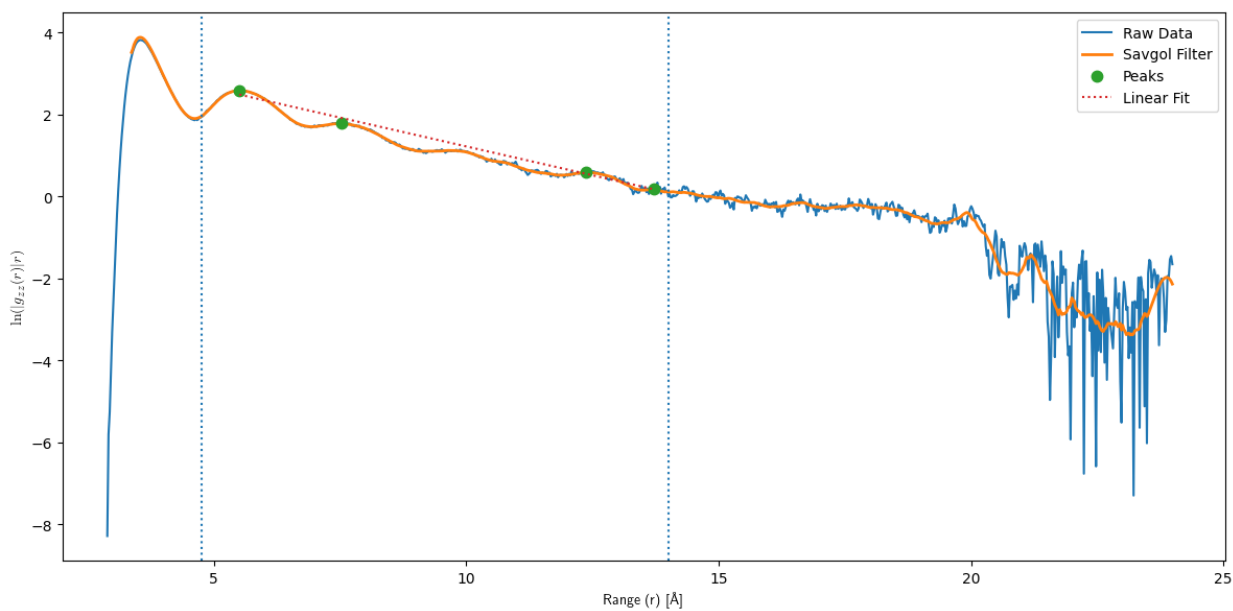

**Figure S56.** Fitting of aqueous CsCl at a concentration of 0.50 M. The envelope of the fit is determined by fitting through points at the top of each of the marked peaks. A Savgol filter was applied to smooth the data to assist peak finding, particularly at higher values of $r$. The vertical lines indicate the extrema that fitting was carried out over.

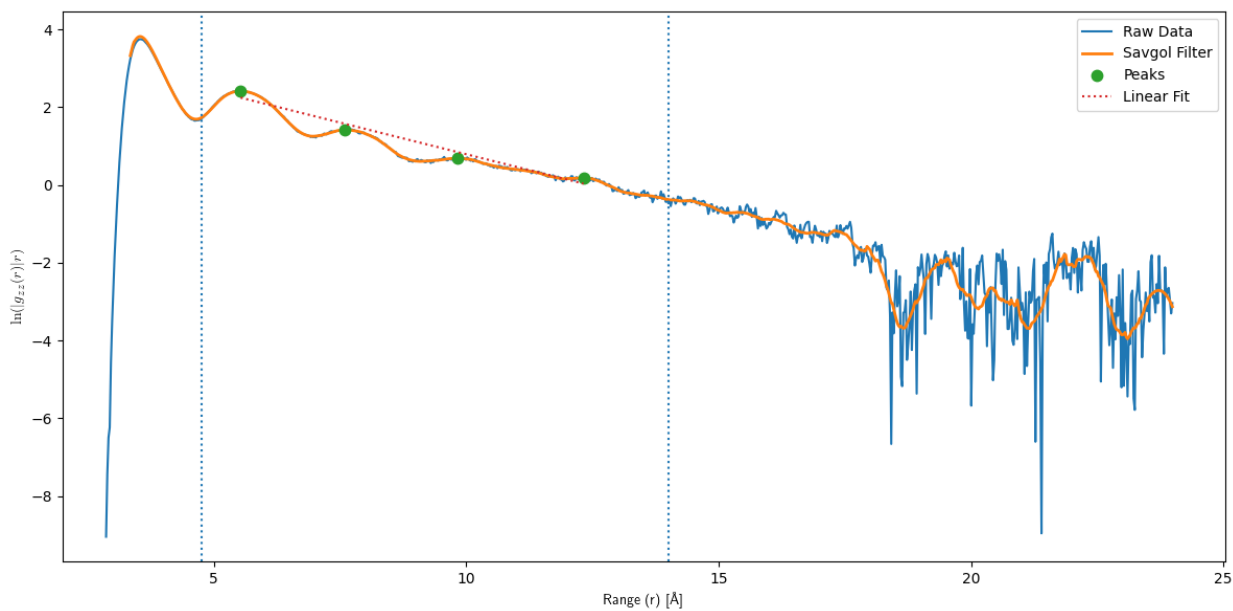

**Figure S57.** Fitting of aqueous CsCl at a concentration of 0.73 M. The envelope of the fit is determined by fitting through points at the top of each of the marked peaks. A Savgol filter was applied to smooth the data to assist peak finding, particularly at higher values of $r$. The vertical lines indicate the extrema that fitting was carried out over.

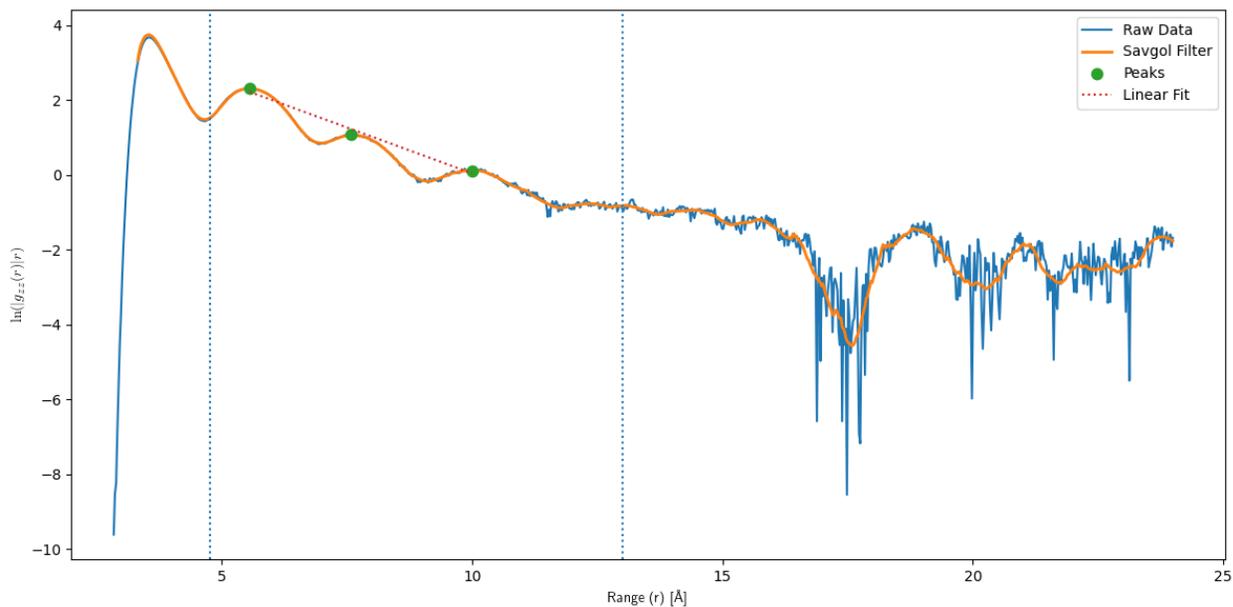

**Figure S58.** Fitting of aqueous CsCl at a concentration of 0.96 M. The envelope of the fit is determined by fitting through points at the top of each of the marked peaks. A Savgol filter was applied to smooth the data to assist peak finding, particularly at higher values of $r$. The vertical lines indicate the extrema that fitting was carried out over.

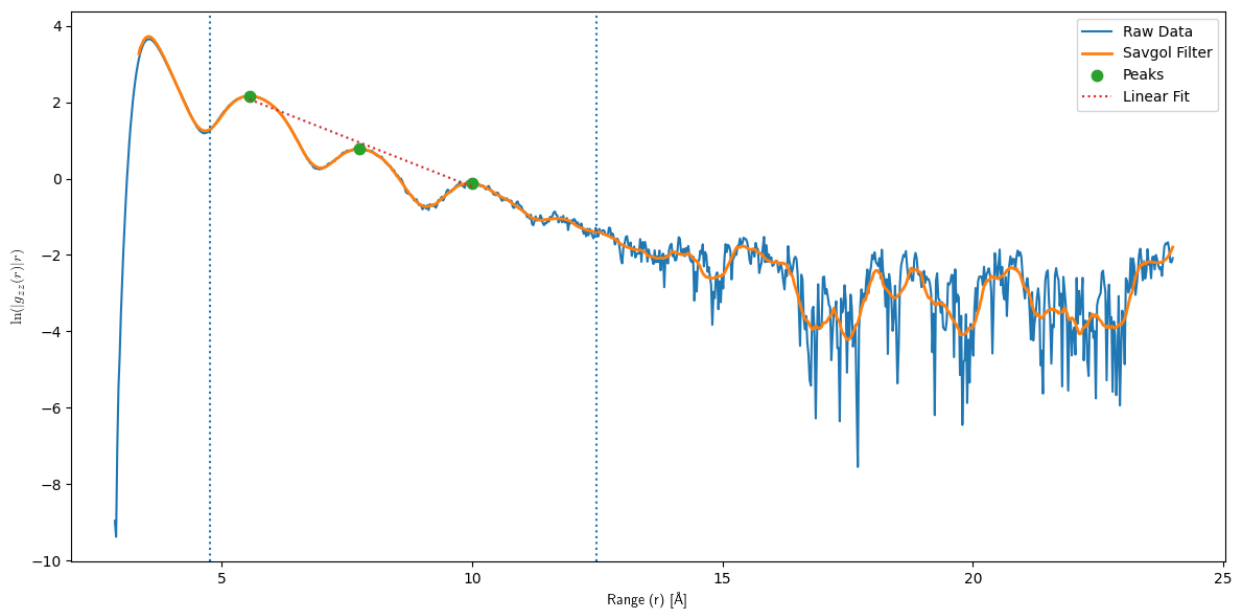

**Figure S59.** Fitting of aqueous CsCl at a concentration of 1.20 M. The envelope of the fit is determined by fitting through points at the top of each of the marked peaks. A Savgol filter was applied to smooth the data to assist peak finding, particularly at higher values of $r$. The vertical lines indicate the extrema that fitting was carried out over.

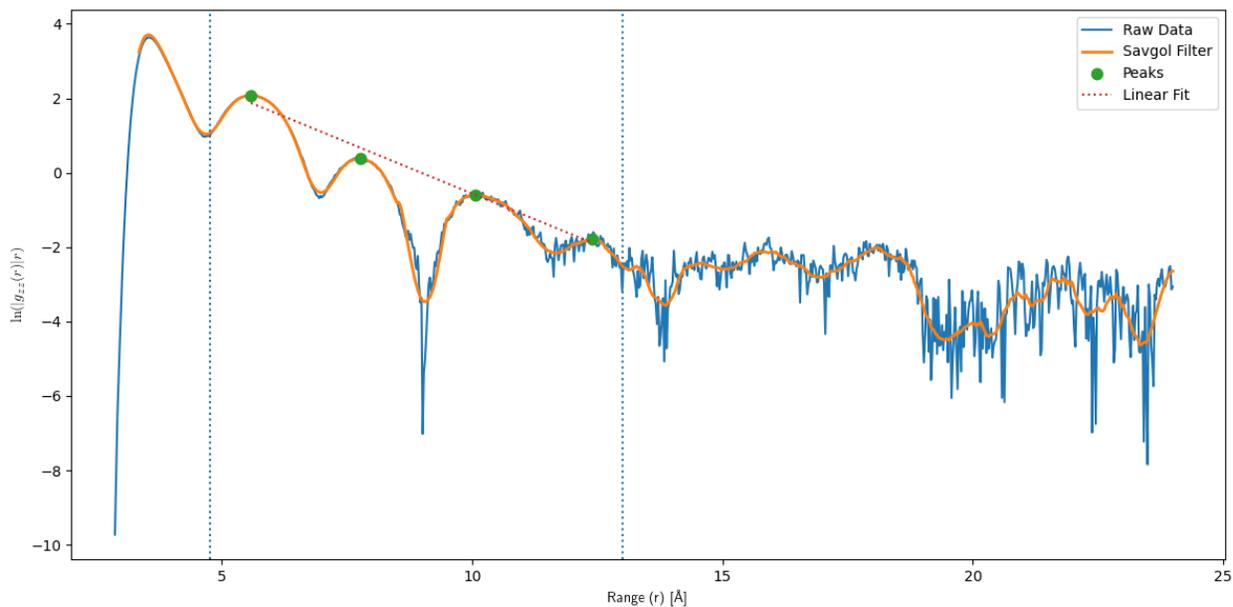

**Figure S60.** Fitting of aqueous CsCl at a concentration of 1.42 M. The envelope of the fit is determined by fitting through points at the top of each of the marked peaks. A Savgol filter was applied to smooth the data to assist peak finding, particularly at higher values of $r$. The vertical lines indicate the extrema that fitting was carried out over.

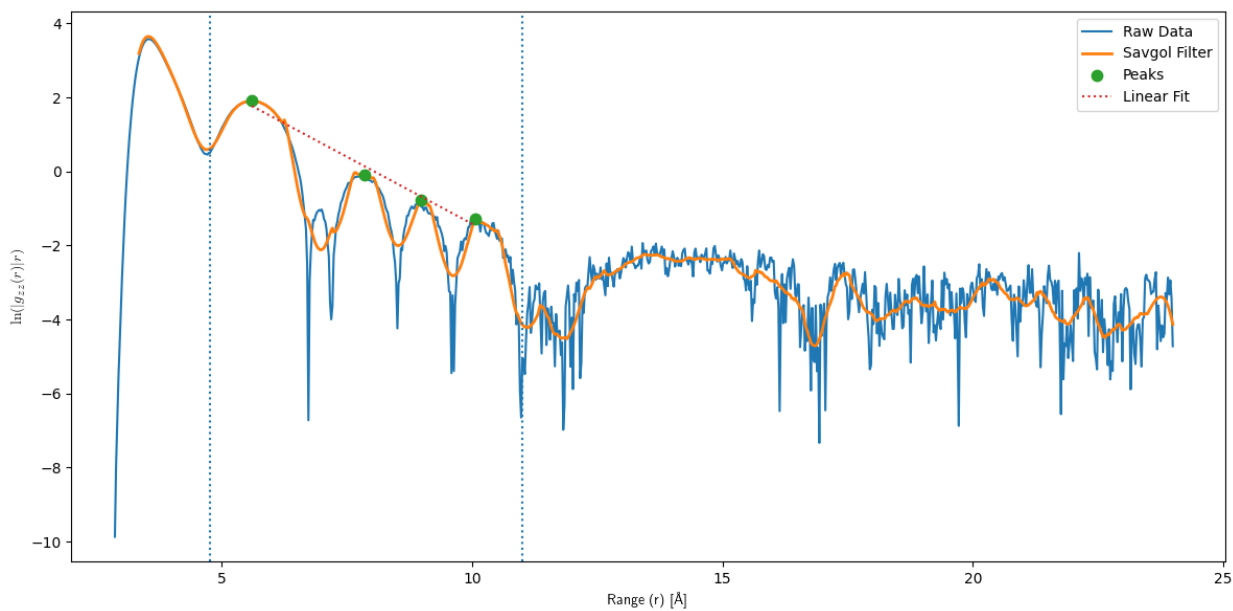

**Figure S61.** Fitting of aqueous CsCl at a concentration of 1.86 M. The envelope of the fit is determined by fitting through points at the top of each of the marked peaks. A Savgol filter was applied to smooth the data to assist peak finding, particularly at higher values of $r$. The vertical lines indicate the extrema that fitting was carried out over.

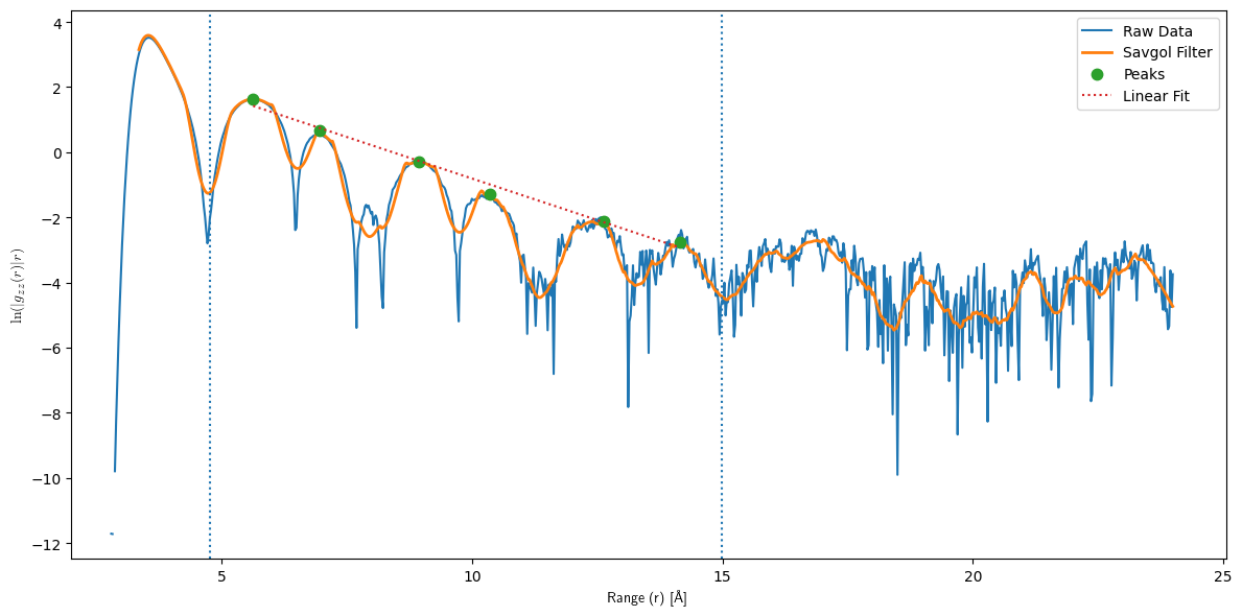

**Figure S62.** Fitting of aqueous CsCl at a concentration of 2.68 M. The envelope of the fit is determined by fitting through points at the top of each of the marked peaks. A Savgol filter was applied to smooth the data to assist peak finding, particularly at higher values of $r$. The vertical lines indicate the extrema that fitting was carried out over.

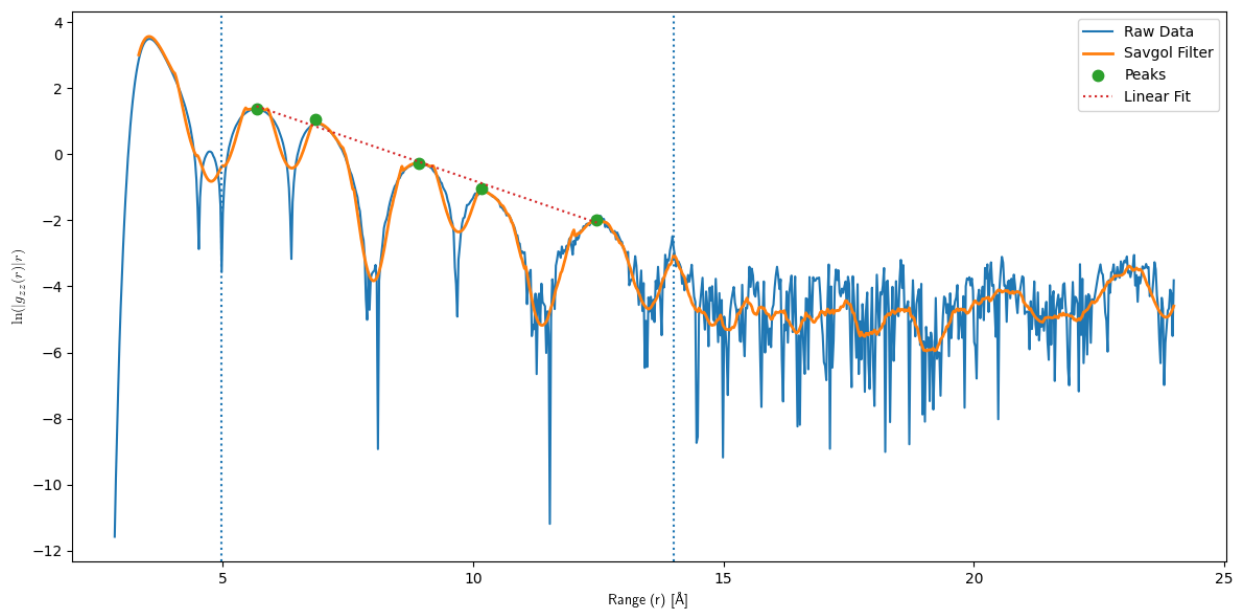

**Figure S63.** Fitting of aqueous CsCl at a concentration of 3.44 M. The envelope of the fit is determined by fitting through points at the top of each of the marked peaks. A Savgol filter was applied to smooth the data to assist peak finding, particularly at higher values of $r$. The vertical lines indicate the extrema that fitting was carried out over.

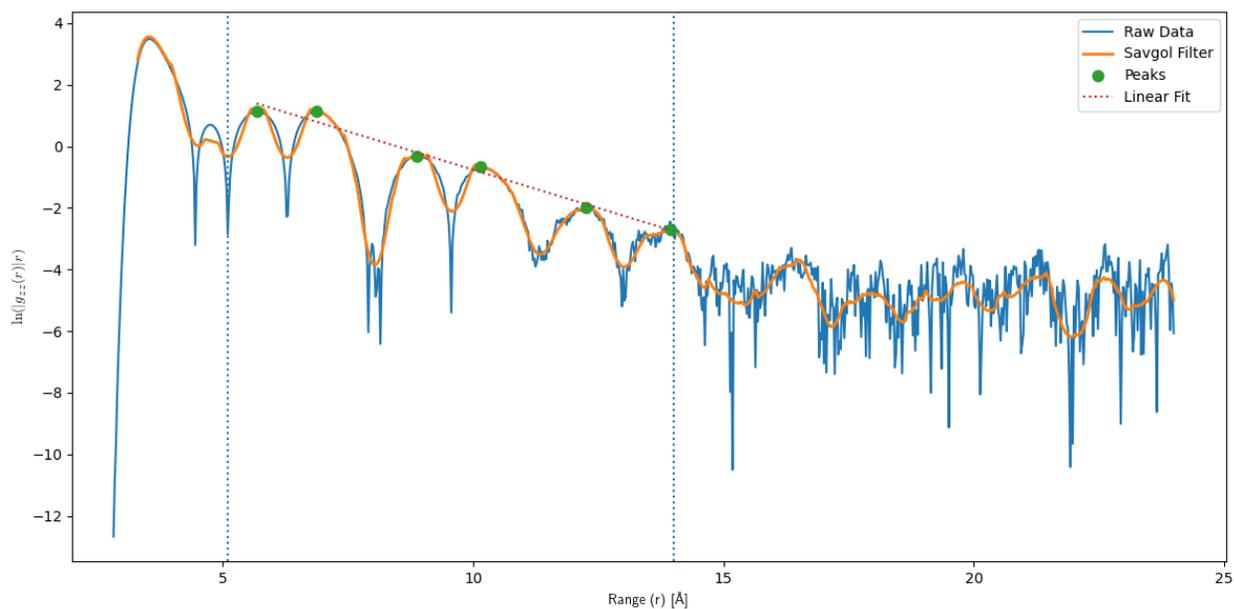

**Figure S64.** Fitting of aqueous CsCl at a concentration of 4.15 M. The envelope of the fit is determined by fitting through points at the top of each of the marked peaks. A Savgol filter was applied to smooth the data to assist peak finding, particularly at higher values of $r$. The vertical lines indicate the extrema that fitting was carried out over.

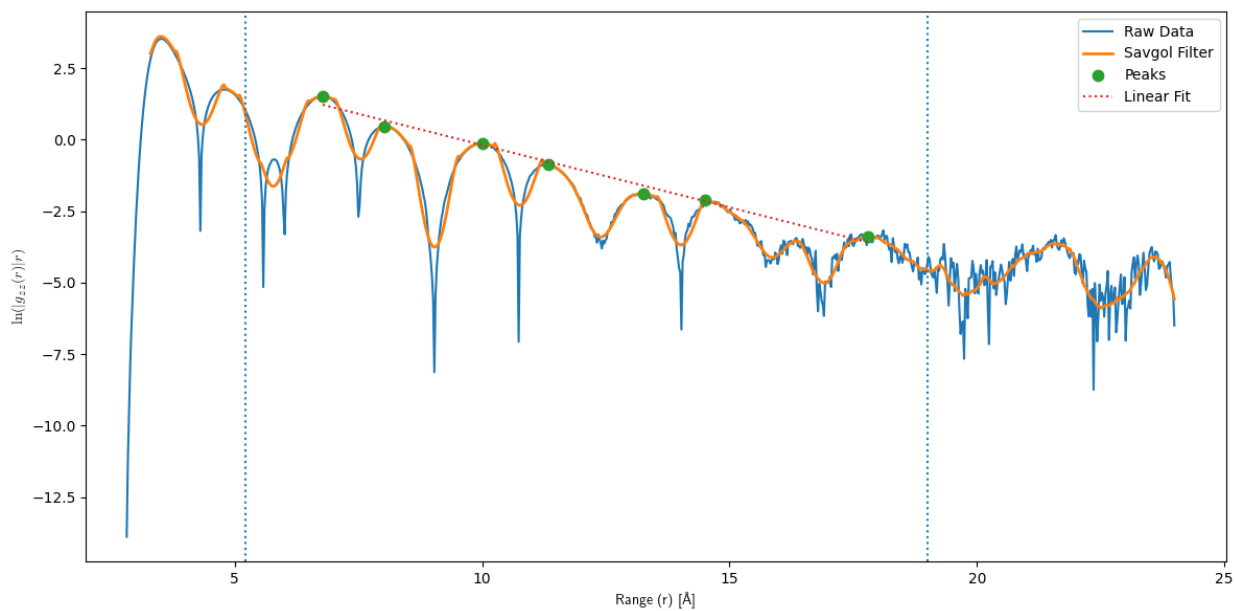

**Figure S65.** Fitting of aqueous CsCl at a concentration of 6.96 M. The envelope of the fit is determined by fitting through points at the top of each of the marked peaks. A Savgol filter was applied to smooth the data to assist peak finding, particularly at higher values of $r$. The vertical lines indicate the extrema that fitting was carried out over.

# Radial Distribution Functions and Discrete Fourier Transforms

**LiCl$_{(aq)}$**

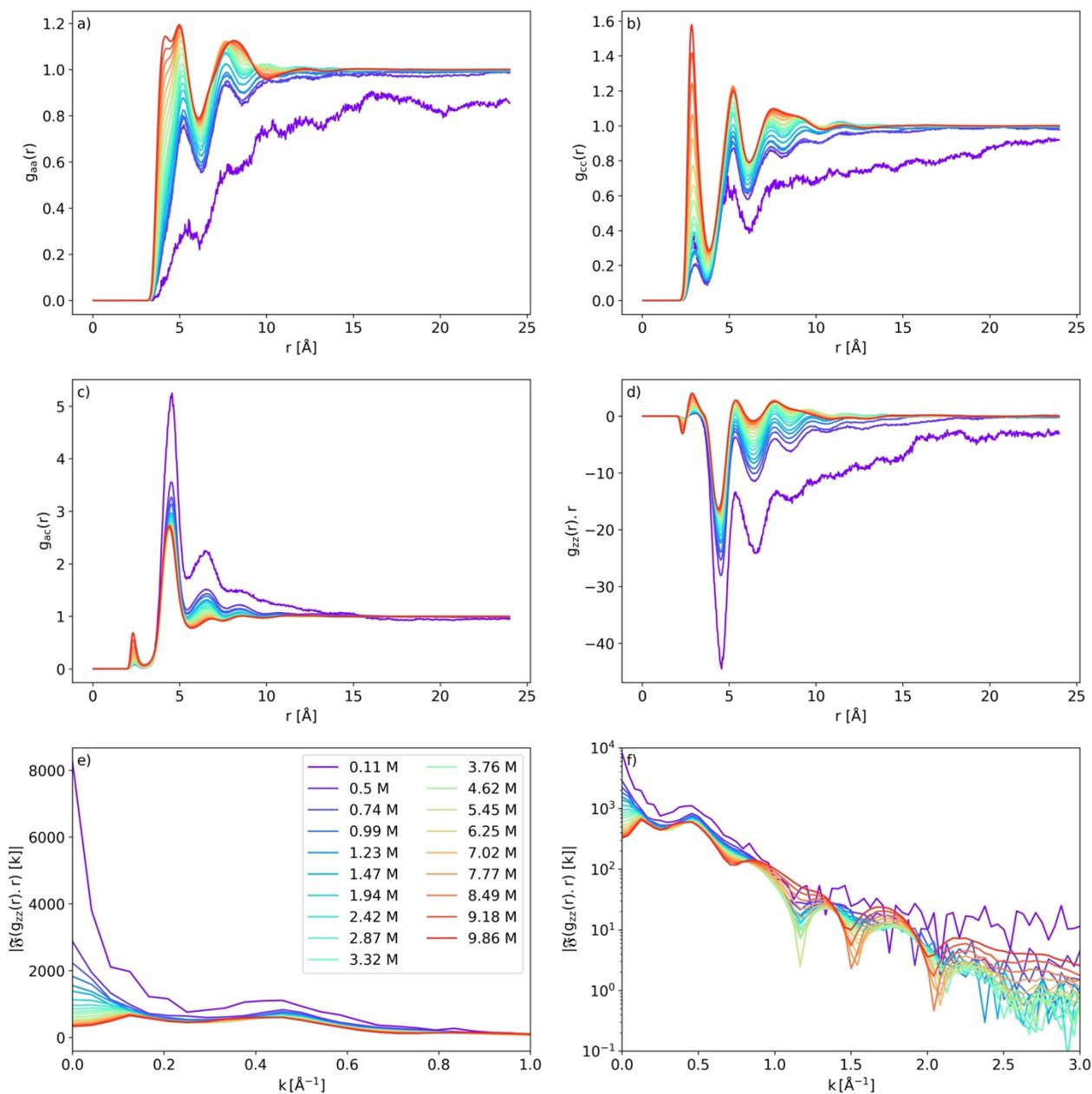

**Figure S66.** Radial distribution functions and discrete Fourier Transforms for LiCl$_{(aq)}$. a) anion-anion b) cation-cation c) anion-cation d) charge-charge correlation function $g_{zz}(r) \cdot r$ e) $|\mathfrak{F}(g_{zz}(r) \cdot r)[k]|$ f) $|\mathfrak{F}(g_{zz}(r) \cdot r)[k]|$ on a log scale.

**NaCl_(aq)**

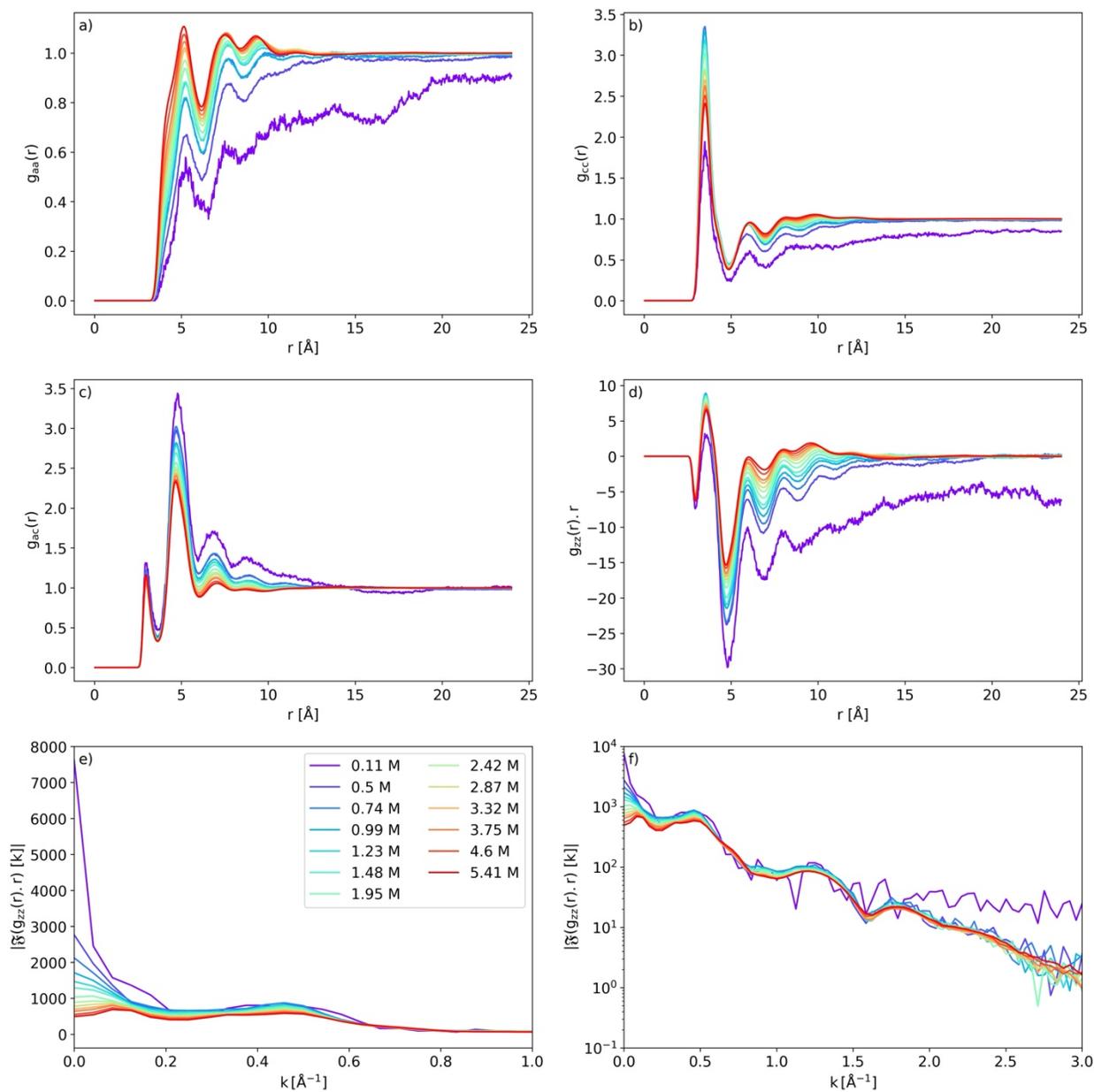

**Figure S67.** Radial distribution functions and discrete Fourier Transforms for NaCl$_{(aq)}$. a) anion-anion b) cation-cation c) anion-cation d) charge-charge correlation function $g_{zz}(r) \cdot r$
e) $|\mathfrak{F}(g_{zz}(r) \cdot r)[k]|$ f) $|\mathfrak{F}(g_{zz}(r) \cdot r)[k]|$ on a log scale.

**KCl$_{(aq)}$**

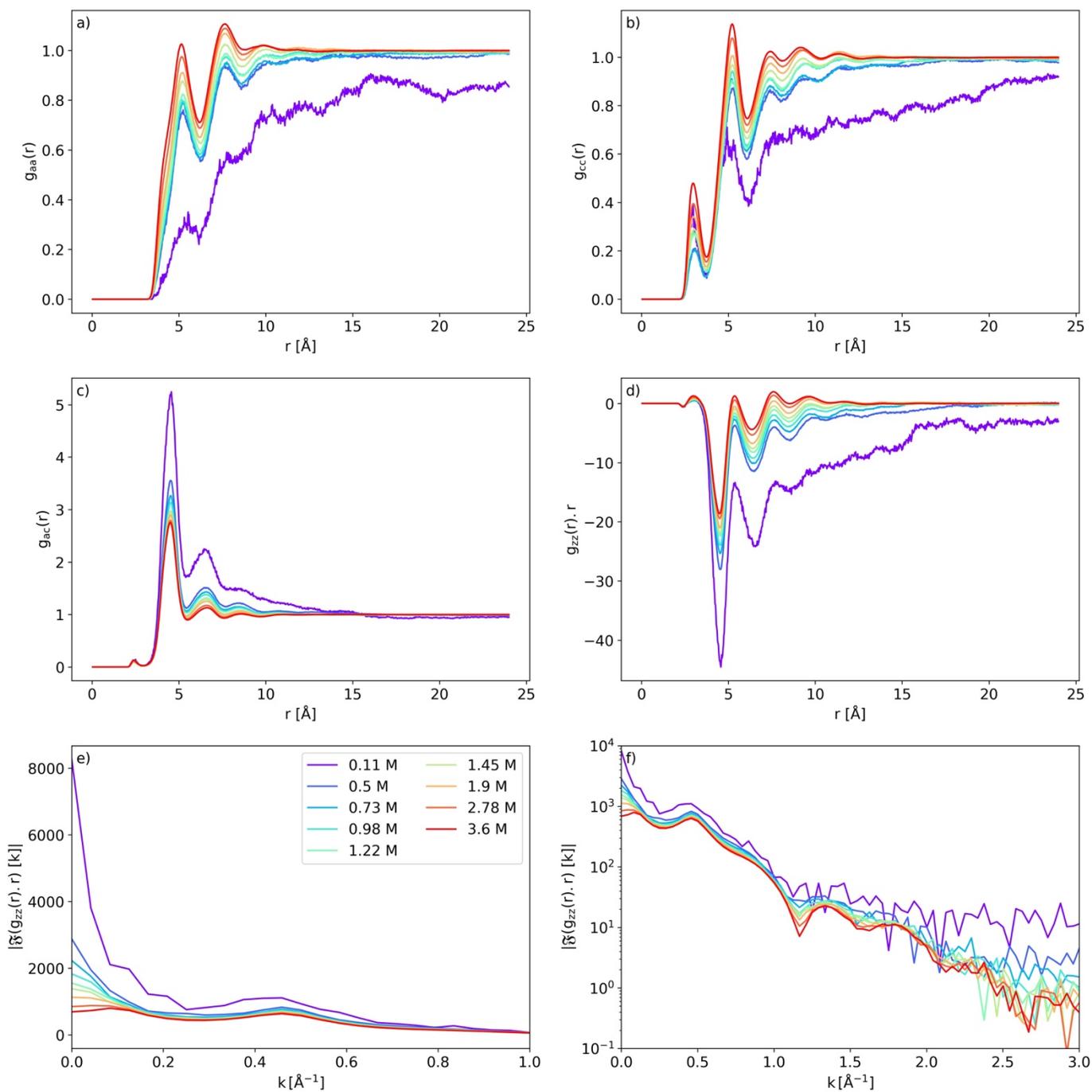

**Figure S68.** Radial distribution functions and discrete Fourier Transforms for KCl$_{(aq)}$. a) anion-anion b) cation-cation c) anion-cation d) charge-charge correlation function $g_{zz}(r) \cdot r$ e) $|\mathfrak{F}(g_{zz}(r) \cdot r)[k]|$ f) $|\mathfrak{F}(g_{zz}(r) \cdot r)[k]|$ on a log scale.

**RbCl$_{(aq)}$**

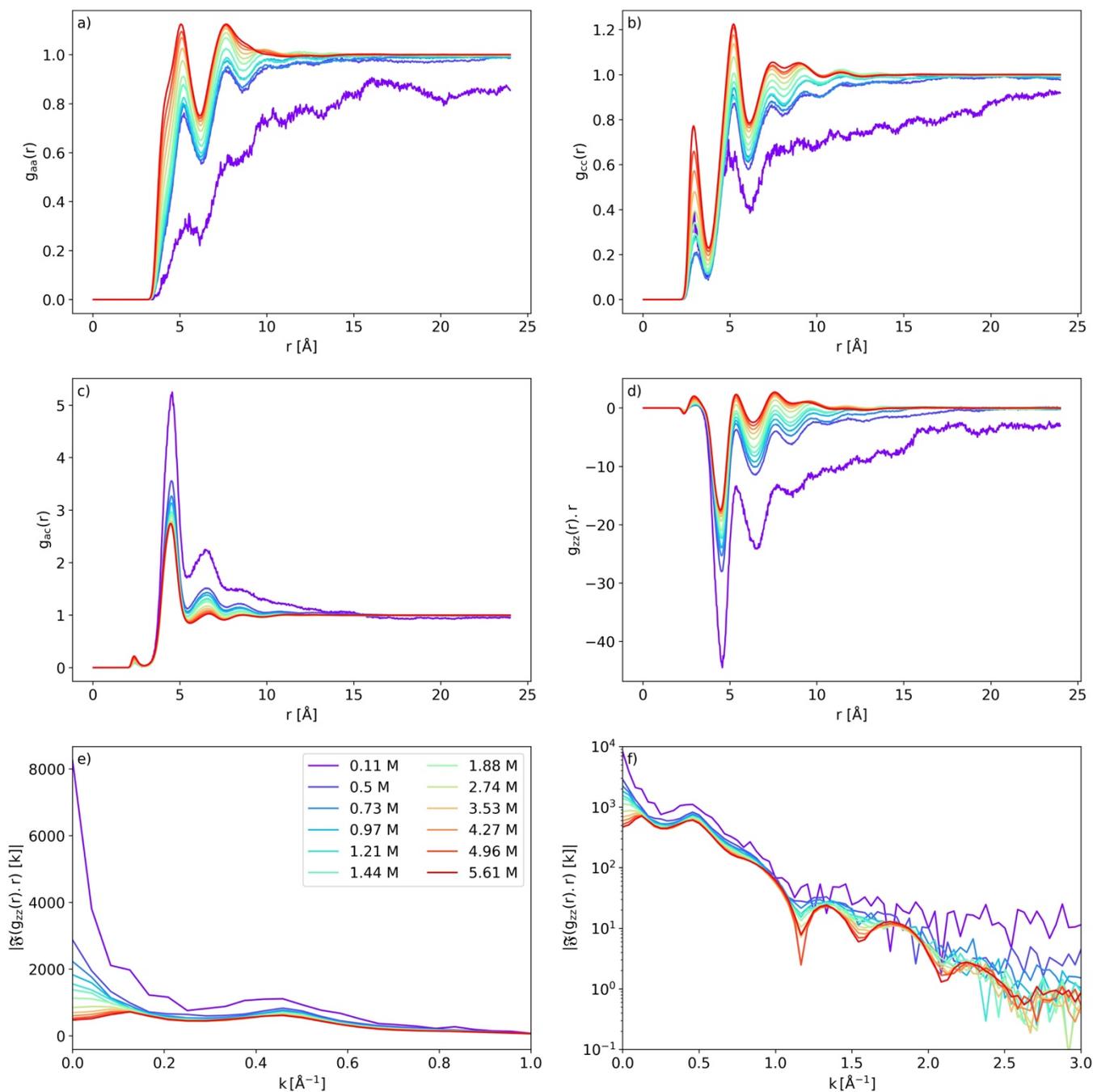

**Figure S69.** Radial distribution functions and discrete Fourier Transforms for RbCl$_{(aq)}$. a) anion-anion b) cation-cation c) anion-cation d) charge-charge correlation function $g_{zz}(r) \cdot r$ e) $|\mathfrak{F}(g_{zz}(r) \cdot r)[k]|$ f) $|\mathfrak{F}(g_{zz}(r) \cdot r)[k]|$ on a log scale.

# CsCl$_{(aq)}$

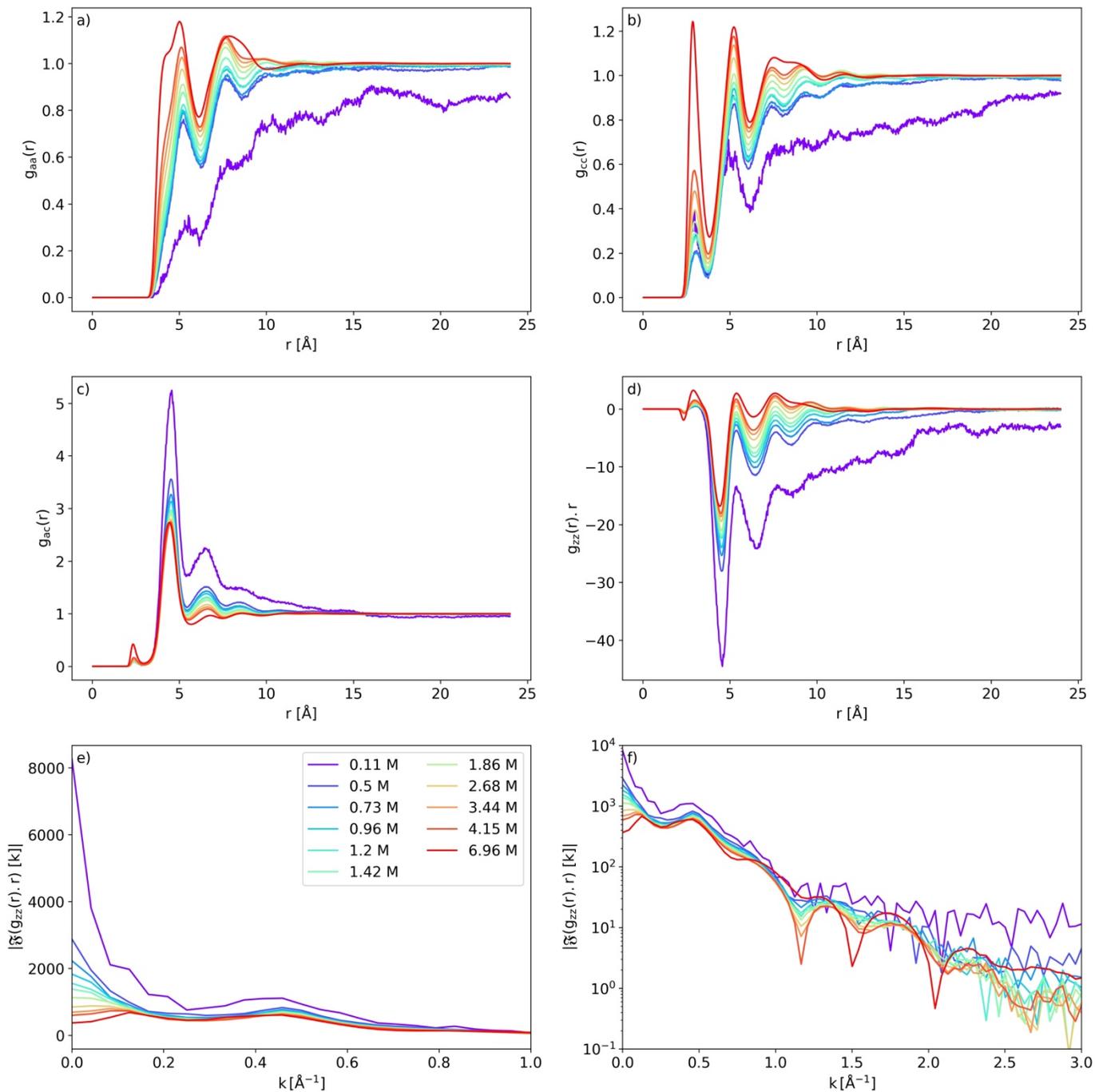

**Figure S70.** Radial distribution functions and discrete Fourier Transforms for CsCl$_{(aq)}$. a) anion-anion b) cation-cation c) anion-cation d) charge-charge correlation function $g_{zz}(r) \cdot r$ e) $|\mathfrak{F}(g_{zz}(r) \cdot r)[k]|$ f) $|\mathfrak{F}(g_{zz}(r) \cdot r)[k]|$ on a log scale.

# Simulation Box Size Comparison

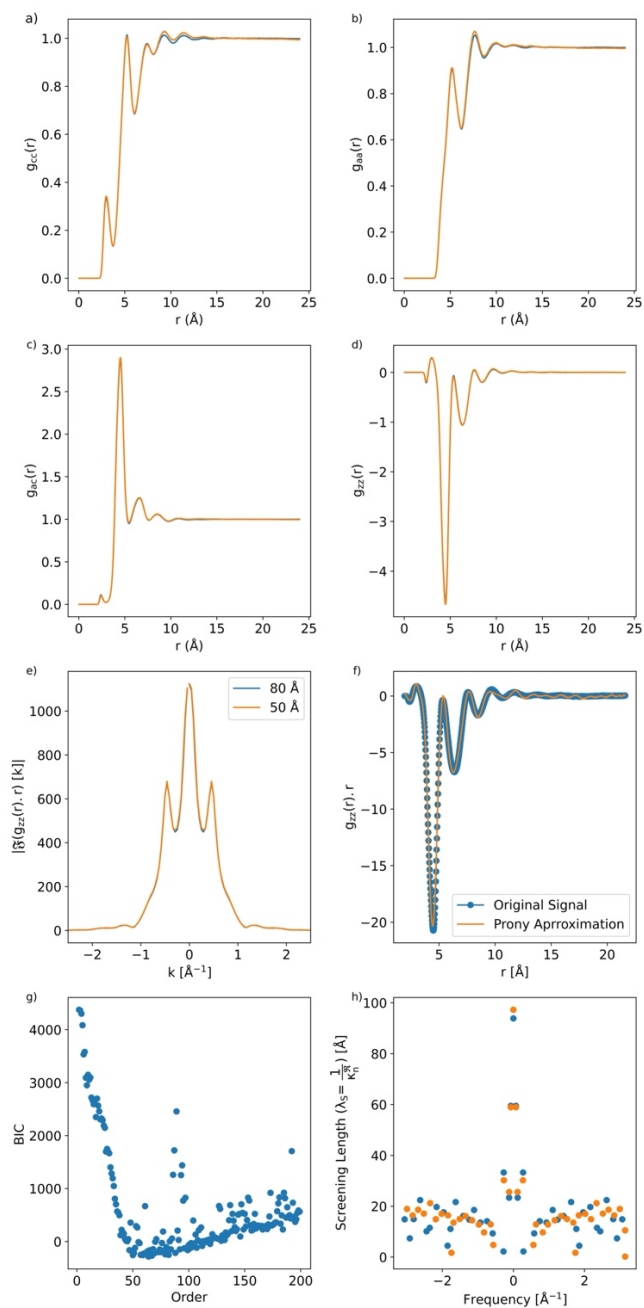

**Figure S71.** Comparison between a box side length of 50 and 80 Å for 1.94 M LiCl$_{(aq)}$ to determine if finite size affects due to periodic boundary conditions impact the Prony's method results. a) – c) radial distribution functions $g_{cc}(r), g_{aa}(r)$ and $g_{ac}(r)$. d) $g_{zz}(r)$, e) the Fourier transform of $g_{zz}(r) \cdot r$ (e)). In all cases these functions are nearly identical for the two box side lengths. f) Prony's method approximation for the 80 Å side length. g) BIC as a function of increasing Prony terms. h) Comparison of extracted screening lengths as a function of frequency from 50 and 80 Å systems, showing only minor deviations and that both anomalous and normal underscreening are presented regardless of the system size.

# Prony's Method Fits

## LiCl$_{(aq)}$

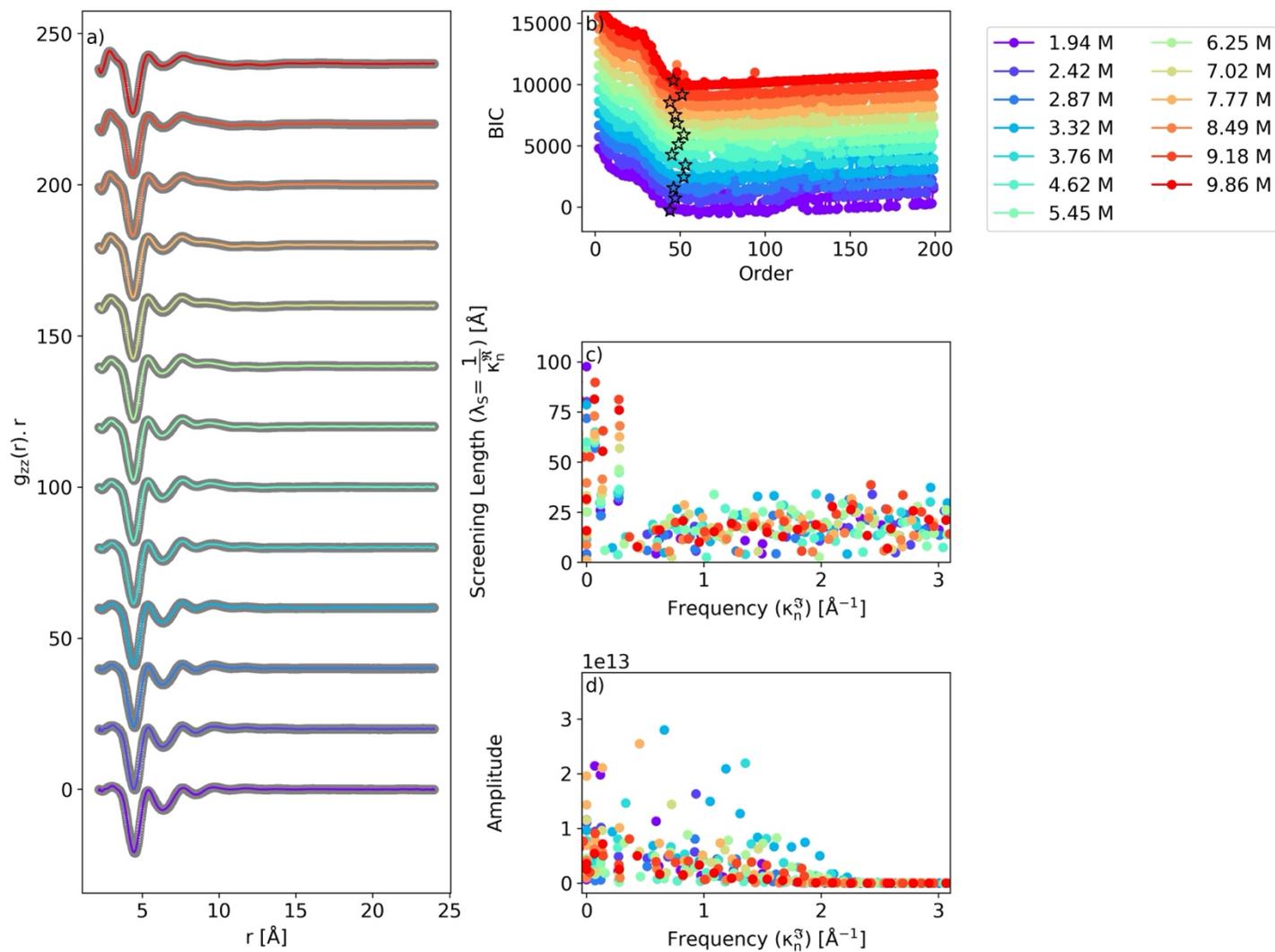

**Figure S72.** Prony's method applied to LiCl$_{(aq)}$ for concentrations ~2.0-10 M. a) The original $g_{zz}(r) \cdot r$ (grey) and Prony's method approximation (coloured), each concentration is offset by 20. b) BIC as a function of increasing Prony terms. Reconstructions using 2-200 Prony terms are considered, and then the resulting BIC curve is fitting using a cubic univariate spline function to find the minimum, shown using a star marker c) Mode screening lengths for all concentrations as a function of mode frequency. d) Mode amplitude as a function of concentration.

**NaCl$_{(aq)}$**

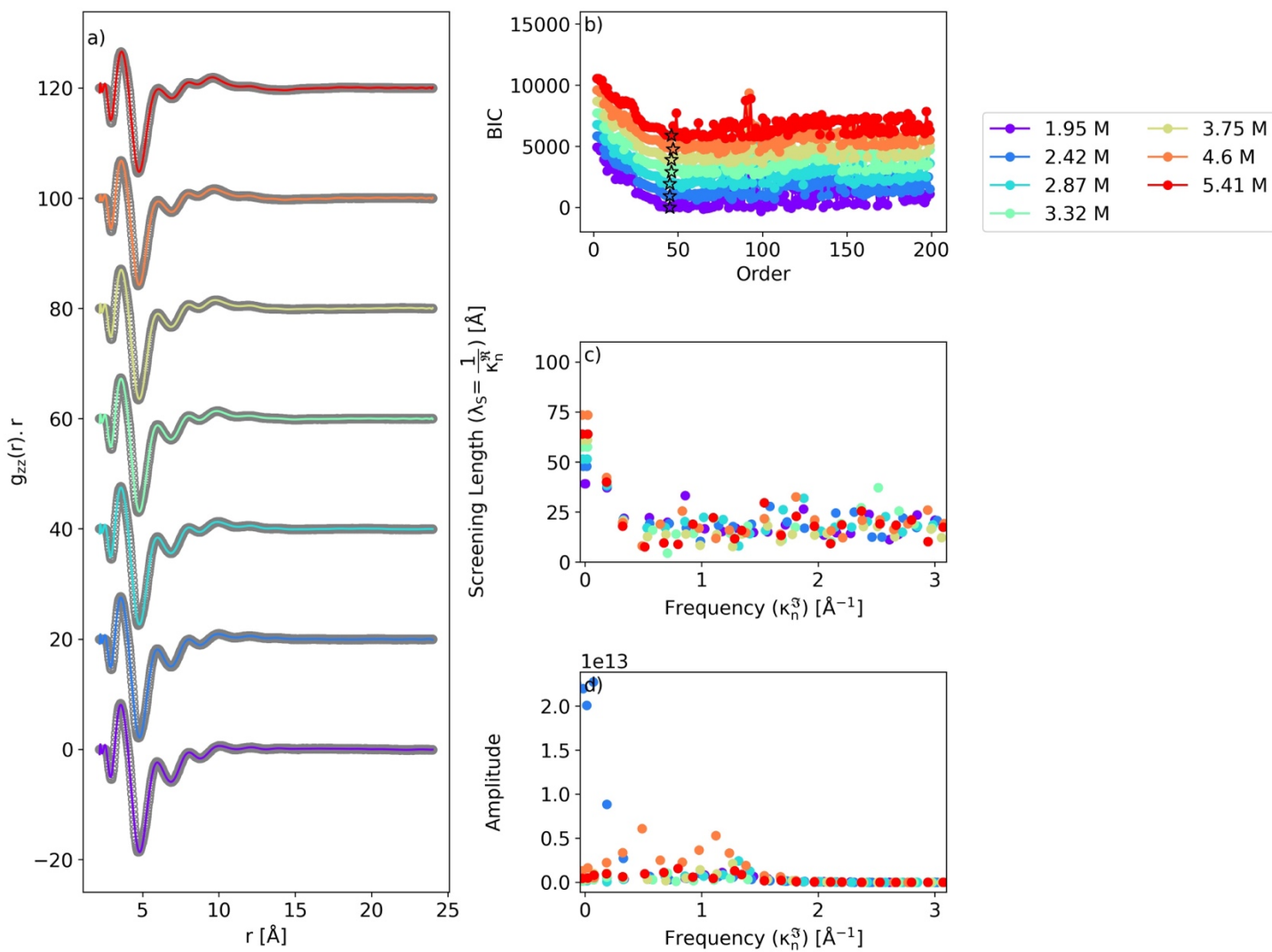

**Figure S73.** Prony's method applied to NaCl$_{(aq)}$ for concentrations >2.0 M. a) The original $g_{zz}(r) \cdot r$ (grey) and Prony's method approximation (coloured), each concentration is offset by 20. b) BIC as a function of increasing Prony terms. Reconstructions using 2-200 Prony terms are considered, and then the resulting BIC curve is fitting using a cubic univariate spline function to find the minimum, shown using a star marker c) Mode screening lengths for all concentrations as a function of mode frequency. d) Mode amplitude as a function of concentration.

**KCl$_{(aq)}$**

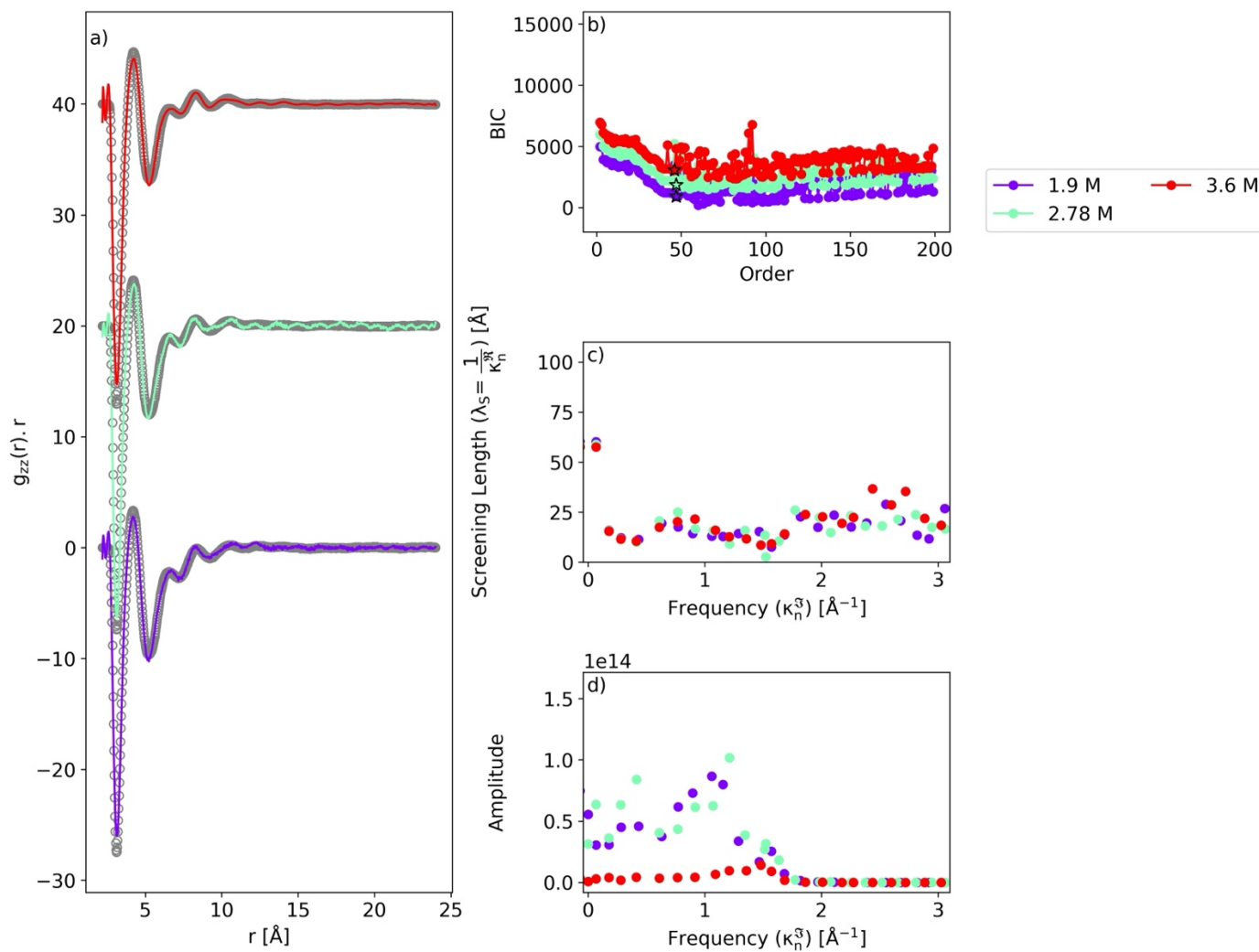

**Figure S74.** Prony's method applied to KCl$_{(aq)}$ for concentrations >2.0 M. a) The original $g_{zz}(r) \cdot r$ (grey) and Prony's method approximation (coloured), each concentration is offset by 20. b) BIC as a function of increasing Prony terms. Reconstructions using 2-200 Prony terms are considered, and then the resulting BIC curve is fitting using a cubic univariate spline function to find the minimum, shown using a star marker c) Mode screening lengths for all concentrations as a function of mode frequency. d) Mode amplitude as a function of concentration.

**RbCl$_{(aq)}$**

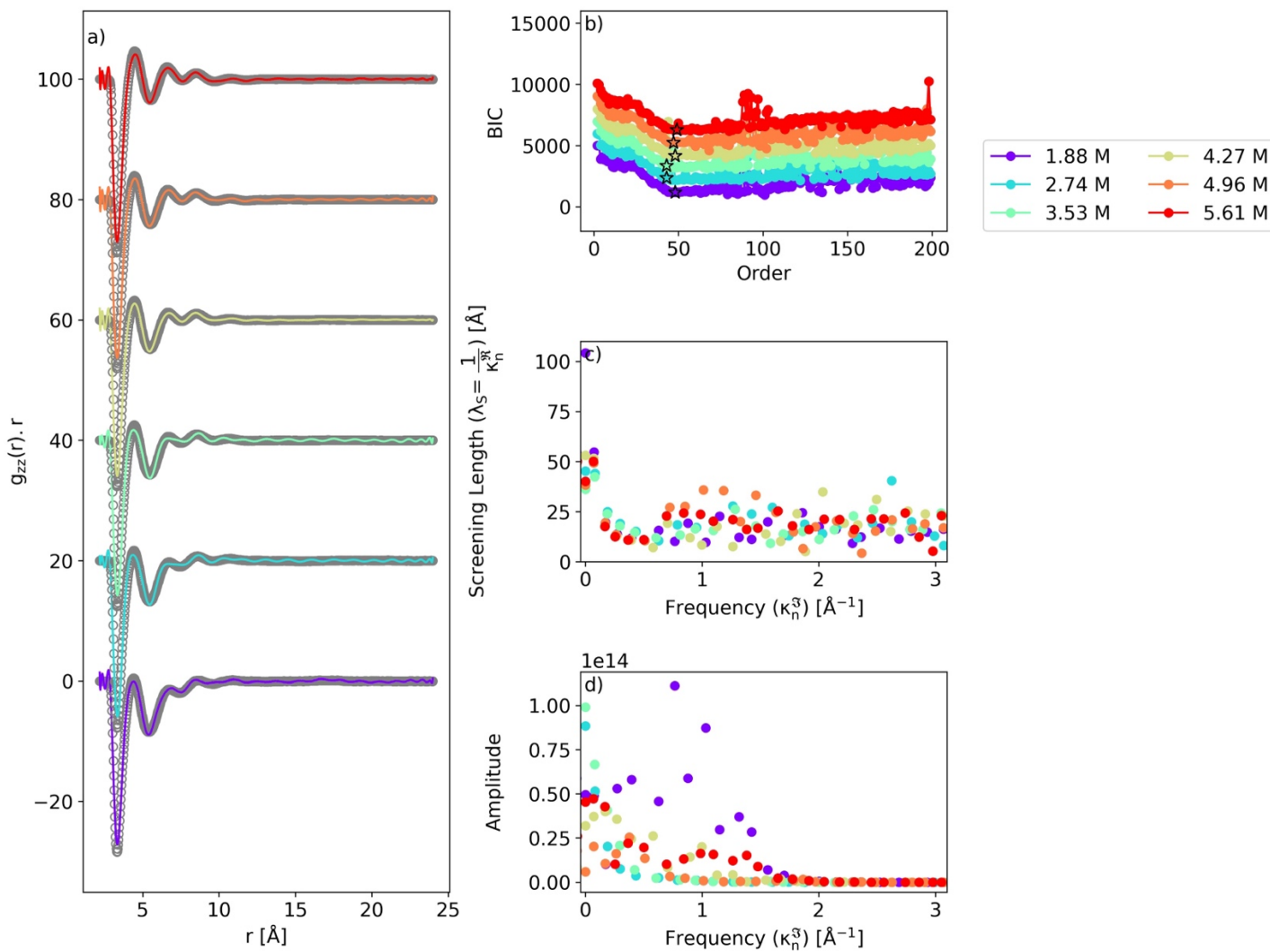

**Figure S75.** Prony's method applied to RbCl$_{(aq)}$ for concentrations >2.0 M. a) The original $g_{zz}(r) \cdot r$ (grey) and Prony's method approximation (coloured), each concentration is offset by 20. b) BIC as a function of increasing Prony terms. Reconstructions using 2-200 Prony terms are considered, and then the resulting BIC curve is fitting using a cubic univariate spline function to find the minimum, shown using a star marker c) Mode screening lengths for all concentrations as a function of mode frequency. d) Mode amplitude as a function of concentration.

**CsCl_(aq)**

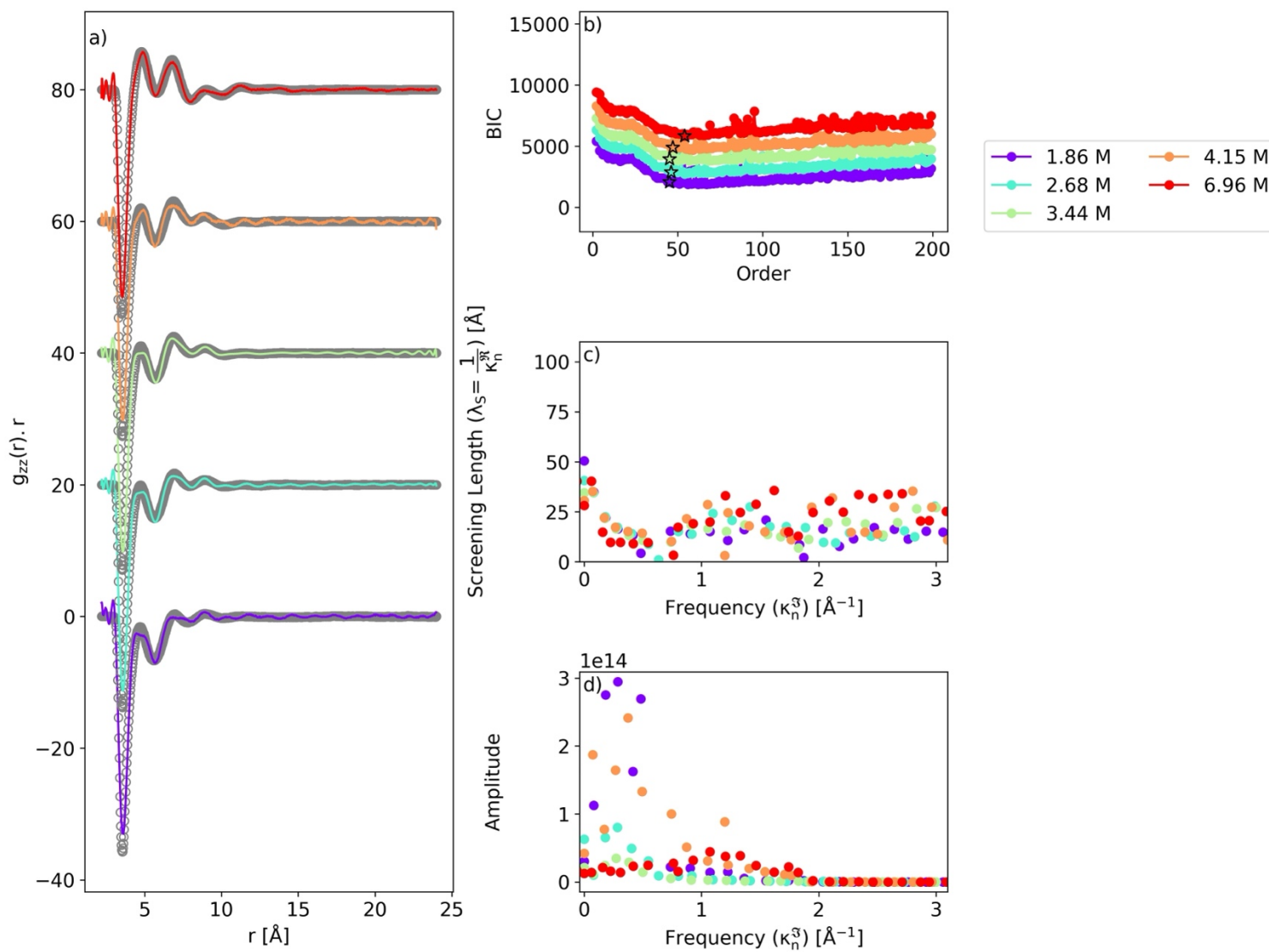

**Figure S76.** Prony's method applied to RbCl_(aq) for concentrations >2.0 M. a) The original $g_{zz}(r) \cdot r$ (grey) and Prony's method approximation (coloured), each concentration is offset by 20. b) BIC as a function of increasing Prony terms. Reconstructions using 2-200 Prony terms are considered, and then the resulting BIC curve is fitting using a cubic univariate spline function to find the minimum, shown using a star marker c) Mode screening lengths for all concentrations as a function of mode frequency. d) Mode amplitude as a function of concentration.